\documentclass[journal]{IEEEtran}
\usepackage{cite}
\usepackage{amsmath,amssymb,amsfonts}
\usepackage{algorithm}
\usepackage{algorithmic}
\usepackage{graphicx}
\usepackage{textcomp}
\usepackage{xcolor}
\usepackage{fixltx2e}
\usepackage{url}
\usepackage{tikz}
\usetikzlibrary{positioning}
\usepackage{diagbox}
\usepackage[subrefformat=parens,labelformat=parens]{subfig}
\usepackage{hyperref}
\usepackage{varwidth}

\begin{document}
%
\title{A Technical Overview of AV1}
%
%
%

\author{Jingning~Han,~\IEEEmembership{Senior~Member,~IEEE,}
        Bohan~Li,~\IEEEmembership{Member,~IEEE,}
        Debargha~Mukherjee,~\IEEEmembership{Senior~Member,~IEEE,}
        Ching-Han~Chiang,
        Adrian~Grange,        
        Cheng~Chen,
        Hui~Su,
        Sarah~Parker,
        Sai~Deng,    
        Urvang~Joshi,
        Yue~Chen,        
        Yunqing~Wang,
        Paul~Wilkins,
        Yaowu~Xu,~\IEEEmembership{Senior~Member,~IEEE,}
        and~James~Bankoski
\thanks{This work has been submitted to the IEEE for possible publication. Copyright may be transferred without notice, after which this version may no longer be accessible. J. Han is with the WebM Codec team, Google LLC, Mountain View, CA, 94043
 USA e-mail: \{jingning\}@google.com. }
 }

%



\maketitle

\begin{abstract}
The AV1 video compression format is developed by the Alliance for Open Media consortium. It achieves more than $30\%$ reduction in bit-rate compared to its predecessor VP9 for the same decoded video quality. This paper provides a technical overview of the AV1 codec design that enables the compression performance gains with considerations for hardware feasibility.
\end{abstract}


%
\IEEEpeerreviewmaketitle

\section{Introduction}

\IEEEPARstart{T}{he} last decade has seen a steady and significant growth of web-based video applications including video-on-demand (VoD) service, live streaming, conferencing, and virtual reality \cite{CISCO}. Bandwidth and storage costs have driven the need for video compression techniques with better compression efficiency. VP9 \cite{VP9} and HEVC \cite{HEVC}, both debuted in 2013, achieved in the range of $50\%$ higher compression performance \cite{codec_comp} than the prior codec H.264/AVC \cite{h264} and were quickly adopted by the industry.

As the demand for high performance video compression continued to grow, the Alliance for Open Media \cite{AOMedia} was formed in 2015 as a consortium for the development of open, royalty-free technology for multimedia delivery. Its first video compression format AV1, released in 2018, enabled about $30\%$ compression gains over its predecessor VP9. The AV1 format is already supported by many web platforms including Android, Chrome, Microsoft Edge, and Firefox and multiple web-based video service providers, including YouTube, Netflix, Vimeo, and Bitmovin, have begun rolling out AV1 streaming services at scale.

Web-based video applications have seen a rapid shift from conventional desktop computers to mobile devices and TVs in recent years. For example, it is quite common to see users watch YouTube and Facebook videos on mobile phones. Meanwhile nearly all the smart TVs after 2015 have native apps to support movie playback from YouTube, Netflix, and Amazon. Therefore, a new generation video compression format needs to ensure that it is decodable on these devices. However, to improve the compression efficiency, it is almost inevitable that a new codec will include coding techniques that are more computationally complex than its predecessors. With the slowdown in the growth of general CPU clock frequency and power constraints on mobile devices in particular, next generation video compression codecs are expected to rely heavily on dedicated hardware decoders. Therefore during the AV1 development process, all the coding tools were carefully reviewed for hardware considerations (e.g., latency, silicon area, etc.), which resulted in a codec design well balanced for compression performance and hardware feasibility.

This paper provides a technical overview of the AV1 codec. Prior literature highlights some major characteristics of the codec and reports preliminary performance results \cite{AV1-PCS,AV1-ATSIP,bitstream-spec}. A description of the available coding tools in AV1 is provided in \cite{AV1-ATSIP}. For syntax element definition and decoder operation logic, the readers are referred to the AV1 specification \cite{bitstream-spec}. Instead, this paper will focus on the design theories of the compression techniques and the considerations for hardware decoder feasibility, which together define the current state of the AV1 codec. For certain coding tools that potentially demand substantial searches to realize the compression gains, it is imperative to complement them with proper encoder strategies that materialize the coding gains at a practical encoder complexity. We will further explore approaches to optimizing the trade off between encoder complexity and the coding performance therein. The AV1 codec includes contributions from the entire AOMedia teams \cite{AOMedia} and the greater eco-system around the globe. An incomplete contributor list can be found at \cite{libaom-contributor}.

The AV1 codec supports input video signals in the 4:0:0 (monochrome), 4:2:0, 4:2:2, and 4:4:4 formats. The allowed pixel representations are 8-, 10-, and 12-bit. The AV1 codec operates on pixel blocks. Each pixel block is processed in a predictive-transform coding scheme, where the prediction comes from either intra frame reference pixels, inter frame motion compensation, or some combinations of the two. The residuals undergo a 2-D unitary transform to further remove the spatial correlations and the transform coefficients are quantized. Both the prediction syntax elements and the quantized transform coefficient indexes are then entropy coded using arithmetic coding. There are 3 optional in-loop post-processing filter stages to enhance the quality of the reconstructed frame for reference by subsequent coded frames. A normative film grain synthesis unit is also available to improve the perceptual quality of the displayed frames.

We will start by considering frame level designs, before progressing on to look at coding block level operations and the entropy coding system applied to all syntax elements. Finally we will discuss in-loop and out-of-loop filtering. The coding performance is evaluated using libaom AV1 encoder \cite{libaom}, which is developed as a production codec for various services including VoD, video conferencing and light field, with encoder optimizations that utilize the AV1 coding tools for compression performance improvements while keeping the computational complexity in check. We note that the libaom AV1 encoder optimization is being actively developed for better compression performance and higher encoding speed. We refer to the webpage \cite{libaom-tracker}  for the related performance statistics update.

\section{High Level Syntax}
\label{sec:hls}

The AV1 bitstream is packetized into open bitstream units (OBUs). An ordered sequence of OBUs are fed into the AV1 decoding process, where each OBU comprises a variable length string of bytes. An OBU contains a header and a payload. The header identifies the OBU type and specifies the payload size. Typical OBU types include:
\begin{itemize}
\item \textbf{Sequence Header} contains information that applies to the entire sequence, e.g., sequence profile (see Section \ref{sec:leveldef}), whether to enable certain coding tools, etc.

\item \textbf{Temporal Delimiter} indicates the frame presentation time stamp. All displayable frames following a temporal delimiter OBU will use this time stamp, until next temporal delimiter OBU arrives. A temporal delimiter and its subsequent OBUs of the same time stamp are referred to as a temporal unit. In the context of scalable coding, the compression data associated with all representations of a frame at various spatial and fidelity resolutions will be in the same temporal unit.

\item \textbf{Frame Header} sets up the coding information for a given frame, including signaling inter or intra frame type, indicating the reference frames, signaling probability model update method, etc.

\item \textbf{Tile Group} contains the tile data associated with a frame. Each tile can be independently decoded. The collective reconstructions form the reconstructed frame after potential loop filtering.

\item \textbf{Frame} contains the frame header and tile data. The frame OBU is largely equivalent to a frame header OBU and a tile group OBU, but allows less overhead cost.

\item \textbf{Metadata} carries information such as high dynamic range, scalability, and timecode.

\item \textbf{Tile List} contains tile data similar to a tile group OBU. However, each tile here has an additional header that indicates its reference frame index and position in the current frame.  This allows the decoder to process a subset of tiles and display the corresponding part of the frame, without the need to fully decode all the tiles in the frame. Such capability is desirable for light field applications \cite{lightfield}.
\end{itemize}
We refer to \cite{bitstream-spec} for bit field definitions and more detailed consideration of high level syntax.

\section{Reference Frame System}

\subsection{Reference Frames}
\label{sec:refframe}
The AV1 codec allows a maximum of 8 frames in its decoded frame buffer. For a coding frame, it can choose any 7 frames from the decoded frame buffer as its reference frames. The bitstream allows the encoder to explicitly assign each reference a unique reference frame index ranging from 1 to 7. In principle, the reference frames index 1-4 are designated for the frames that precede the current frame in terms of display order whilst index 5-7 are for reference frames coming after the current one. For compound inter prediction, two references can be combined to form the prediction (see Section \ref{sec:compound}). If both reference frames either precede or follow the current frame, this is considered to be uni-directional compound prediction. This contrasts with bi-directional compound prediction where there is one previous and one future reference frame. In practice, the codec can link a reference frame index to any frame in the decoded frame buffer, which allows it to fill all the reference frame indexes when there are not enough reference frames on either side.

In estimation theory, it is commonly known that extrapolation (uni-directional compound) is usually less accurate than interpolation (bi-directional compound) prediction \cite{stat-learning}. The allowed uni-directional reference frame combinations are hence limited to only 4 possible pairs, i.e., (1, 2), (1, 3), (1, 4), and (5, 7), but all the 12 combinations in the bi-directional case are supported. This limitation reduces the total number of compound reference frame combinations from 21 to 16. It follows the assumption that if the numbers of the reference frames on both sides of the current frame in natural display order are largely balanced, the bi-directional predictions are likely to provide better prediction. When most reference frames are on one side of the current frame, the extrapolations that involve the nearest one are more relevant to the current frame. 

When a frame coding is complete, the encoder can decide which reference frame in the decoded frame buffer to replace and explicitly signals this in the bit-stream. The mechanism also allows one to bypass updating the decoded frame buffer. This is particularly useful for high motion videos where certain frames are less relevant to neighboring frames.

\subsection{Alternate Reference Frame}
\label{sec:alt}
The alternate reference frame (ARF) is a frame that will be coded and stored in the decoded frame buffer with the option of not being displayed. It serves as a reference frame for subsequent frames to be processed. To transmit a frame for display, the AV1 codec can either code a new frame or directly use a frame in the decoded frame buffer -- this is called ``show existing frame''. An ARF that is later being directly displayed can be effectively used to code a future frame in a pyramid coding structure \cite{hierarchical}. 

Moreover, the encoder has the option to synthesize a frame that can potentially reduce the collective prediction errors among several display frames. One example is to apply temporal filtering along the motion trajectories of consecutive original frames to build an ARF, which retains the common information \cite{info-theory} with the acquisition noise on each individual frame largely removed. The encoder typically uses a relatively lower quantization step size to code the common information (i.e., ARF) to optimize the overall rate-distortion performance \cite{denoise-ICIP}. A potential downside here is that this results in an additional frame for decoders to process, which could potentially stretch throughput capacity on some hardware. To balance the compression performance and decoder throughput, each level definition defines an upper bound on the permissible decoded sample rate, namely maximum decode rate. Since the decoded sample rate is calculated based on the total number of samples in both displayable frames and ARFs that will not be used as a ``show existing frame", it effectively limits the number of allowable synthesized ARF frames.

\subsection{Frame Scaling}
\label{sec:frame_scaling}
The AV1 codec supports the option to scale a source frame to a lower resolution for compression, and re-scale the reconstructed frame to the original frame resolution. This design is particularly useful when a few frames are overly complex to compress, and hence cannot fit in the target streaming bandwidth range. The down scaling factor is constrained to be within the range of  8/16 to 15/16. The reconstructed frame is first linearly upscaled to the original size, followed by a loop restoration filter as part of the post processing stage. Both the linear upscaling filter and the loop restoration filter operations are normatively defined. We will discuss it with more details in Section \ref{sec:superres}. In order to maintain a cost-effective hardware implementation where no additional expense on line buffers is required beyond the size for regular frame decoding, the re-scaling process is limited to the horizontal direction. The up-scaled and filtered version of the decoded frame will be available as a reference frame for coding subsequent frames.

\section{Superblock and Tile}
\subsection{Superblock}
A superblock is the largest coding block the AV1 codec can process. The superblock size can be either $128 \times128$ luma samples or $64 \times 64$ luma samples, which is signaled by the sequence header. A superblock can be further partitioned into smaller coding blocks, each with their own prediction and transform modes. A superblock coding is only dependent on its above and left neighboring superblocks.

\subsection{Tile}
A tile is a rectangular array of superblocks whose spatial referencing, including intra prediction reference and the probability model update, is limited to be within the tile boundary. As a result, the tiles within a frame can be independently coded, which facilitates simple and effective multi-threading for both encoder and decoder implementations. The minimum tile size is 1 superblock. The maximum tile width corresponds to 4096 luma samples and the maximum tile size corresponds to $4096 \times 2304$ luma samples. A maximum of 512 tiles are allowed in a frame.

\begin{figure}[!t]
\centering
\includegraphics[width=0.98\columnwidth]{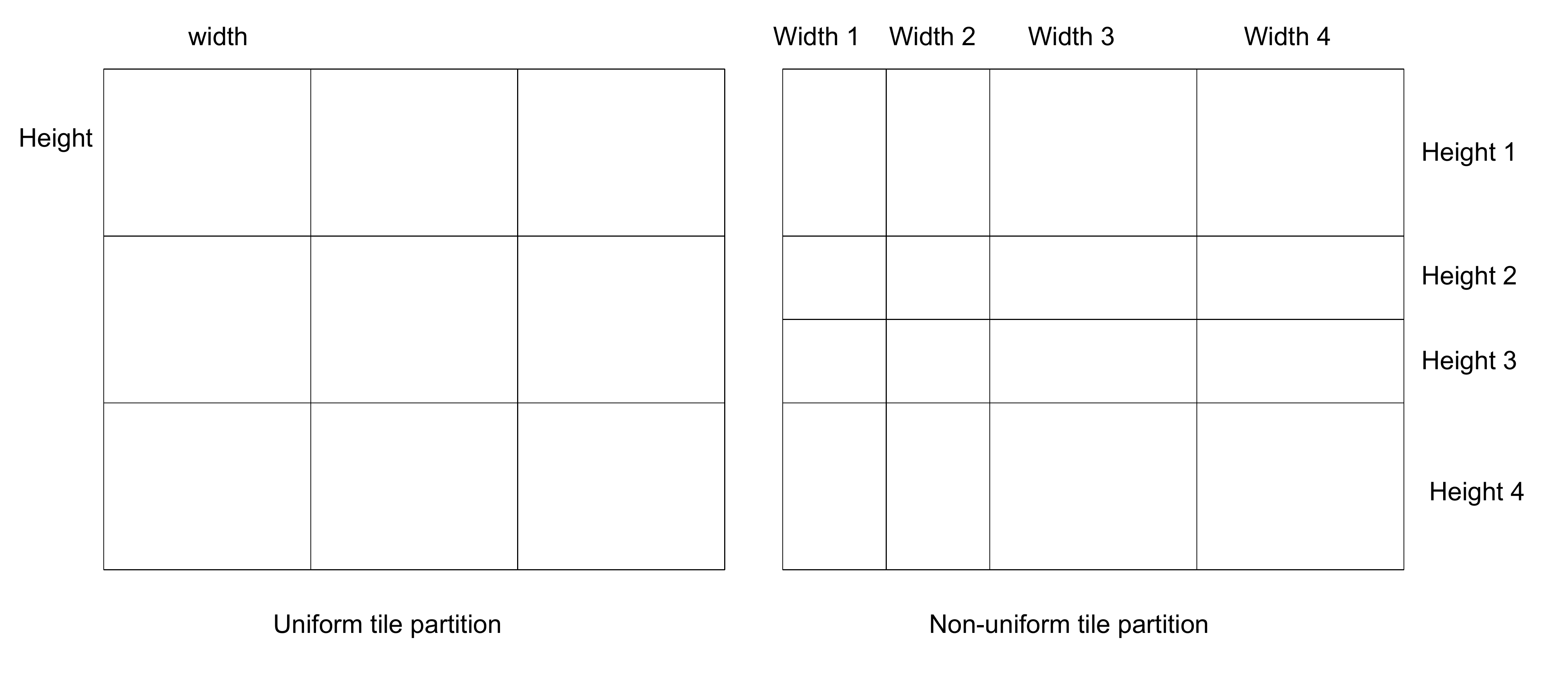}
\caption{An illustration of the uniform and non-uniform tile sizes. The uniform tile size option uses the same tile dimension across the frame. The non-uniform tile size option requires a series of width and height values to determine the lattice.}
\label{fig:tile_size}
\end{figure}

AV1 supports two ways to specify the tile size for each frame. The uniform tile size option follows the VP9 tile design and assumes all the tiles within a frame are of the same dimension, except those sitting at the bottom or right frame boundary. It allows one to identify the number of tiles vertically and horizontally in the bit-stream and derives the tile dimension based on the frame size. A second option, the non-uniform tile size, assumes a lattice form of tiles. The spacing is non-uniform in both vertical and horizontal directions and tile dimensions must be specified in the bit-stream in units of superblocks. It is designed to recognize the fact that the computational complexity differs across superblocks within a frame, due to the variations in video signal statistics. The non-uniform tile size option allows one to use smaller tile sizes for regions that require higher computational complexity, thereby balancing the workload among threads. This is particularly useful when one has ample computing resource in terms of multiple cores and needs to minimize the frame coding latency. An example is provided in Figure \ref{fig:tile_size} to demonstrate the two tile options.

The uniform/non-uniform tile size options and the tile sizes are decided on a frame by frame basis. It is noteworthy that the post-processing filters are applied across the tile boundaries to avoid potential coding artifacts (e.g., blocking artifacts) along the tile edges.

\section{Coding Block Operations}
\subsection{Coding Block Partitioning}
\label{sec:cbsize}
\begin{figure}[!t]
\centering
\includegraphics[width=3.5in]{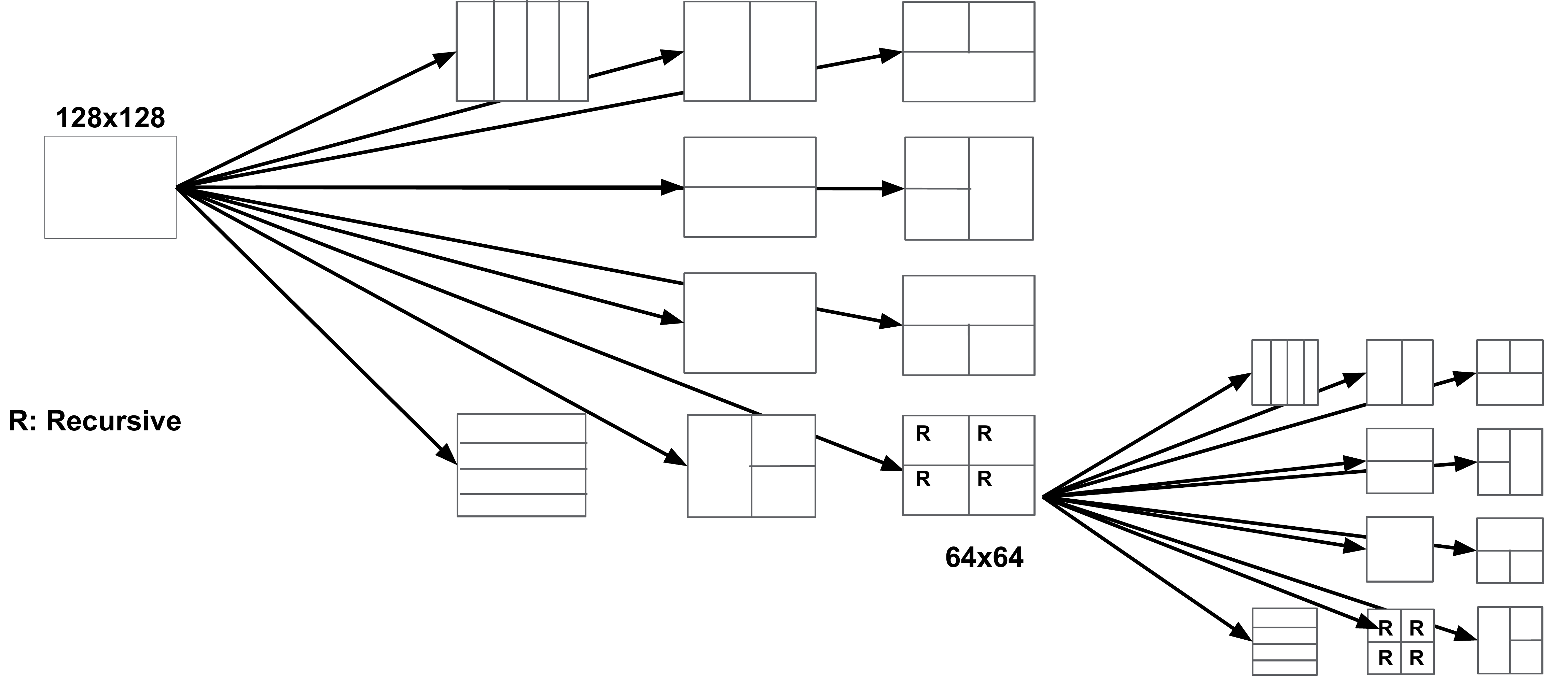}
\caption{The recursive block partition tree in AV1.}
\label{fig:block_partition}
\end{figure}

A superblock can be recursively partitioned into smaller block sizes for coding. AV1 inherits the recursive block partitioning design used in VP9. To reduce the overhead cost on prediction mode coding for video signals that are highly correlated, a situation typically seen in 4K videos, AV1 supports a maximum coding block size of $128\times128$ luma samples. The allowed partition options at each block level include 10 possibilities as shown in Figure \ref{fig:block_partition}. To improve the prediction quality for complex videos,  the minimum coding block size is extended to $4\times 4$ luma samples. While such extensions provide more coding flexibility, they have implications for hardware decoders. Certain block size dependent constraints are specifically designed to circumvent such complications.

\subsubsection{Block Size Dependent Constraints}
\label{sec:block_size_constraint}
The core computing unit in a hardware decoder is typically designed around a superblock. Increasing the superblock size from $64\times64$ to $128\times 128$ would require about 4 times silicon area for the core computing unit. To resolve this issue, we constrain the decoding operations to be conducted in $64\times 64$ units even for larger block sizes. For example, to decode a $128\times 128$ block in YUV420 format, one needs to decode the luma and chroma components corresponding to the first $64\times64$ block, followed by those corresponding to the next $64\times 64$ block, etc, in contrast to processing the luma component for the entire $128\times 128$ block, followed by the chroma components. This constraint effectively re-arranges the entropy coding order for the luma and chroma components, and has no penalty on the compression performance. It allows a hardware decoder to process a $128\times 128$ block as a series of $64\times 64$ blocks, and hence retain the same silicon area.

At the other end of the spectrum, the use of $4\times 4$ coding blocks increases the worst-case latency in YUV420 format, which happens when all the coding blocks are $4\times 4$ luma samples and are coded using intra prediction modes. To rebuild an intra coded block, one needs to wait for its above and left neighbors to be fully reconstructed, because of the spatial pixel referencing. In VP9, the $4\times 4$ luma samples within a luma $8\times 8$ block are all coded in either inter or intra mode. If in intra mode, the collocated $4\times4$ chroma components will use an intra prediction mode followed by a $4\times4$ transform. An unconstrained $4\times4$ coding block size would require each $2\times2$ chroma samples to go through prediction and transform coding, which creates a dependency in the chroma component decoding. Note that inter modes do not have such spatial dependency issues.

AV1 adopts a constrained chroma component coding for $4\times4$ blocks in YUV420 format to resolve this latency issue. If all the luma blocks within an $8\times8$ block are coded in inter mode, the chroma component will be predicted in $2\times2$ units using the motion information derived from the corresponding luma block. If any luma block is coded in an intra mode, the chroma component will follow the bottom-right $4\times4$ luma block's coding mode and conduct the prediction in $4\times4$ units. The prediction residuals of chroma components then go through a $4\times4$ transform.

These block size dependent constraints enable the extension of the coding block partition system with limited impact on hardware feasibility. However, an extensive rate-distortion optimization search is required to translate this increased flexibility into compression gains.

\subsubsection{Two-Stage Block Partitioning Search}
Observing that the key flexibility in variable coding block size is provided by the recursive partition that goes through the square coding blocks, one possibility is to employ a two-stage partition search approach. The first pass starts from the largest coding block size and goes through square partitions only. For each coding block, the rate-distortion search is limited, e.g. only using the largest transform block and 2-D DCT kernel. Its partition decisions can be analyzed to determine the most likely operating range, in which the second block partition search will conduct an extensive rate-distortion optimization search for all the 10 possible partitions. Changing the allowed block size search range drawn from the first pass partition results would give different trade-offs between the compression performance and the encoding speed. We refer to \cite{multipass} for more experimental results.

We will next discuss the compression techniques available at a coding block level within a partition.

\subsection{Intra Frame Prediction}
\label{sec:intra_pred}
For a coding block in intra mode, the prediction mode for the luma component and the prediction mode for both chroma components are signaled separately in the bitstream. The luma prediction mode is entropy coded using a probability model based on its above and left coding blocks' prediction context. The entropy coding of the chroma prediction mode is conditioned on the state of the luma prediction mode. The intra prediction operates in units of transform blocks (as introduced in Section \ref{sec:tx_coding}) and uses previously decoded boundary pixels as a reference.

\subsubsection{Directional Intra Prediction}
\begin{figure}[!t]
\centering
\includegraphics[width=3in]{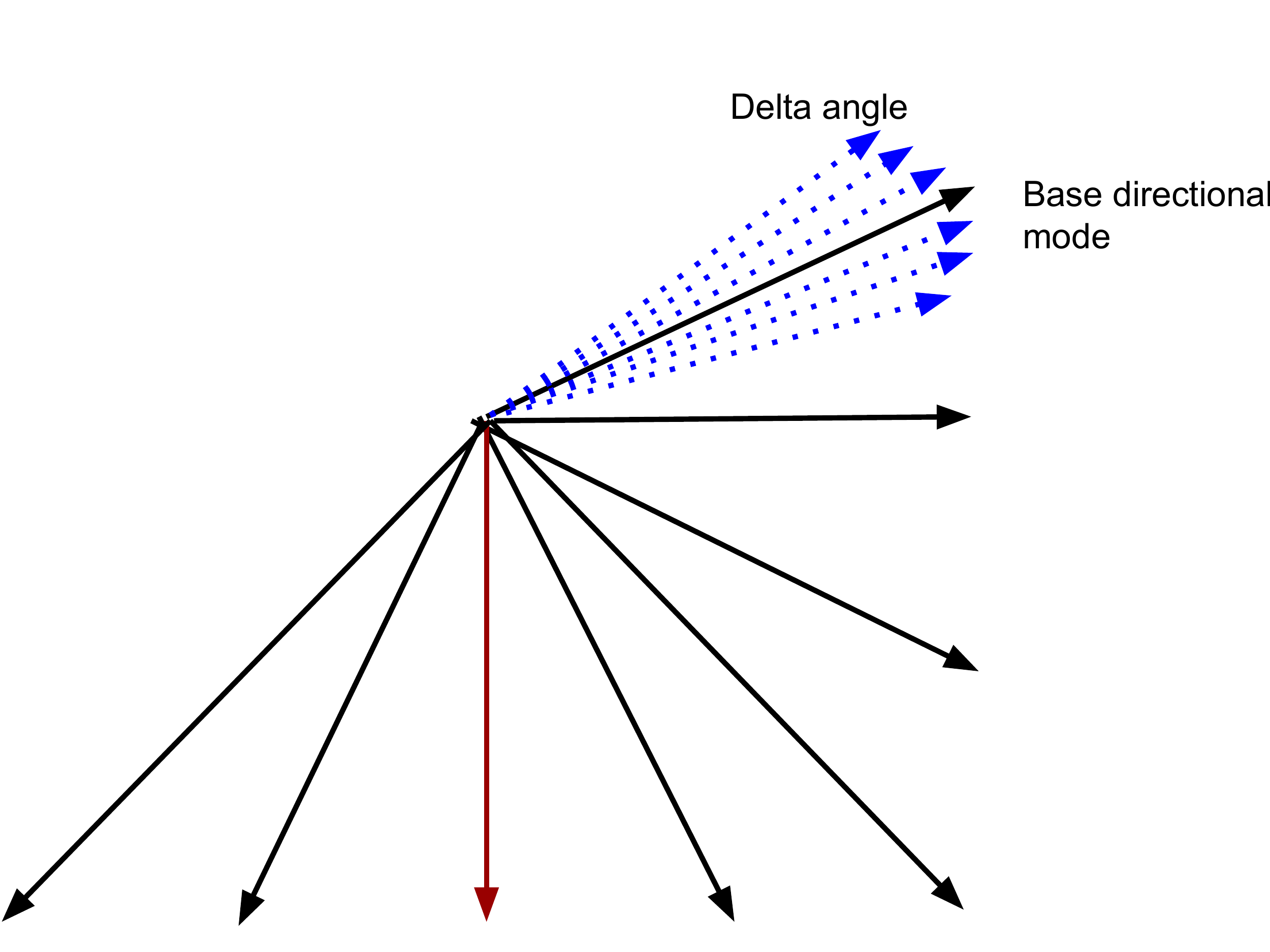}
\caption{Directional intra prediction modes. The original 8 directions in VP9 are used as a base. Each allows a supplementary signal to tune the prediction angle in units of 3\textdegree.}
\label{fig:directional_intra}
\end{figure}

AV1 extends the directional intra prediction options in VP9 to support higher granularity. The original 8 directional modes in VP9 are used as a base in AV1, with a supplementary signal to fine tune the prediction angle. This comprises up to 3 steps clockwise or counter clockwise, each of 3\textdegree~as shown in Figure \ref{fig:directional_intra}. A 2-tap bilinear filter is used to interpolate the reference pixels when a prediction points to a sub-pixel position. For coding block size of less than $8\times8$, only the 8 base directional modes are allowed, since the small number of pixels to be predicted does not justify the overhead cost of the additional granularity.

\subsubsection{Non-directional Smooth Intra Prediction}
\label{sec:smooth_intra}
\begin{figure}[!t]
\centering
\includegraphics[width=2in]{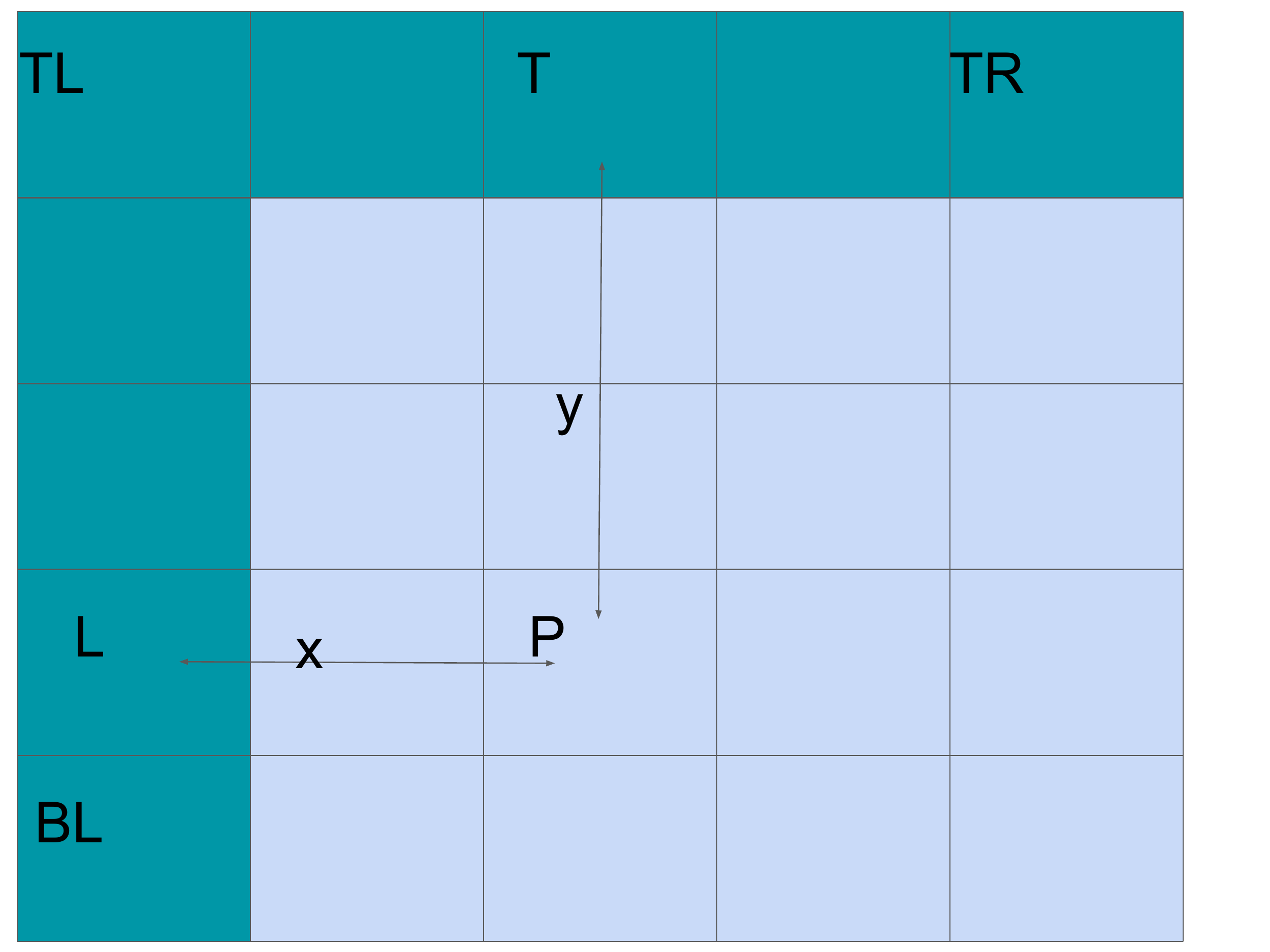}
\caption{An illustration of the distance weighted smooth intra prediction. The dark green pixels are the reference and the light blue ones are the prediction. The variables $x$ and $y$ are the distance from left and top boundaries, respectively.}
\label{fig:intra_smooth}
\end{figure}

VP9 has 2 non-directional intra smooth prediction modes: DC\_PRED and TM\_PRED. AV1 adds 3 new smooth prediction modes that estimate pixels using a distance weighted linear combination, namely SMOOTH\_V\_PRED, SMOOTH\_H\_PRED, and SMOOTH\_PRED. They use the bottom-left (BL) and top-right (TR) reference pixels to fill the right-most column and bottom-row, thereby forming a closed loop boundary condition for interpolation. We use the notations in Figure \ref{fig:intra_smooth} to demonstrate their computation procedures:
\begin{itemize}
\item SMOOTH\_H\_PRED: $P_H = w(x) L + (1 - w(x)) TR$;
\item SMOOTH\_V\_PRED: $P_V = w(y) T + (1 - w(y)) BL$;
\item SMOOTH\_PRED: $P = (P_H + P_V) / 2$.
\end{itemize}
where $w(x)$ represents the weight based on distance $x$ from the boundary, whose values are preset.

AV1 replaces the TM\_PRED mode which operates as 
\begin{equation*}
P = T + L - TL
\end{equation*}
with a PAETH\_PRED mode that follows:
\begin{equation*}
P = argmin |x - (T + L - TL)|, \forall x \in \{T, L, TL\}.
\end{equation*}
The non-linearity in the PAETH\_PRED mode allows the prediction to steer the referencing angle to align with the direction that exhibits highest correlation.

\subsubsection{Recursive Intra Prediction}
The inter pixel correlation is modeled as a 2-D first-order Markov field. Let $X(i, j)$ denote a pixel at position $(i, j)$. Its prediction is formed by
\begin{equation}
\hat{X}(i, j) = \alpha \hat{X}(i -1, j) + \beta \hat{X}(i, j - 1) + \gamma \hat{X}(i - 1, j - 1),
\end{equation}
where $\hat{X}$s on the right-hand side are the available reconstructed boundary pixels or the prediction of the above and left pixels. The coefficient set $\{\alpha, \beta, \gamma\}$  forms a linear predictor based on the spatial correlations. A total of 5 different sets of linear predictors are defined in AV1, each represents a different spatial correlation pattern.

\begin{figure}
\centering
\includegraphics[width=0.8\columnwidth]{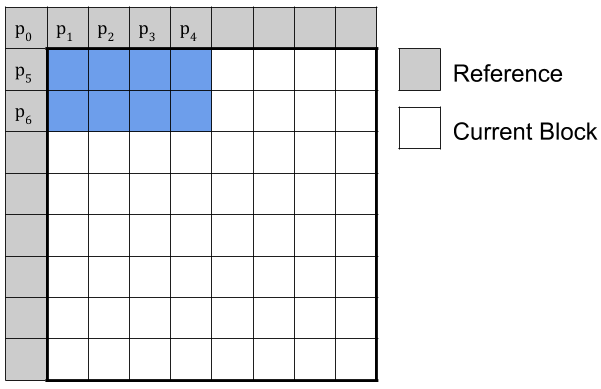}
\caption{Recursive-filter-based intra predictor. Reference pixels $p_0$-$p_6$ are used to linearly predict the $4\times2$ patch in blue. The predicted pixels will be used as reference for next $4\times2$ patch in the current block.}
\label{fig:filter_intra}
\end{figure}

To improve hardware throughput, instead of recursively predicting each pixel, AV1 predicts a $4\times2$ pixel patch from its adjacent neighbors, e.g. $p_0 - p_6$ for the blue patch $x_0 - x_7$ in Fig. \ref{fig:filter_intra}, whose coefficients can be directly derived from $\{\alpha, \beta, \gamma\}$ by expanding the recursion:
\begin{align*}
x_0 &= \alpha p_1 + \beta p_5 + \gamma p_0 \\
x_1 &= \alpha p_2 + \beta x_0 + \gamma p_1 \\
 &= \beta \gamma p_0 + (\alpha \beta + \gamma)p_1 + \alpha p_2 + \beta^2 p_5  \\
x_2 &= \alpha p_3 + \beta x_1 + \gamma p_2 \\
&= \gamma \beta^2 p_0 + (\alpha \beta + \gamma)\beta p_1 + (\alpha \beta + \gamma) p_2 + \alpha p_3 + \beta^3 p_5 \\
&\vdots
\end{align*}
Such expansion avoids the inter pixel dependency within a $4\times2$ patch, thereby allowing hardware decoders to process the predictions in parallel.

\subsubsection{Chroma from Luma Prediction}
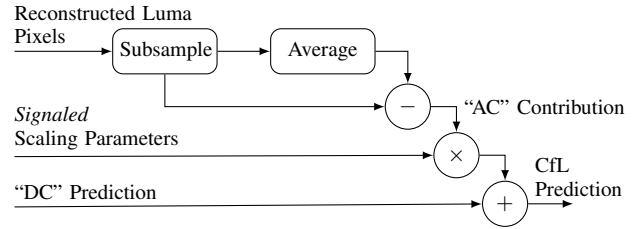
\begin{figure}[htb]
  \centering
  \begin{tikzpicture}[node distance = 2.1cm, auto]
\footnotesize
    \tikzstyle{block} = [rectangle, draw, text width=1.2cm, text centered,
    rounded corners, minimum height=0.6cm]
    \tikzstyle{op} = [circle, draw, text width=0.27cm, text centered, minimum height=0.6cm]
    \node [] (start) {};
    \node [below = 1.2cm of start] (scaling) {};
    \node [below = 0.468cm of scaling] (dc) {};
    \node [block, right of = start] (sub) {Subsample};
    \node [block, right of = sub] (avg) {Average};
    \node [op, below right = 0.3cm of avg] (minus) {$-$};
    \node [op, below right = 0.3cm of minus] (times) {$\times$};
    \node [op, below right = 0.3cm of times] (plus) {$+$};

    \node [right = 0.6cm of plus] (end) {};
    \draw[-latex] (start) node[above right, text width=3cm] {Reconstructed Luma Pixels} -> (sub);
    \draw[-latex] (sub) -> (avg);
    \draw[-latex] (avg) -| (minus);
    \draw[-latex] (sub) |- (minus);
    \draw[-latex] (minus) -| node{``AC'' Contribution} (times);
    \draw[-latex] (times) -| (plus);
    \draw[-latex] (scaling) node[above right, text width=3.6cm] {\textit{Signaled}\\Scaling Parameters} -> (times);
    \draw[-latex] (dc) node[above right, text width=3.6cm] {``DC'' Prediction} -> (plus);
    \draw[-latex] (plus) node[above right = 0cm and 0.3cm, text width=1.8cm] {CfL \\Prediction} -> (end);
\normalsize
  \end{tikzpicture}
  \caption{Outline of the operations required to build the CfL prediction \cite{chroma-Trudeau}.}
\label{fig:cfl}
\end{figure}

Chroma from luma prediction models chroma pixels as a linear function of corresponding reconstructed luma pixels. As depicted in Figure \ref{fig:cfl}, the predicted chroma pixels are obtained by adding the DC prediction of the chroma block to a scaled AC contribution, which is the result of multiplying the AC component of the downsampled luma block by a scaling factor explicitly signaled in the bitstream \cite{chroma-Trudeau}.

\subsubsection{Intra Block Copy}
AV1 allows intra-frame motion compensated prediction, which uses to the previously coded pixels within the same frame, namely Intra Block Copy (IntraBC). A motion vector at full pixel resolution is used to locate the reference block. This may imply a half-pixel accuracy motion displacement for the chroma components, in which context a bilinear filter is used to conduct sub-pixel interpolation. The IntraBC mode is only available for intra coding frames, and can be turned on and off by frame header.

Typical hardware decoders pipeline the pixel reconstruction and the post-processing filter stages, such that the post-processing filters are applied to the decoded superblocks, while later superblocks in the same frame are being decoded. Hence an IntraBC reference block is retrieved from the pixels after post-processing filters. In contrast, a typical encoder would process all the coding blocks within a frame, then decide the post-processing filter parameters that minimize the reconstruction error. Therefore, the IntraBC mode most likely accesses the coded pixels prior to the post-processing filters for rate-distortion optimization. Such discrepancy hinders the efficiency of IntraBC mode. To circumvent this issue, all the post-processing filters are disabled if the IntraBC mode is allowed in an intra only coded frame.

In practice, the IntraBC mode is most likely useful for images that contain substantial amount of text content, or similar repeated patterns, in which setting post-processing filters are less effective. For natural images where pixels largely form an auto-regressive model, the encoder needs to be cautious regarding the use of IntraBC mode, as the absence of post-processing filters, may trigger visual artifacts at coarse quantization.

\subsubsection{Color Palette}
In this mode, a color palette ranging between 2 to 8 base colors (i.e. pixel value) is built for each luma/chroma plane, where each pixel gets assigned a color index. The number of base colors is an encoder decision that determines the trade-off between fidelity and compactness. The base colors are predictively coded in the bit-stream using those of neighboring blocks as reference. The color indexes are coded pixel-by-pixel using a probability model conditioned on previously coded color indexes. The luma and chroma channels can decide whether to use the palette mode independently. This mode is particularly suitable for a pixel block that contains limited pixel variations.

\subsection{Inter Frame Prediction}
AV1 supports a rich toolsets to exploit the temporal correlation in video signals. These include adaptive filtering in translational motion compensation, affine motion compensation, and highly flexible compound prediction modes.

\subsubsection{Translational Motion Compensation}
\label{sec:trans_MC}
\begin{figure}[!t]
\centering
\includegraphics[width=3.5in]{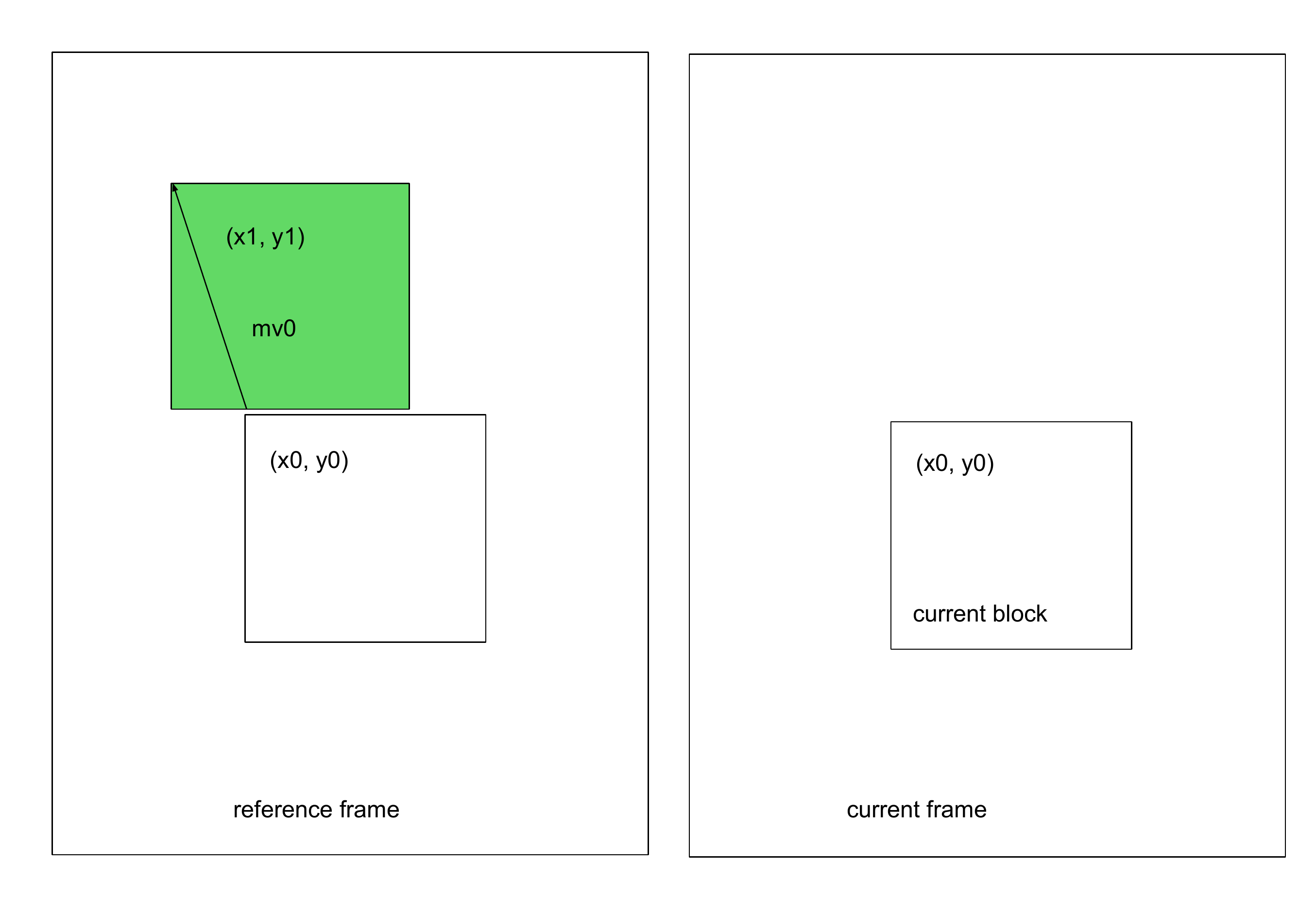}
\caption{Translational motion compensated prediction.}
\label{fig:translational_mc}
\end{figure}

A coding block uses a motion vector to find its prediction in a reference frame. It first maps its current position, e.g. top-left pixel position $(x_0, y_0)$ in Figure \ref{fig:translational_mc}, in the reference frame. It is then displaced by the motion vector to the target reference block whose top-left pixel is located at $(x_1, y_1)$.

\begin{figure}[!t]
\centering
\includegraphics[width=3.5in]{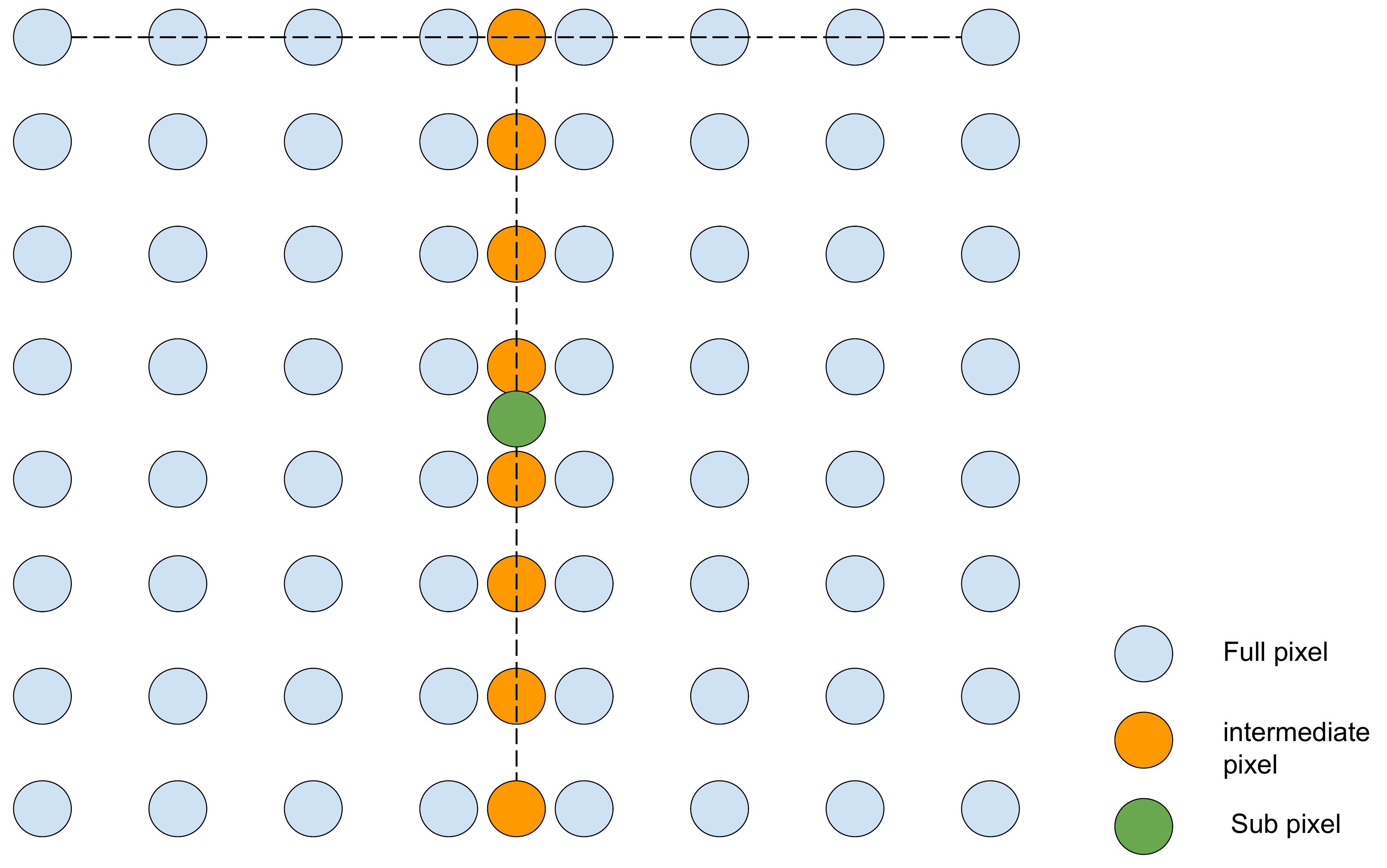}
\caption{Sub-pixel generation through separable interpolation filter.}
\label{fig:sub_pixel}
\end{figure}

AV1 allows 1/8 pixel motion vector accuracy. A sub-pixel is generated through separable interpolation filters. A typical procedure is shown in Figure \ref{fig:sub_pixel}, where one first computes the horizontal interpolation through all the related rows. A second vertical filter is applied to the resulting intermediate pixels to produce the final sub-pixel. Clearly the intermediate pixels (orange) can be reused to produce multiple final sub-pixels (green).

Common block-based encoder motion estimations are conducted via measurements of the sum of absolute difference (SAD) or the sum of squared error (SSE) \cite{BME-survey,BME-conf,SAD-fast}, which tend to favor a reference block that resembles the DC and lower AC frequency components well, whereas the high frequency components are less reliably predicted. An interpolation filter with a high cutoff frequency would allow more high frequency components from the reference region to form the prediction, and is suitable for cases where the high frequency components between the reference and the current block are highly correlated. Conversely an interpolation filter with a low cutoff frequency would largely remove high frequency components that are less relevant to the current block.

\begin{figure}
\centering
\includegraphics[width=0.98\columnwidth]{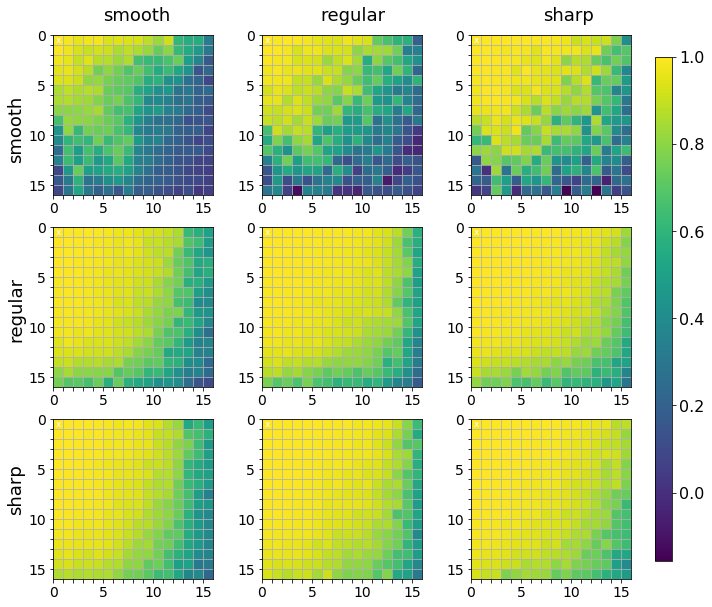}
\caption{A heat map of the correlations between the prediction and the source signals in the transform domain. The motion estimation here is in units of $16\times16$ block. The prediction and source blocks are grouped based on their optimal interpolation filters. The test clip is old\_town\_cross\_480p. The top and left labels mark the interpolation filters used in the horizontal and vertical directions, respectively. It is evident that the groups using SHARP filter tend to have higher correlation in high frequency components along the corresponding direction.}
\label{fig:dct_correlation}
\end{figure}

An adaptive interpolation filter scheme is used in VP9, where an inter-coded block in VP9 can choose from three 8-tap interpolation filters that correspond to different cutoff frequencies in a Hamming window in the frequency domain. The selected interpolation filter is applied to both vertical and horizontal directions. AV1 inherits the interpolation filter selection design and extends it to support independent filter selection for the vertical and horizontal directions, respectively. It exploits the potential temporal statistical discrepancy between the vertical and horizontal directions for improved prediction quality. Each direction can choose from 3 finite impulse response (FIR) filters, namely SMOOTH,  REGULAR, and SHARP in ascending order of cutoff frequencies. A heat map of the correlations between the prediction and the source signals in the transform domain is shown in Figure \ref{fig:dct_correlation}, where the prediction and source block pairs are grouped according to their optimal 2-D interpolation filters. It is evident that the signal statistics differ in vertical and horizontal directions and an independent filter selection in each direction captures such discrepancy well. 

\begin{figure}
\includegraphics[width=0.8\columnwidth]{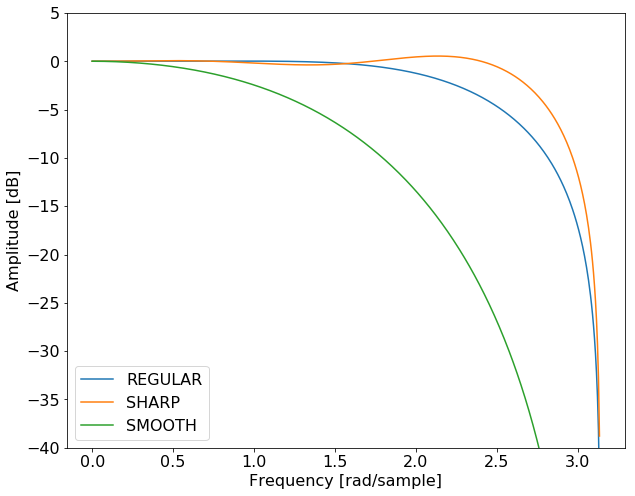}
\centering
\caption{The frequency responses of the 3 interpolation filters at half-pixel position.}
\label{fig:freq_response}
\end{figure}

To reduce the decoder complexity, the SMOOTH and REGULAR filters adopt a 6-tap FIR design, which appears to be sufficient for a smooth and flat baseband. The SHARP filter continues to use an 8-tap FIR design to mitigate the ripple effect near the cutoff frequency. The filter coefficients that correspond to half-pixel interpolation are
\begin{align*}
SMOOTH  & ~~ [-2, 14, 52, 52, 14, -2] \\
REGULAR &~~ [2, -14, 76, 76, -14, 2] \\
SHARP  & ~~[-4, 12, -24, 80, 80, -24, 12, -4].
\end{align*}
whose frequency responses are shown in Figure \ref{fig:freq_response}. To further reduce the worst case complexity when all coding blocks are in $4\times4$ luma samples, there are two additional 4-tap filters that are used when the coding block has dimensions of 4 or less. The filter coefficients for half-pixel interpolation are
\begin{align*}
SMOOTH & ~~ [12, 52, 52, 12] \\
REGULAR & ~~ [-12, 76, 76, -12].
\end{align*}
The SHARP filter option is not applicable due to the short filter taps.

\subsubsection{Affine Model Parameters}
\label{sec:affine_model}
Besides conventional translational motion compensation, AV1 also supports the affine transformation model that projects a current pixel at $(x, y)$ to a prediction pixel at $(x', y')$ in a reference frame through
\begin{equation}
\begin{bmatrix}
x' \\
y'
\end{bmatrix}
 = 
\begin{bmatrix}
h_{11} & h_{12} & h_{13} \\
h_{21} & h_{22} & h_{23}
\end{bmatrix}
\begin{bmatrix}
x \\
y \\
1
\end{bmatrix}.
\label{eq:affine_model}
\end{equation}
The tuple $(h_{13}, h_{23})$ corresponds to a conventional motion vector used in the translational model. Parameters $h_{11}$ and $h_{22}$ control the scaling factors in the vertical and horizontal axes, and in conjunction with the pair $h_{12}$ and $h_{21}$ decide the rotation angle.

A global affine model is associated with each reference frame, where each of the four non-translational parameters has 12-bit precision and the translational motion vector is coded in 15-bit precision. A coding block can choose to use it directly provided the reference frame index. The global affine model captures the frame level scaling and rotation, and hence primarily focuses on the settings of rigid motion over the entire frame. 
In addition, a local affine model at coding block level would be desirable to adaptively track the non-translational motion activities that vary across the frame. However the overhead cost of sending the affine model parameters per coding block also introduces additional side-information \cite{affine-twostage}. 
As a result, various research efforts focus on the estimation of the affine model parameters without the extra overhead \cite{affine-264, affine-pred}.
A local affine parameter estimation scheme based on the regular translational motion vectors from spatial neighboring blocks has also been developed for AV1.

The translational motion vector $(h_{13}, h_{23})$ in the local affine model is explicitly transmitted in the bit-stream. To estimate the other four parameters, it hypothesizes that the local scaling and rotation factors can be reflected by the pattern of the spatial neighbors' motion activities. The codec scans through a block's nearest neighbors and finds the ones whose motion vector points toward the same reference frame. A maximum of 8 candidate reference blocks are allowed. For each selected reference block, its center point will first be offset by the center location of the current block to create an original sample position. This offset version will then add the motion vector difference between the two blocks to form the destination sample position after the affine transformation. A least square regression is conducted over the available original and destination sample position pairs to calculate the affine model parameters.

\begin{figure}[!t]
\centering
\includegraphics[width=3.5in]{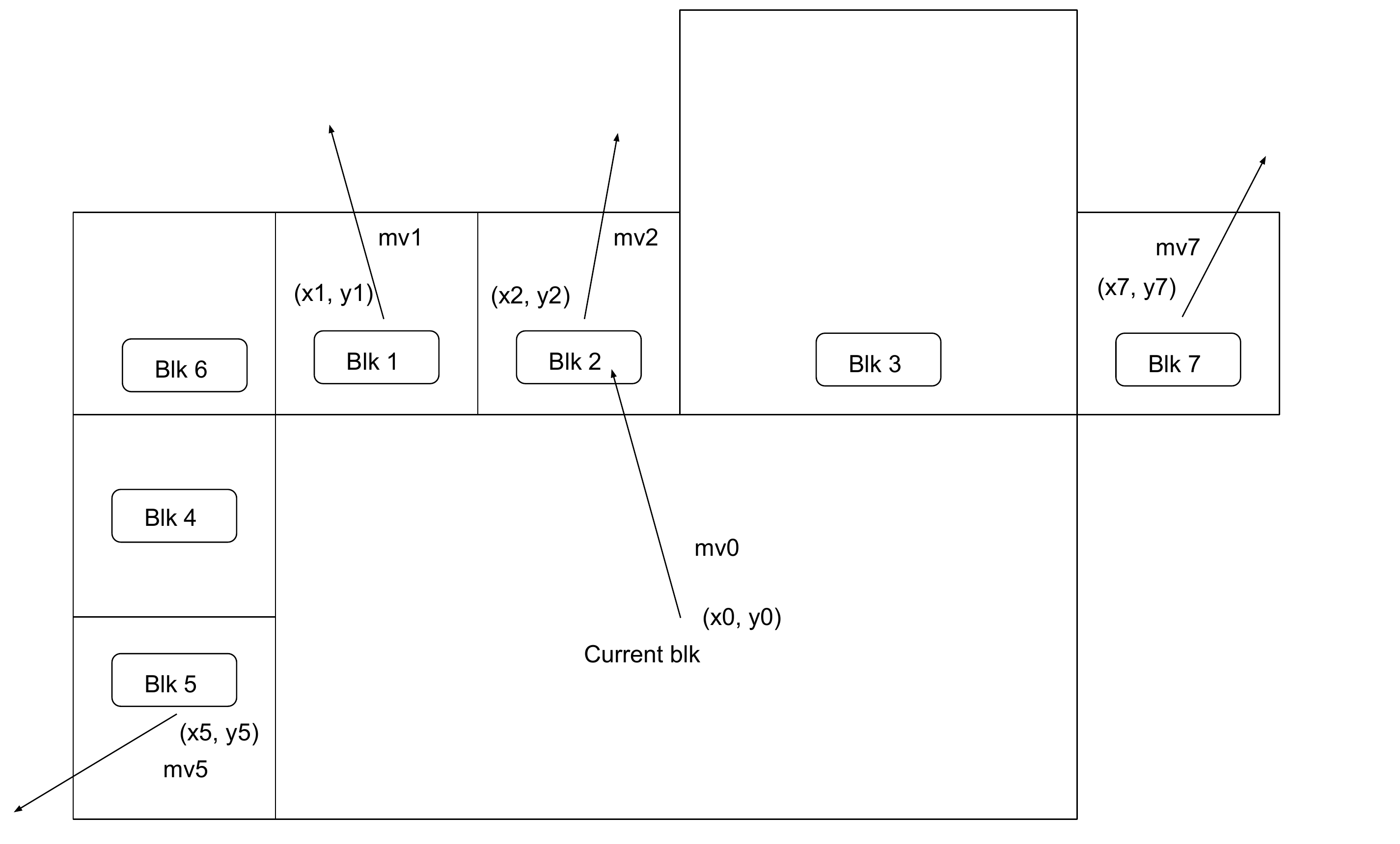}
\caption{An illustration of the local affine parameter estimation.}
\label{fig:local_warp}
\end{figure}

We use Figure \ref{fig:local_warp} as an example to demonstrate the affine parameter estimation process. 
The nearest neighbor blocks are marked by the scan order. For Block $k$, its center position is denoted by $(x_k, y_k)$ and the motion vector is denoted by $mv_k$. The current block is denoted by $k=0$. Assume in this case Block 1, 2, 5, and 7 share the same reference as the current block and are selected as the reference blocks. An original sample position is formed as
\begin{equation}
(a_k, b_k) = (x_k, y_k) - (x_0, y_0),
\end{equation}
where $k\in\{1, 2, 5, 7\}$. The corresponding destination sample position is obtained by further adding the motion vector difference.
\begin{equation}
(a_k', b_k') = (a_k, b_k) + (mv_k.x, mv_k.y) - (mv_0.x, mv_0.y).
\end{equation}
To formulate the least square regression, we denote the sample data as
\begin{equation}
P =
\begin{bmatrix}
a_1, b_1 \\
a_2, b_2 \\
a_5, b_5 \\
a_7, b_7
\end{bmatrix},
q =
\begin{bmatrix}
a_1' \\
a_2' \\
a_5' \\
a_7'
\end{bmatrix}, \text{and~}
r =
\begin{bmatrix}
b_1' \\
b_2' \\
b_5' \\
b_7'
\end{bmatrix}.
\end{equation}
The least square regression gives the affine parameter in \eqref{eq:affine_model} as
\begin{equation}
\begin{bmatrix}
h_{11} \\
h_{12}
\end{bmatrix}
= (P^TP)^{-1}P^Tq, \text{~and~}
\begin{bmatrix}
h_{21} \\
h_{22}
\end{bmatrix}
= (P^TP)^{-1}P^Tr.
\end{equation}
In practice, one needs to ensure that the spatial neighboring block is relevant to the current block. Hence we discard the reference block if any component of the motion vector difference is above 8 pixels in absolute value. Furthermore, if the number of available reference blocks is less than 2, the least square regression problem is ill posed, hence the local affine model is disabled.

\subsubsection{Affine Motion Compensation}
\begin{figure}[!t]
\centering
\includegraphics[width=3.2in]{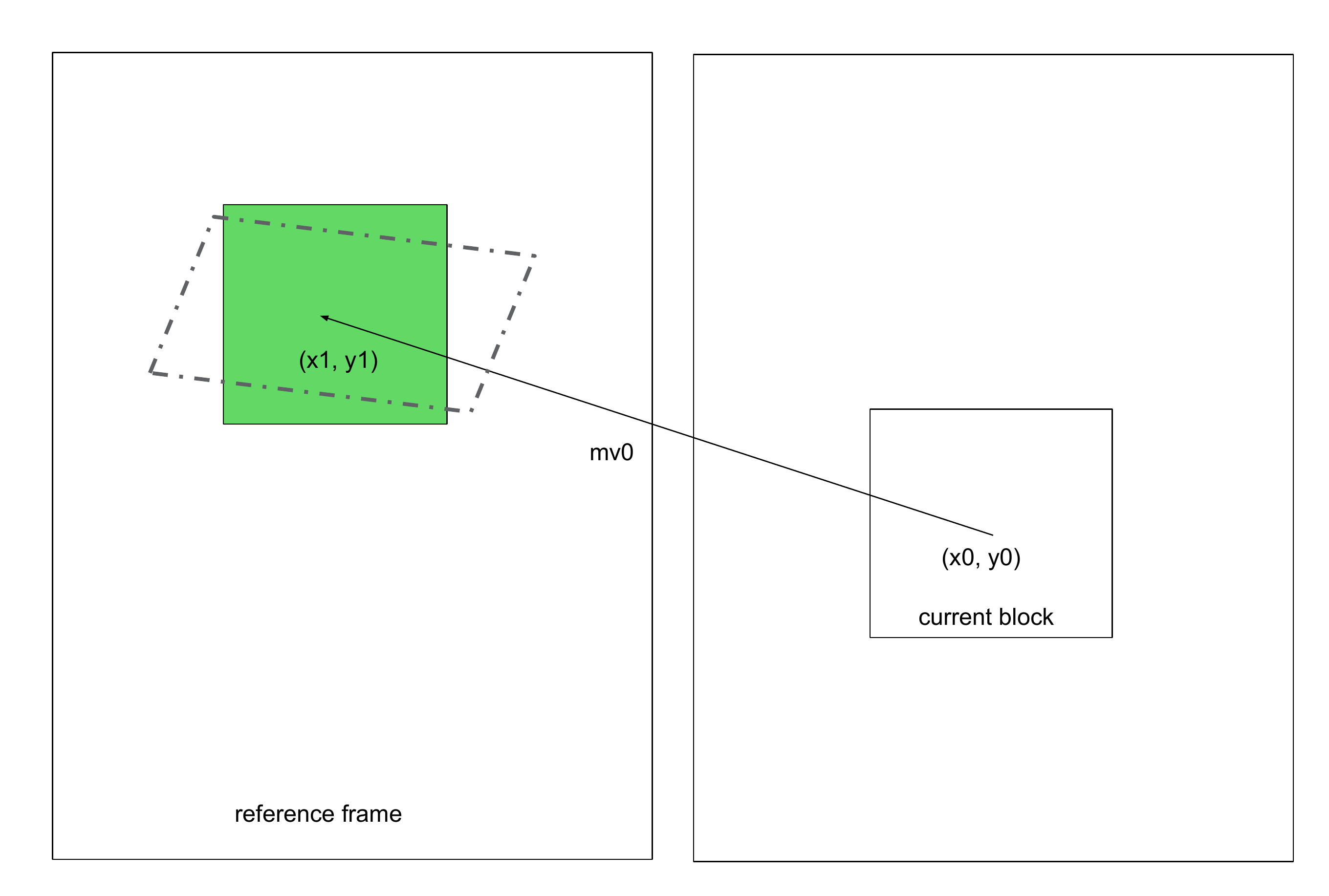}
\caption{Build the affine prediction.}
\label{fig:translational_mv}
\end{figure}

With the affine model established, we next discuss techniques in AV1 for efficient prediction construction \cite{warp-motion}. The affine model is allowed for block size at $8\times8$ and above. A prediction block is decomposed into $8\times8$ units. The center pixel of each $8\times8$ prediction unit is first determined by the translational motion vector $(h_{13}, h_{23})$, as shown in Figure \ref{fig:translational_mv}. The rest of the pixels at position $(x, y)$ in the green square in Figure \ref{fig:translational_mv}, are scaled and rotated around the center pixel at $(x_1, y_1)$ to form the affine projection $(x', y')$ in the dash line following
\begin{equation}
\begin{bmatrix}
x' \\
y'
\end{bmatrix}
=
\begin{bmatrix}
h_{11} & h_{12} \\
h_{21} & h_{22}
\end{bmatrix}
\begin{bmatrix}
x - x_1 \\
y - y_1
\end{bmatrix}
+
\begin{bmatrix}
x_1 \\
y_1
\end{bmatrix}.
\label{eq:rotation_scale}
\end{equation}

The affine projection allows 1/64 pixel precision. A set of 8-tap FIR filters (6-tap in certain corner cases) are designed to construct the sub-pixel interpolations. A conventional translational model has a uniform sub-pixel offset across the entire block, which allows one to effectively ``reuse'' most intermediate outcomes to reduce the overall computation. Typically as introduced in Section \ref{sec:trans_MC}, to interpolate an $8\times 8$ block, a horizontal filter is first applied to generate an intermediate $15\times8$ array from a $15\times15$ reference region. A second vertical filter is then applied to the intermediate $15\times8$ array to produce the final $8\times8$ prediction block. Hence a translational model requires $(15\times8)\times8$ multiplications for the horizontal filter stage, and $(8\times8)\times8$ multiplications for the vertical filter stage, 1472 multiplications in total.

Unlike the translational model, it is reasonable to assume that each pixel in an affine model has a different sub-pixel offset, due to the rotation and scaling effect. Directly computing each pixel would require $64\times8\times8=4096$ multiplications. Observe, however, that the rotation and scaling matrix in \eqref{eq:rotation_scale} can be decomposed into two shear matrices:
\begin{equation}
\begin{bmatrix}
h_{11} & h_{12} \\
h_{21} & h_{22}
\end{bmatrix}
=
\begin{bmatrix}
1 & 0 \\
\gamma & 1 + \delta
\end{bmatrix}
\begin{bmatrix}
1 + \alpha & \beta \\
0 & 1
\end{bmatrix},
\label{eq:shear}
\end{equation}
where the first term on the right side corresponds to a vertical interpolation and the second term corresponds to a horizontal interpolation. This translates building an affine reference block into a two-stage interpolation operation. A $15\times8$ intermediate array is first obtained through horizontal filtering over a $15\times15$ reference region, where the horizontal offsets are computed as:
\begin{equation}
\text{horz offset} = (1 + \alpha)(x - x_1) + \beta(y - y_1).
\label{eq:horz_offset}
\end{equation}
The intermediate array then undergoes vertical filtering to interpolate vertical offsets:
\begin{equation}
\text{vert offset} = \gamma(x- x_1)+(1+\delta)(y - y_1)
\label{eq:vert_offset}
\end{equation}
and generates the $8\times8$ prediction block. It thus requires a total of 1472 multiplications, the same as the translational case. However, it is noteworthy that the actual computational cost of affine model is still higher, since the filter coefficients change at each pixel, whereas the translational model uses a uniform filter in the horizontal and vertical stage, respectively.

To improve the cache performance AV1 requires the horizontal offset in \eqref{eq:horz_offset} to be within 1 pixel away from $(x - x_1)$ and the vertical offset in \eqref{eq:vert_offset} to be within 1 pixel away from $(y - y_1)$, which constrains the reference region within a $15\times15$ pixel array. Consider the first stage that generates a $15\times8$ intermediate pixel array. The displacements from its center are $(x - x_1) \in [-4, 4)$ and $(y - y_1) \in [-7,8)$. Hence we have the constraint on the maximum horizontal offset as
\begin{equation}
\text{max}~ \alpha (x - x_1) + \beta (y - y_1) = 4|\alpha| + 7|\beta| < 1.
\label{eq:warp_valid1}
\end{equation}
Similarly $(x - x_1) \in [-4, 4)$ and $(y - y_1) \in [-4,4)$ in the second stage, which leads to
\begin{equation}
4|\gamma| + 4|\delta| < 1.
\label{eq:warp_valid2}
\end{equation}
A valid affine model in AV1 needs to satisfy both conditions in \eqref{eq:warp_valid1} and \eqref{eq:warp_valid2}.

\subsubsection{Compound Predictions}
\label{sec:compound}
The motion compensated predictions from two reference frames (see supported reference frame pairs in Section \ref{sec:refframe}) can be linearly combined through various compound modes. The compound prediction is formulated by
\begin{equation*}
P(x, y) = m(x, y) * R_1(x, y) + (64 - m(x, y)) * R_2(x, y),
\end{equation*}
where the weight $m(x, y)$ is scaled by 64 for integer computation, $R_1(x, y)$ and $R_2(x, y)$ represent the pixels at position $(x, y)$ in the two reference blocks. $P(x, y)$ will be scaled down by $1/64$ to form the final prediction.

\textbf{Distance Weighted Predictor} Let $d_1$ and $d_2$ denote the temporal distance between the current frame and its two reference frames, respectively. The weight $m(x, y)$ is determined by the relative values of $d_1$ and $d_2$. Assuming that $d_1 \leq d_2$, the weight coefficient is defined by
\begin{equation}
m(x, y) = 
\begin{cases}
36, & d_2 < 1.5d_1 \\
44, & d_2 < 2.5d_1 \\
48, & d_2 < 3.5d_1 \\
52, & otherwise
\end{cases}
\end{equation}
The distribution is symmetric for the case $d_1 \ge d_2$.

\textbf{Average Predictor} A special case of the distance weighted predictor, where the two references are equally weighted, i.e., $m(x, y) = 32$.

\textbf{Difference Weighted Predictor} The weighting coefficient is computed per pixel based on the difference between the two reference pixels. A binary sign is sent per coding block to decide which reference block prevails when the pixel difference is above a certain threshold.
\begin{equation}
m(x, y) =
\begin{cases}
38 + \frac{|R_1(x, y) - R_2(x, y)|}{16}, & sign = 0 \\
64 - (38 + \frac{|R_1(x, y) - R_2(x, y)|}{16}), & sign = 1
\end{cases}
\end{equation}
Note that $m(x, y)$ is further capped by $[0, 64]$.

\textbf{Wedge Mode} A set of 16 coefficient arrays have been preset for each eligible block size. They effectively split the coding block into two sections along various oblique angles. The $m(x, y)$ is mostly set to 64 in one section, and 0 in the other, except near the transition edge, where there is a gradual change from 64 to 0 with 32 at the actual edge.

\begin{figure}[!t]
\centering
\includegraphics[width=3.5in]{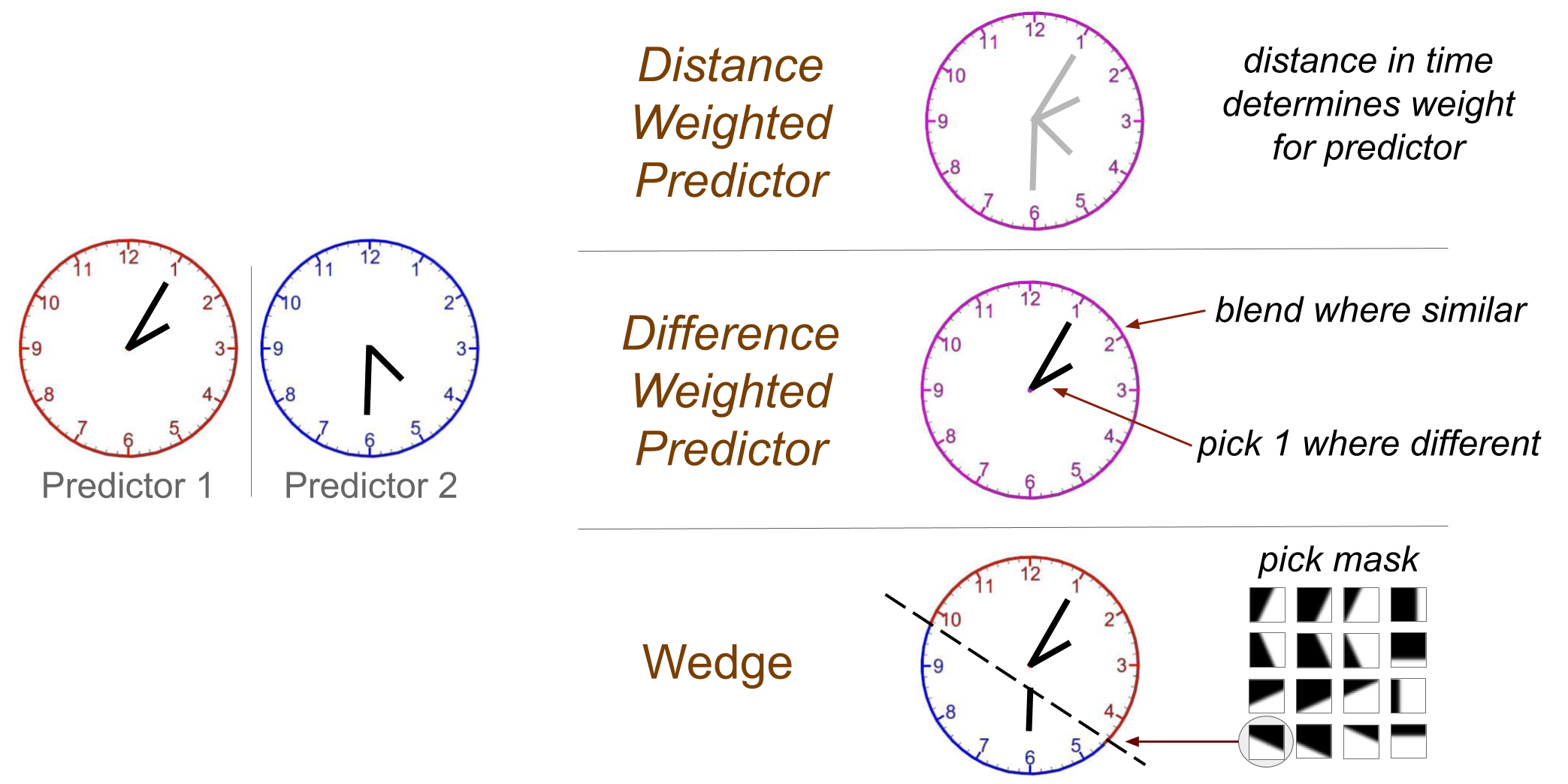}
\caption{Illustration of the compound prediction modes. The distance weighted predictor uniformly combines the two reference blocks. The difference weighted predictor combines the pixels when their values are close (e.g., the dial plate and the numbers), and picks one reference when the difference is large (e.g. the clock hands). The wedge predictor uses one of the preset masks to split the block into two sections, each filled with one reference block's pixels. In the above example, it stitches the lower part predictor 1 and the higher upper part of predictor 2 as the compound prediction.}
\label{fig:compound_modes}
\end{figure}

We use Figure \ref{fig:compound_modes} to demonstrate the compound options and their effects. The numerous compound modes add substantial encoding complexity in order to realize their potential coding gains. A particular hotspot lies in the motion estimation process, because each reference block is associated with its own motion vector. Simultaneously optimizing both motion vectors for a given compound mode makes the search space grow exponentially. Prior research \cite{jointME} proposes a joint search approach that iteratively fixes one motion vector and searches the other motion vector until the results converge, which can significantly reduces the number of motion vector search points for a compound mode.

Other prediction modes supported by AV1 that blend multiple reference blocks include an overlapped block motion compensation and a combined inter-intra prediction mode, both of which operate on a single reference frame and allow only one motion vector.

\textbf{Overlapped Block Motion Compensation}
The overlapped block motion compensation mode modifies the original design in \cite{OBMC-Sullivan}  to account for variable block sizes \cite{OBMC-AV1}. It exploits the immediate spatial neighbors' motion information to improve the prediction quality for pixels near its top and left boundaries, where the true motion trajectory correlates with the motion vectors on both sides.

\begin{figure}[!t]
\centering
\subfloat[]{\includegraphics[width=2.5in]{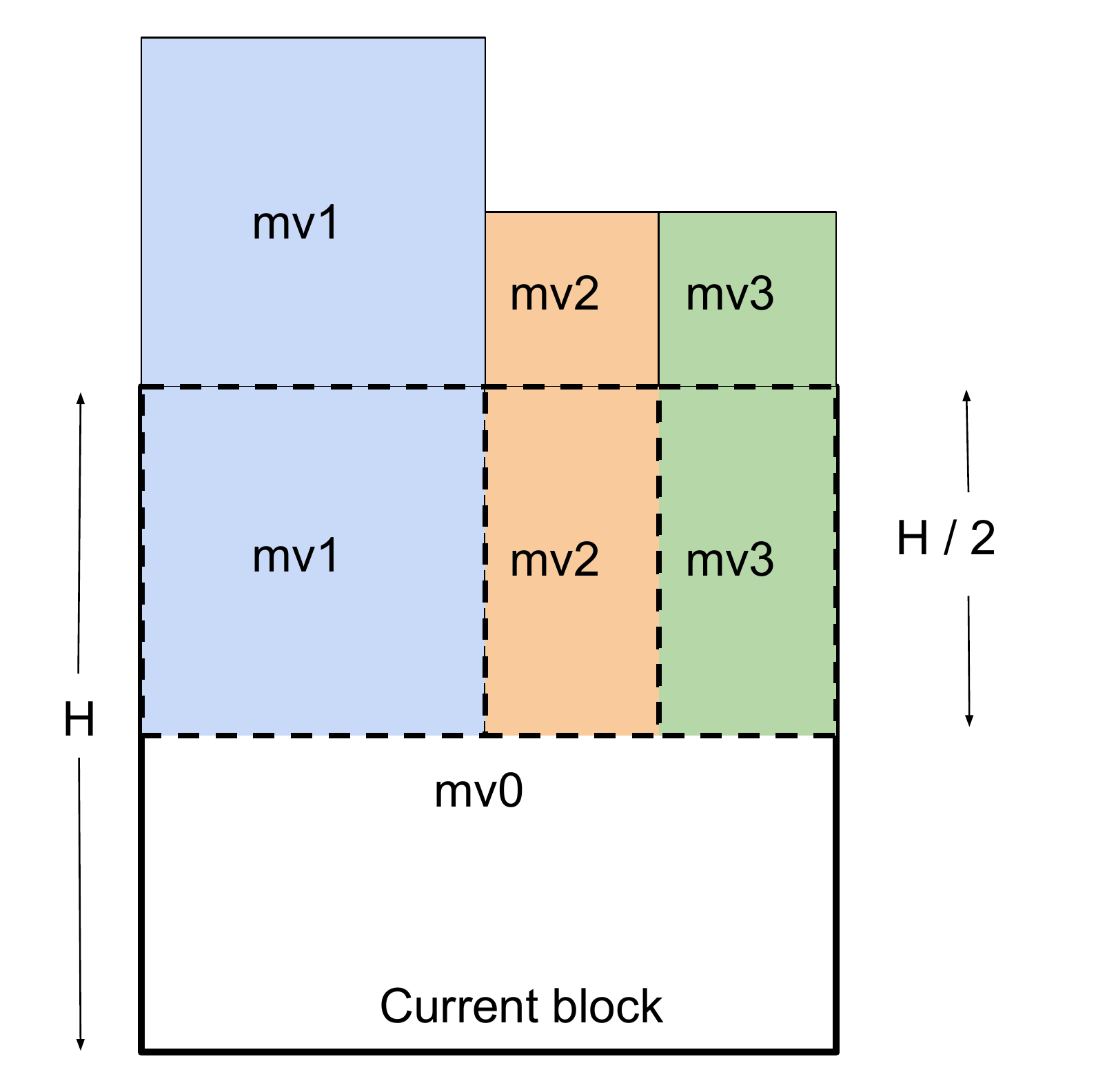}%
\label{fig:obmc_top}}
\hfil
\subfloat[]{\includegraphics[width=2.5in]{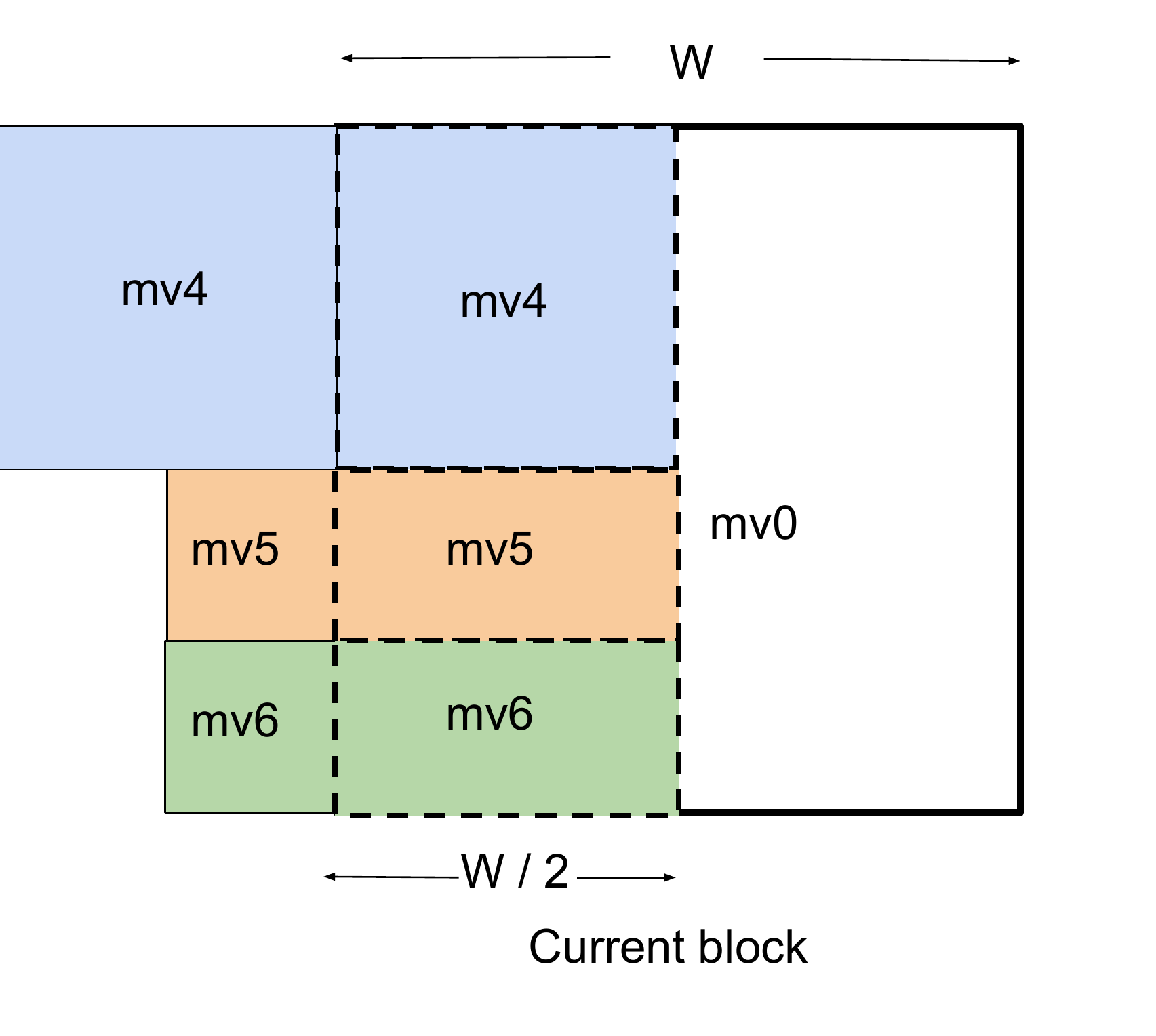}%
\label{fig:obmc_left}}
\caption{Overlapped block motion compensation using top and left neighboring blocks' motion information, shown in (a) and (b) respectively.}
\label{fig:obmc_blend}
\end{figure}

\begin{figure}[!t]
\centering
\includegraphics[width=0.98\columnwidth]{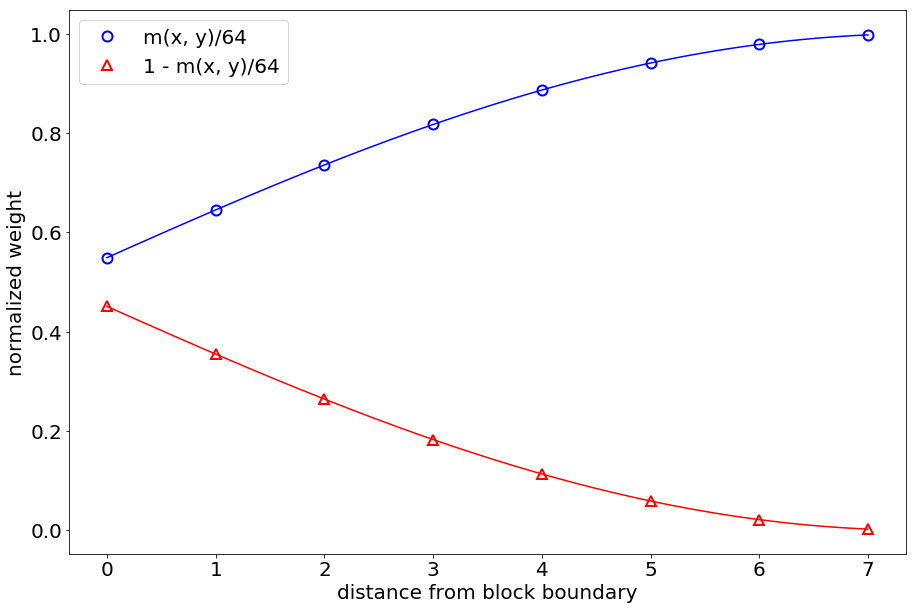}
\caption{Normalized weights for OBMC with $H=16$ or $W=16$. }
\label{fig:raised_cos}
\end{figure}

It first scans through the immediate neighbors above and finds up to 4 reference blocks that have the same reference frame as the current block. An example is shown in Figure \subref*{fig:obmc_top}, where the blocks are marked according to their scan order. The motion vector of each selected reference block is employed to generate a motion compensated block that extends from the top boundary towards the center of the current block. Its width is the same as the reference block's width and its height is half of the current block's height, as shown in Figure \subref*{fig:obmc_top}. An intermediate blending result is formed as:
\begin{equation}
P_{int}(x, y) = m(x, y)R_1(x, y) + (64 - m(x, y))R_{above}(x, y),
\end{equation}
where $R_1(x, y)$ is the original motion compensated pixel at position $(x, y)$ using current block's motion vector $mv_0$, and $R_{above}(x, y)$ is the pixel from the overlapped reference block. The weight $m(x, y)$ follows a raised cosine function:
\begin{equation}
m(x, y) = 64 * ( \frac{1}{2}sin(\frac{\pi}{H}(y + \frac{1}{2})) + \frac{1}{2}),
\end{equation}
where $y = 0,~1,~\cdots,~H/2 - 1$ is the row index, $H$ is the current block height. The weight distribution for $H=16$ is shown in Figure \ref{fig:raised_cos}.

The scheme next processes the immediate left neighbors to extract the available motion vectors and build overlapped reference blocks extending from the left boundary towards the center, as shown in Figure \subref*{fig:obmc_left}. The final prediction is calculated by:
\begin{equation}
P(x, y) = m(x, y)P_{int}(x, y) + (64 - m(x, y))R_{left}(x, y),
\end{equation}
where $R_{left}(x, y)$ is the pixel from the left-side overlapped reference block. The weight $m(x, y)$ is a raised cosine function of the column index $x$:
\begin{equation}
m(x, y) = 64 * ( \frac{1}{2}sin(\frac{\pi}{W}(x + \frac{1}{2})) + \frac{1}{2}),
\end{equation}
where $x=0,~1,~\cdots,~W/2 - 1$ and $W$ is the current block width.

\begin{figure}[!t]
\centering
\includegraphics[width=0.98\columnwidth]{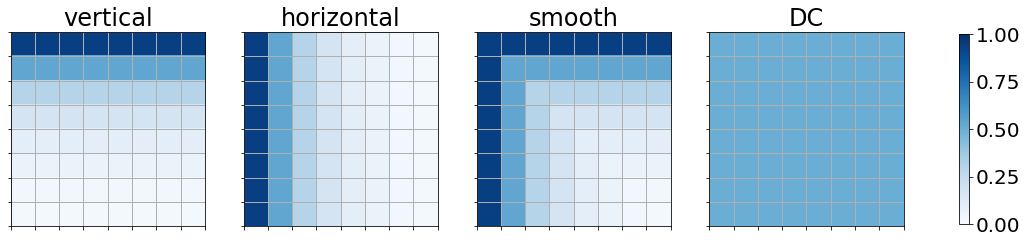}
\caption{Normalized weight masks of compound inter-intra prediction for 8x8 blocks. }
\label{fig:interintra_mask}
\end{figure}

\textbf{Compound Inter-Intra Predictor} This mode combines an intra prediction block and a translational inter prediction block. The intra prediction is selected among the DC, vertical, horizontal, and smooth modes (see Section \ref{sec:smooth_intra}). The combination can be achieved through either a wedge mask similar to the compound inter case above, or a preset coefficient set that gradually reduces the intra prediction weight along its prediction direction. Examples of the preset coefficients for each intra mode are shown in Figure \ref{fig:interintra_mask}.

As discussed above, AV1 supports a large variety of compound prediction tools. Exercising each mode in the rate-distortion optimization framework fully realizes their potential, at the cost of bringing a significant complexity load for the encoder. Efficient selection of the appropriate compound coding modes without extensive rate-distortion optimization searches remains a challenge.

\subsection{Dynamic Motion Vector Referencing Scheme}
Motion vector coding accounts for a sizable portion of the overall bit-rate. Modern video codecs typically adopt predictive coding for motion vectors and code the difference using entropy coding \cite{MVpred-RD, MVpred-hevc}. The prediction accuracy has a large impact on the coding efficiency. AV1 employs a dynamic motion vector referencing scheme that obtains candidate motion vectors from the spatial and temporal neighbors and ranks them for efficient entropy coding.

\subsubsection{Spatial Motion Vector Reference}
\begin{figure}[!t]
\centering
\includegraphics[width=0.98\columnwidth]{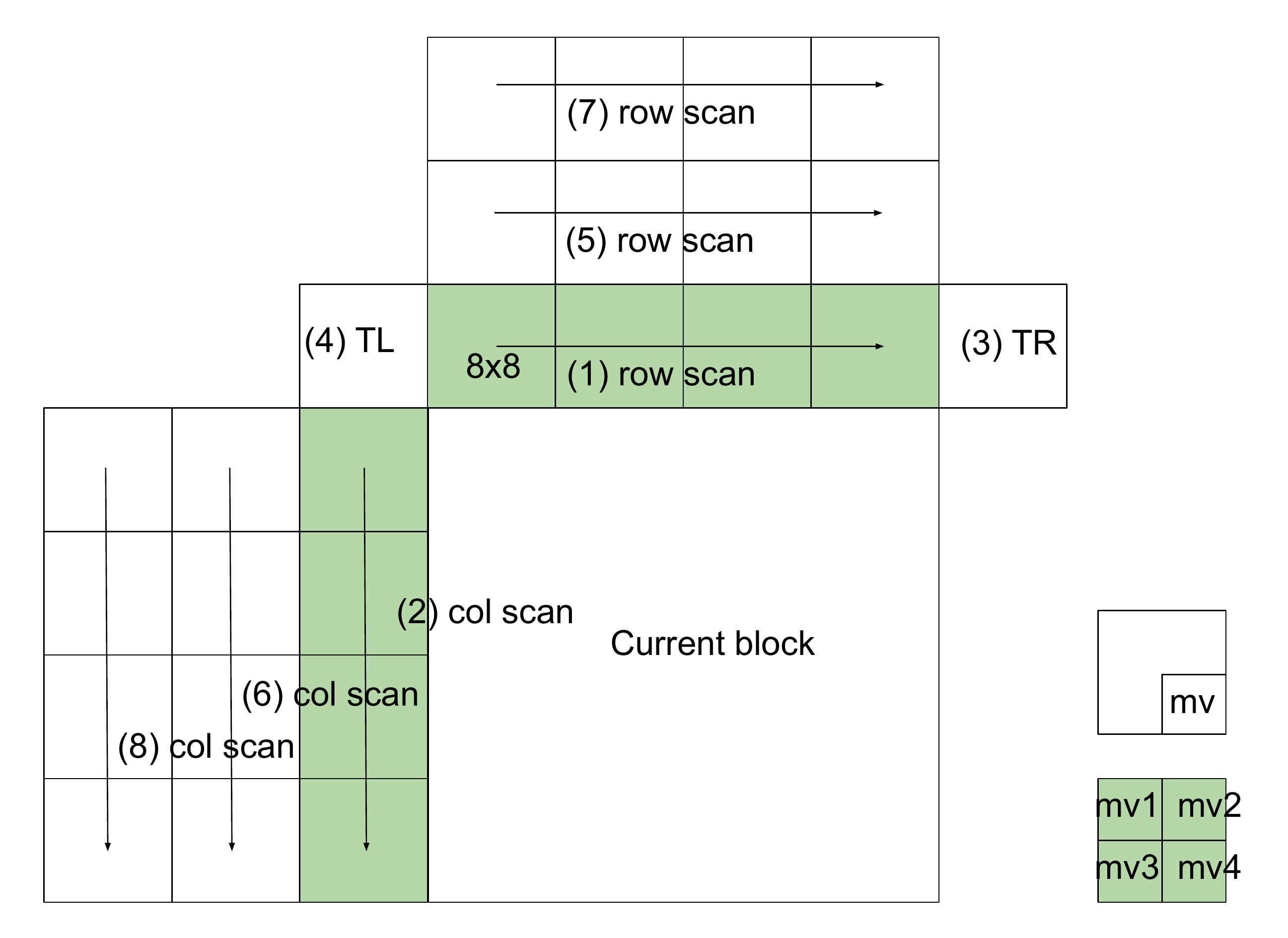}
\caption{Spatial reference motion vector search pattern. The index ahead of each operation represents the processing order. TL stands for the top-left $8\times8$ block. TR stands for the top-right $8\times8$ block.}
\label{fig:spatial_scan}
\end{figure}

A coding block will search its spatial neighbors in the unit of $8\times8$ luma samples to find the ones that have the same reference frame index as the current block. For compound inter prediction modes, this means the same reference frame pairs. The search region contains three $8\times8$ block rows above the current block and three $8\times8$ block columns to the left. The process is shown in Figure \ref{fig:spatial_scan}, where the search order is shown by the index. It starts from the nearest row and column, and interleaves the outer rows and columns. The top-right $8\times8$ block is included if available. The first 8 different motion vectors encountered will be recorded, along with a frequency count and whether they appear in the nearest row or column. They will then be ranked as discussed in Section \ref{sec:drl}.

\begin{figure}[!t]
\centering
\includegraphics[width=0.98\columnwidth]{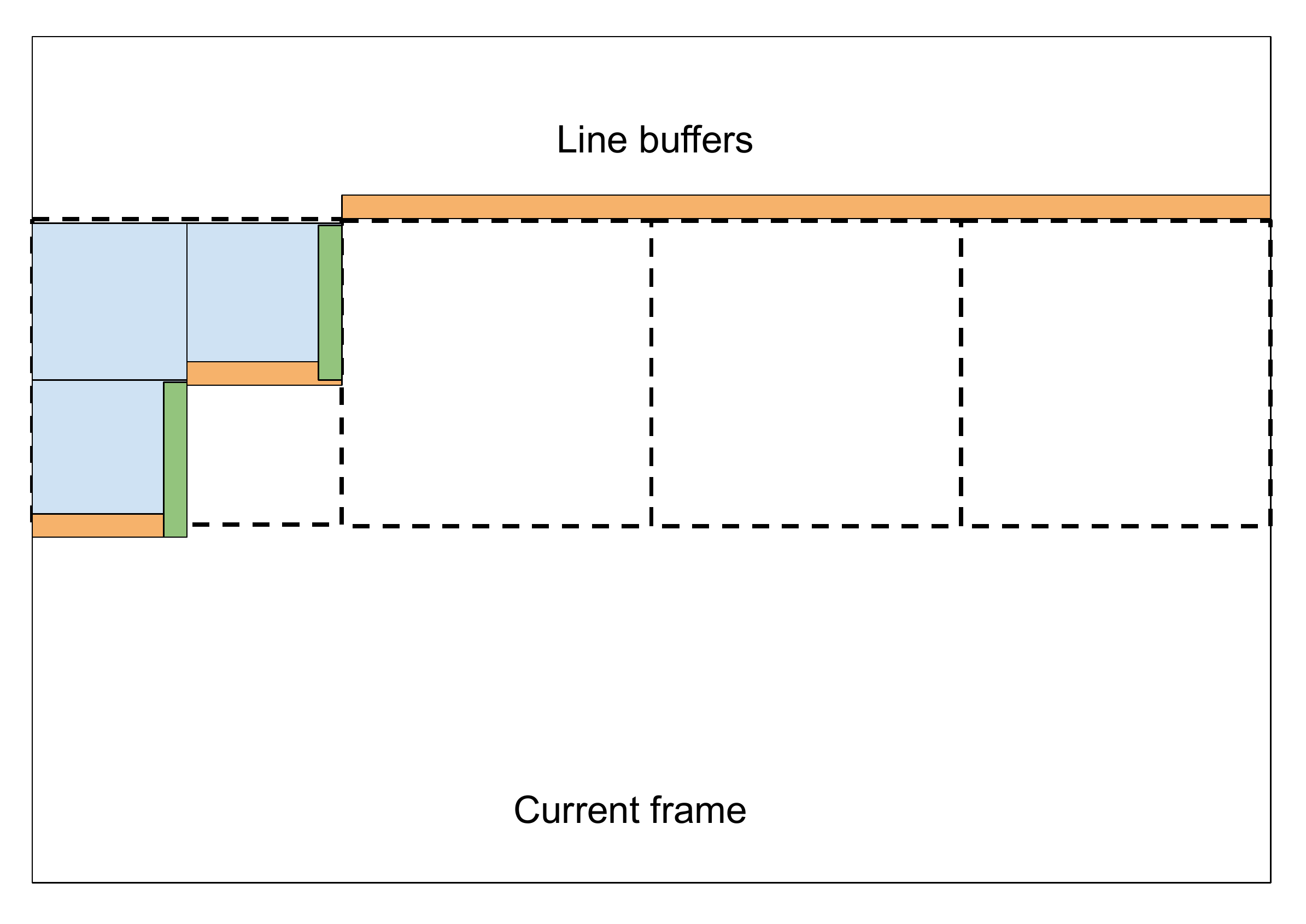}
\caption{The line buffer, shown in the orange color, stores the coding information associated with an entire row of a frame. The dash line shows superblocks. The information in the line buffer will be used as above context by later coding blocks across superblock boundaries. The line buffer is updated as new coding blocks (in blue color) are processed. In contrast, the green color shows coding information to be used as left context by later blocks, the length of which corresponds to the size of a superblock.}
\label{fig:line_buffer}
\end{figure}

Note that the minimum coding block size in AV1 is $4\times4$. Hence an $8\times8$ unit has up to 4 different motion vectors and reference frame indexes to search through. This would require a hardware decoder to store all the motion information at $4\times4$ unit precision for the three $8\times8$ block rows above. Hardware decoders typically use a line buffer concept, which is a dedicated buffer in the static random access memory (SRAM), a fast and expensive unit. The line buffer maintains coding information corresponding to an entire row of a frame, which will be used as context information for later coding blocks. An example of the line buffer concept is shown in Figure \ref{fig:line_buffer}. The line buffer size is designed for the worst case that corresponds to the maximum frame width allowed by the specification. To make the line buffer size economically feasible, AV1 adopts a design that only accesses $4\times4$ block motion information in the immediate above row (the green region in Figure \ref{fig:spatial_scan}). For the rest of the rows, the codec only uses the motion information for $8\times8$ units. If an $8\times8$ block is coded using $4\times4$ blocks, the bottom-right $4\times4$ block's information will be used to represent the entire $8\times8$ block, as shown in Figure \ref{fig:spatial_scan}. This halves the amount of space needed for motion data in the line buffer.

The storage of the coding context to the left, on the other hand, depends on the superblock size and is agnostic to the frame size. It has far less impact on the SRAM space. However, we keep its design symmetric to the above context to avoid the motion vector ranking system described in Section \ref{sec:drl} favoring either side.

\subsubsection{Motion Field Motion Vector Reference}
Common practice extracts the temporal motion vector by referring to the collocated blocks in the reference frames \cite{MVpred-hevc}\cite{direct_mode}. Its efficacy however is largely limited to capture motion trajectories at low velocities.  To reliably track the motion trajectory for efficient motion vector prediction, AV1 uses a motion field approach \cite{MVref}. 

A motion field is created for each reference frame ahead of processing the current frame. First we build motion trajectories between the current frame and the previously coded frames by exploiting motion vectors from previously coded frames through either linear interpolation or extrapolation. The motion trajectories are associated with $8\times8$ blocks in the current frame. Next the motion field between the current frame and a given reference frame can be formed by extending the motion trajectories from the current frame towards the reference frame.

\begin{figure}[!t]
\centering
\includegraphics[width=0.98\columnwidth]{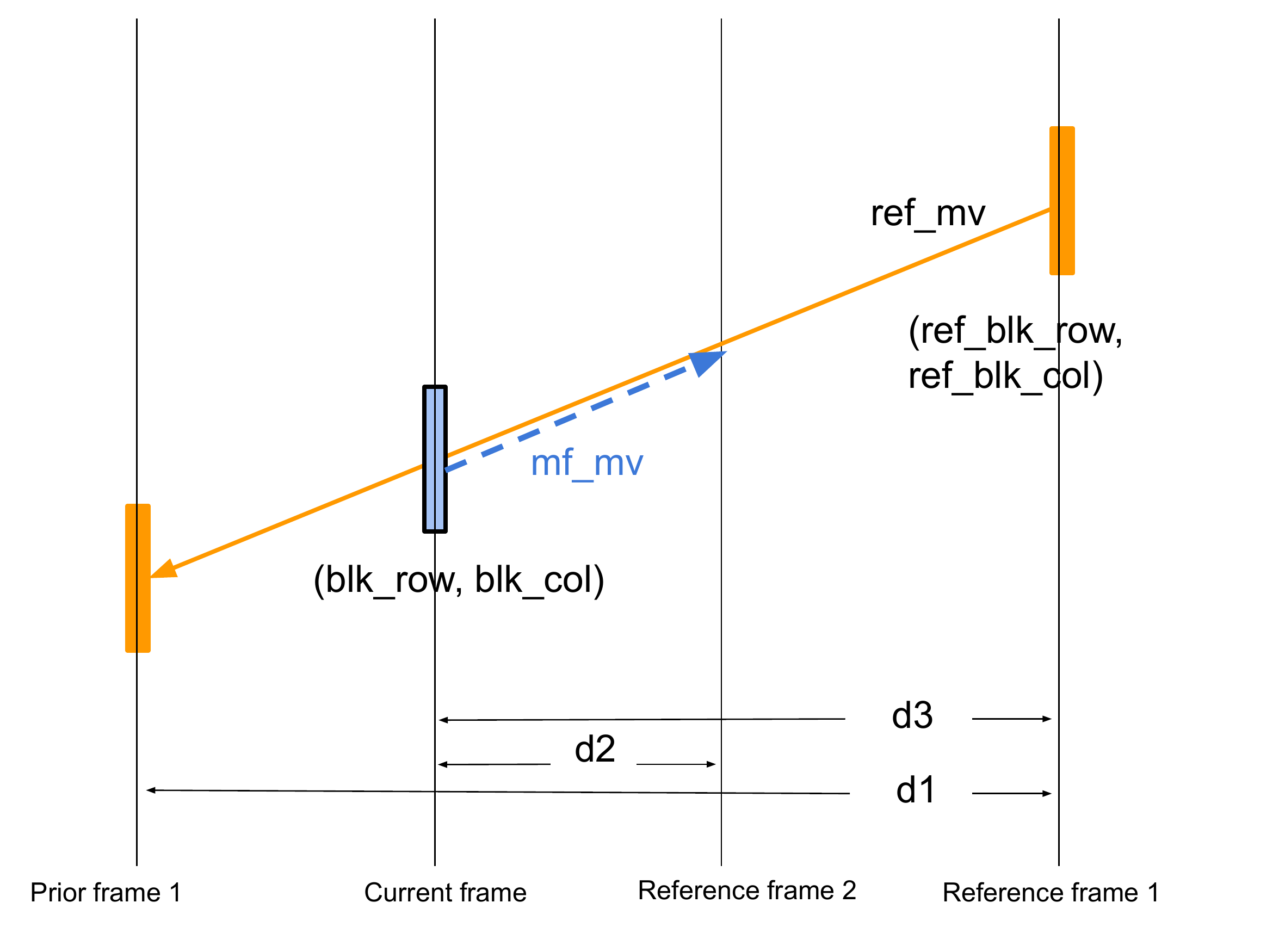}
\caption{Building motion trajectory through motion vector interpolation.}
\label{fig:mfmv_interpolation}
\end{figure}

\textbf{Interpolation} 
The motion vector pointing from a reference frame to a prior frame crosses the current frame. An example is shown in Figure \ref{fig:mfmv_interpolation}. The frames are drawn in display order. The motion vector ref\_mv at block (ref\_blk\_row, ref\_blk\_col) in the reference frame (shown in orange) goes through the current frame. The distance that ref\_mv spans is denoted by $d_1$. The distance between the current frame and the reference frame where ref\_mv originates is denoted by $d_3$. The intersection is located at block position:
\begin{equation}
\begin{bmatrix}
\text{blk\_row} \\
\text{blk\_col}
\end{bmatrix}
=
\begin{bmatrix}
\text{ref\_blk\_row} \\
 \text{ref\_blk\_col} 
\end{bmatrix}
+
\begin{bmatrix}
\text{ref\_mv.row} \\
\text{ref\_mv.col}
\end{bmatrix}
\cdot
\frac{d_3}{d_1}
\label{eq:mfmv_pos1}
\end{equation}
The motion field motion vector that extends from block (blk\_row, blk\_col) in the current frame towards a reference frame along the motion trajectory, e.g., mf\_mv in blue color, is calculated as
\begin{equation}
\begin{bmatrix}
\text{mf\_mv.row} \\
\text{mf\_mv.col}
\end{bmatrix}
=
-
\begin{bmatrix}
\text{ref\_mv.row} \\
\text{ref\_mv.col}
\end{bmatrix}
\cdot \frac{d_2}{d_1}
\label{eq:mfmv_interpolation}
\end{equation}
where $d_2$ is the distance between the current frame and the target reference frame that the motion field is built for.

\begin{figure}[!t]
\centering
\includegraphics[width=0.98\columnwidth]{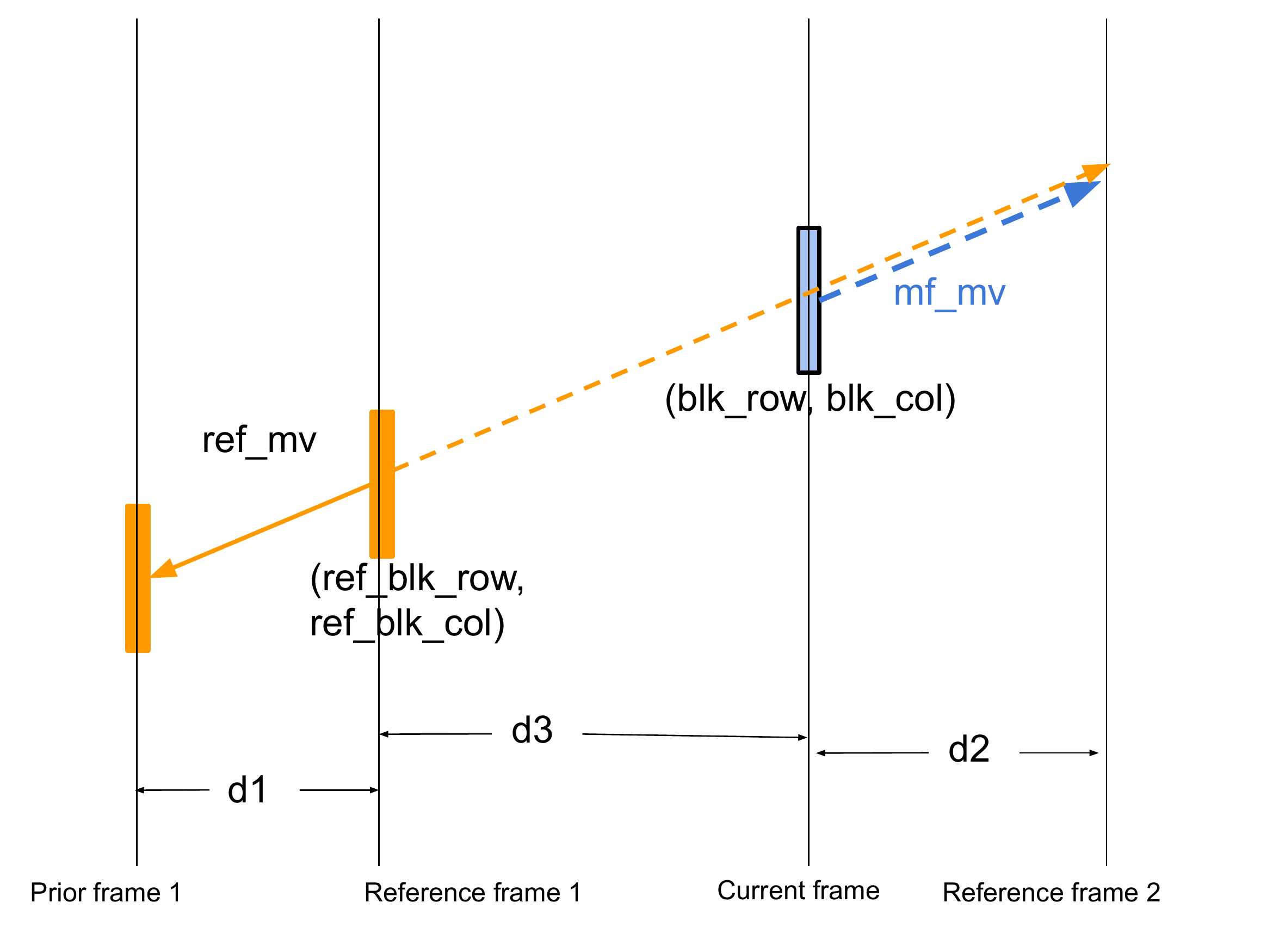}
\caption{Building motion trajectory through motion vector extrapolation.}
\label{fig:mfmv_extrapolation}
\end{figure}

\textbf{Extrapolation}
The motion vector from a reference does not cross the current frame. An example is shown in Figure \ref{fig:mfmv_extrapolation}. The motion vector ref\_mv (in orange) points from reference frame 1 to a prior frame 1. It is extended towards the current frame and they meet at block position:
\begin{equation}
\begin{bmatrix}
\text{blk\_row} \\
\text{blk\_col}
\end{bmatrix}
=
\begin{bmatrix}
\text{ref\_blk\_row} \\
\text{ref\_blk\_col}
\end{bmatrix}
-
\begin{bmatrix}
\text{ref\_mv.row} \\
\text{ref\_mv.col}
\end{bmatrix}
\cdot \frac{d_3}{d_1}
\label{eq:mfmv_pos2}
\end{equation}
Its motion field motion vector towards reference frame 2, mf\_mv (in blue), is given by
\begin{equation}
\begin{bmatrix}
\text{mf\_mv.row} \\
\text{mf\_mv.col}
\end{bmatrix}
=
-
\begin{bmatrix}
\text{ref\_mv.row} \\
\text{ref\_mv.col}
\end{bmatrix}
\cdot \frac{d_2}{d_1}
\label{eq:mfmv_extrapolation}
\end{equation}
where $d_2$ is the distance between the current frame and reference frame 2 in Figure \ref{fig:mfmv_extrapolation}. Note that the signs in both \eqref{eq:mfmv_interpolation} and \eqref{eq:mfmv_extrapolation} depend on whether the two reference frames are on the same side of the current frame.

Typically interpolation provides better estimation accuracy than the extrapolation. Therefore when a block has possible motion trajectories originated from both,  the extrapolated one will be discarded. A coding block uses the motion field of all its $8\times8$ sub-blocks as its temporal motion vector reference.

\subsubsection{Hardware Constraints}
The motion information, including the motion vector and the reference frame index, needs to be stored for later frames to build their motion fields. To reduce the memory footprint, the motion information is stored in units of $8\times8$ blocks. If a coding block is using compound modes, only the first motion vector is saved. The reference frame motion information is commonly stored in the dynamic random access memory (DRAM), a relatively cheaper and slower unit as compared to SRAM, in hardware decoders. It needs, however, to be transferred to SRAM for computing purposes. The bus between DRAM and SRAM is typically 32-bit wide. To facilitate efficient data transfer, a number of data format constraints are employed. We limit the codec to use motion information from up to 4 reference frames (out of 7 available frames) to build the motion field. Therefore only 2 bits are needed for the reference frame index. Furthermore, a motion vector with any component magnitude above $2^{12}$ will be discarded. As a result, the motion vector and reference frame index together can be represented by a 32-bit unit.  

As mentioned in Section \ref{sec:block_size_constraint}, hardware decoders process frames in $64\times64$ block units, which makes the hardware cost invariant to the frame size. In contrast, the above motion field construction can potentially involve any motion vector in the reference frame to build the motion field for a $64\times64$ block, which makes the hardware cost grow as the frame resolution scales up.

\begin{figure}[!t]
\centering
\includegraphics[width=0.98\columnwidth]{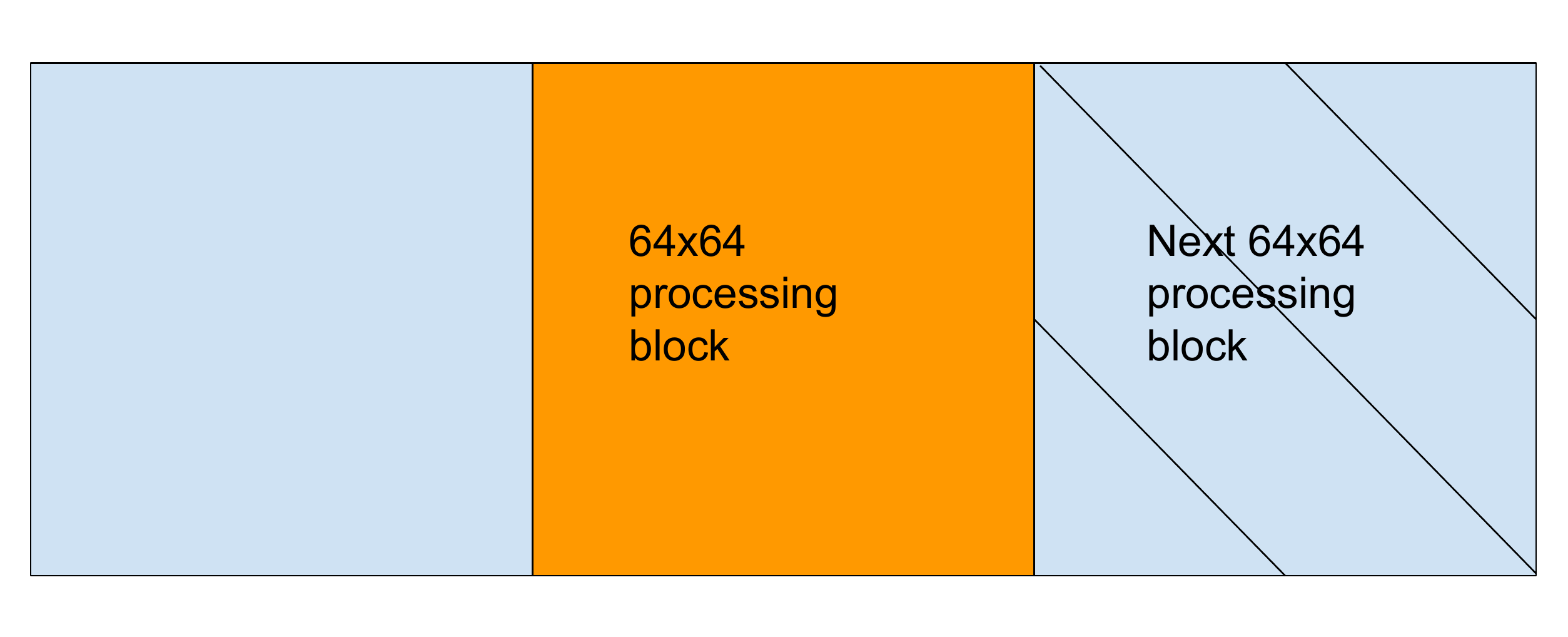}
\caption{The constrained projection localizes the referencing region needed to produce the motion field for a $64\times64$ block. The colocated block in the reference frame is at the same location as the processing block in the current frame. The blue region is the extended block whose motion vectors are used to estimate the motion field for the current $64\times64$ block.}
\label{fig:mfmv_region}
\end{figure}

To solve this problem, we constrain the maximum displacement between (ref\_blk\_row, ref\_blk\_col) and (blk\_row, blk\_col) during the motion vector projection. Let (base\_row, base\_col) denote the top-left block position of the $64\times64$ block that contains (ref\_blk\_row, ref\_blk\_col):
\begin{align}
&\text{base\_row} = (\text{ref\_blk\_row} >> 3) << 3 \\
&\text{base\_col} = (\text{ref\_blk\_col} >> 3) << 3.
\end{align}
The maximum displacement constraints are:
\begin{align}
&\text{blk\_row} \in [\text{base\_row}, ~ \text{base\_row} + 8) \\
&\text{blk\_col} \in [\text{base\_col} - 8, ~\text{base\_col} + 16).
\end{align}
Note that all the indexes here are in $8\times8$ luma sample block units. Any projection in \eqref{eq:mfmv_pos1} or \eqref{eq:mfmv_pos2} that goes beyond this limit will be discarded. This design localizes the reference region in the reference frame used to produce the motion field for a $64\times64$ pixel block to be a $64\times(64 + 2\times 64)$ block, as shown in Figure \ref{fig:mfmv_region}. It allows the codec to load the necessary reference motion vectors per $64\times64$ block from DRAM to SRAM, and process the linear projection ahead of decoding each 64x64 block. Note that we allow the width value to be larger than the height, since the shaded portion of the reference motion vector array can be readily re-used for decoding the next $64\times64$.

\subsubsection{Dynamic Motion Vector Reference List}
\label{sec:drl}
Having established the spatial and temporal reference motion vectors, we will next discuss the scheme to use them for efficient motion vector coding. The spatial and temporal reference motion vectors are classified into two categories based on where they appear: the nearest spatial neighbors and the rest. Statistically the motion vectors from immediate above, left, and top-right blocks tend to have higher correlation with the current block than the rest, and hence are considered with higher priority. Within each category, the motion vectors are ranked in descending order of their appearance counts within the spatial and temporal search range. A motion vector candidate with higher appearance count is considered to be ``popular'' in the local region, i.e., a higher prior probability. The two categories are concatenated to form a ranked list.

The first 4 motion vectors in this ranked list will be used as candidate motion vector predictors. The encoder will pick the one that is closest to the desired motion vector and send its index to the decoder. It is not uncommon for coding blocks to have fewer than 4 candidate motion vectors, due to either the high flexibility in the reference frame selection, or a highly consistent motion activity in the local region. In such context, the candidate motion vector list will be shorter than 4, which allows the codec to save bits spent on identifying the selected index. The dynamic candidate motion vector list is in contrast to the design in VP9, where one always constructs 2 candidate motion vectors. If not enough candidates are found, the VP9 codec will fill the list with zero vectors. AV1 also supports a special inter mode 
that makes the inter predictor use the frame level affine model as discussed in Section \ref{sec:affine_model}.

The motion vector difference will be entropy coded. Since a significant portion of the coding blocks will find a zero motion vector difference, the probability model is designed to account for such bias. AV1 allows a coding block to use 1 bit to indicate whether to directly use the selected motion vector predictor as its final motion vector, or to additionally code the difference. The probability model for this entropy coded bit is conditioned on two factors: whether its spatial neighbors have a non-zero motion vector difference and whether a sufficient number of motion vector predictors are found. For compound modes, where two motion vectors need to be specified, this extends to 4 cases that cover where either block, both, or neither one have a zero difference motion vector. The non-zero difference motion vector coding is consistent in all cases.

\subsection{Transform Coding}
\label{sec:tx_coding}
Transform coding is applied to the prediction residual to remove the potential spatial correlations. VP9 uses a uniform transform block size design, where all the transform blocks within a coding block share the same transform size. Four square transform sizes are supported by VP9, $4\times4$, $8\times8$, $16\times16$, and $32\times32$. A set of separable 2-D transform types, constructed by combinations of 1-D discrete cosine transform (DCT) and asymmetric discrete sine transform (ADST) kernels \cite{DCT,ADST} , are selected based on the prediction mode. AV1 inherits the transform coding scheme in VP9 and extends its flexibility in terms of both the transform block sizes and the kernels. 

\subsubsection{Transform Block Size}
AV1 extends the maximum transform block size to $64\times64$. The minimum transform block size remains $4\times4$. In addition, rectangular transform block sizes at $N\times N/2$, $N/2 \times N$, $N\times N/4$, and $N/4 \times N$ are supported to complement the rectangular coding block sizes in Section \ref{sec:cbsize}.

\begin{figure}[!t]
\centering
\includegraphics[width=0.98\columnwidth]{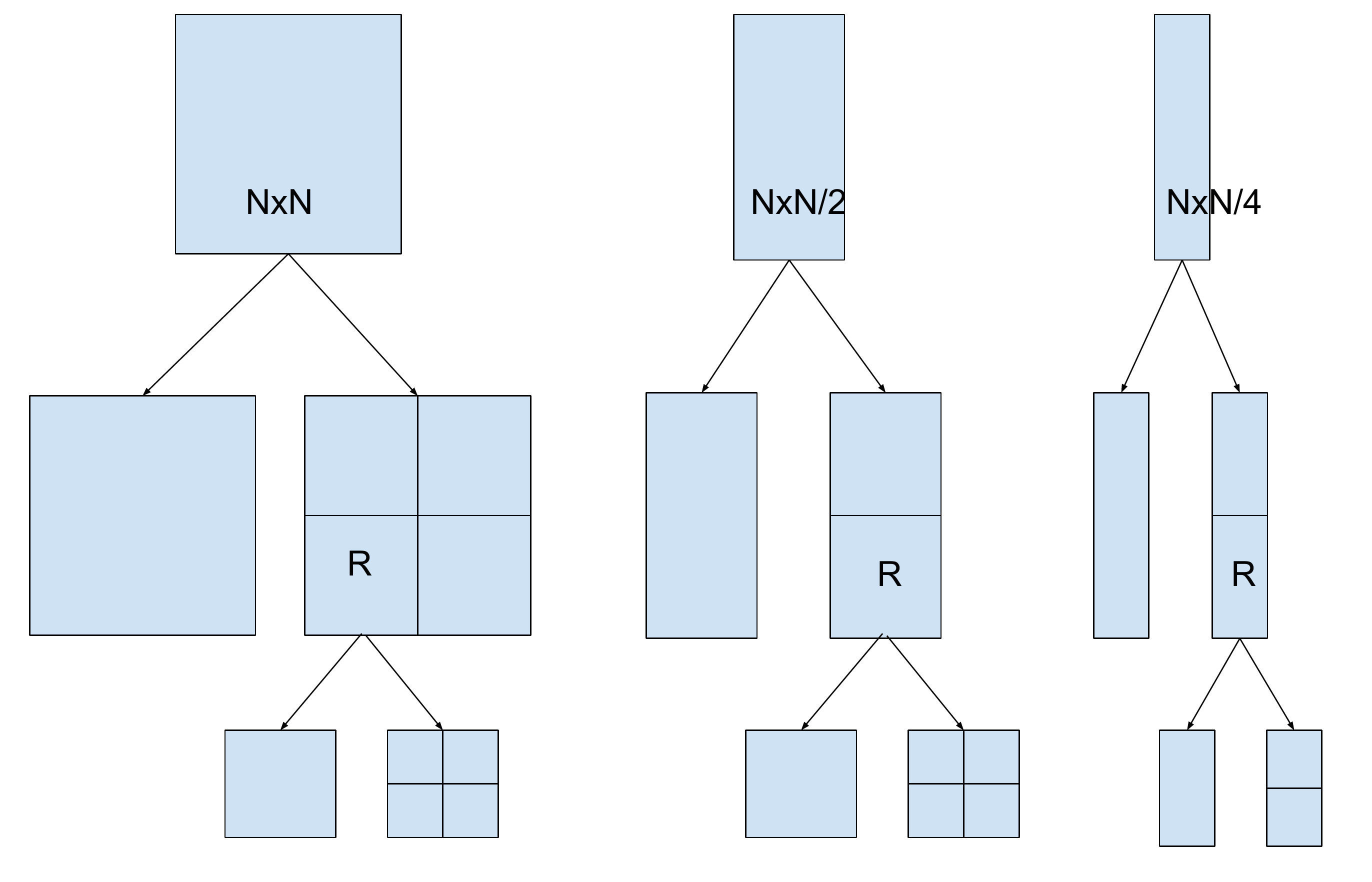}
\caption{The transform block partition for square and rectangular inter blocks. R denotes the recursive partition point. Each coding block allows a maximum 2 level recursive partition.}
\label{fig:txfm_partition}
\end{figure}

A recursive transform block partition approach is adopted in AV1 for all the inter coded blocks to capture localized stationary regions for transform coding efficiency. The initial transform block size matches the coding block size, unless the coding block size is above $64\times64$, in which case the $64\times64$ transform block size is used. For the luma component, up to 2 levels of transform block partitioning are allowed. The recursive partition rules for $N\times N$, $N\times N/2$, and $N\times N/4$ coding blocks are shown in Figure \ref{fig:txfm_partition}. 

\begin{figure}[!t]
\centering
\includegraphics[width=0.98\columnwidth]{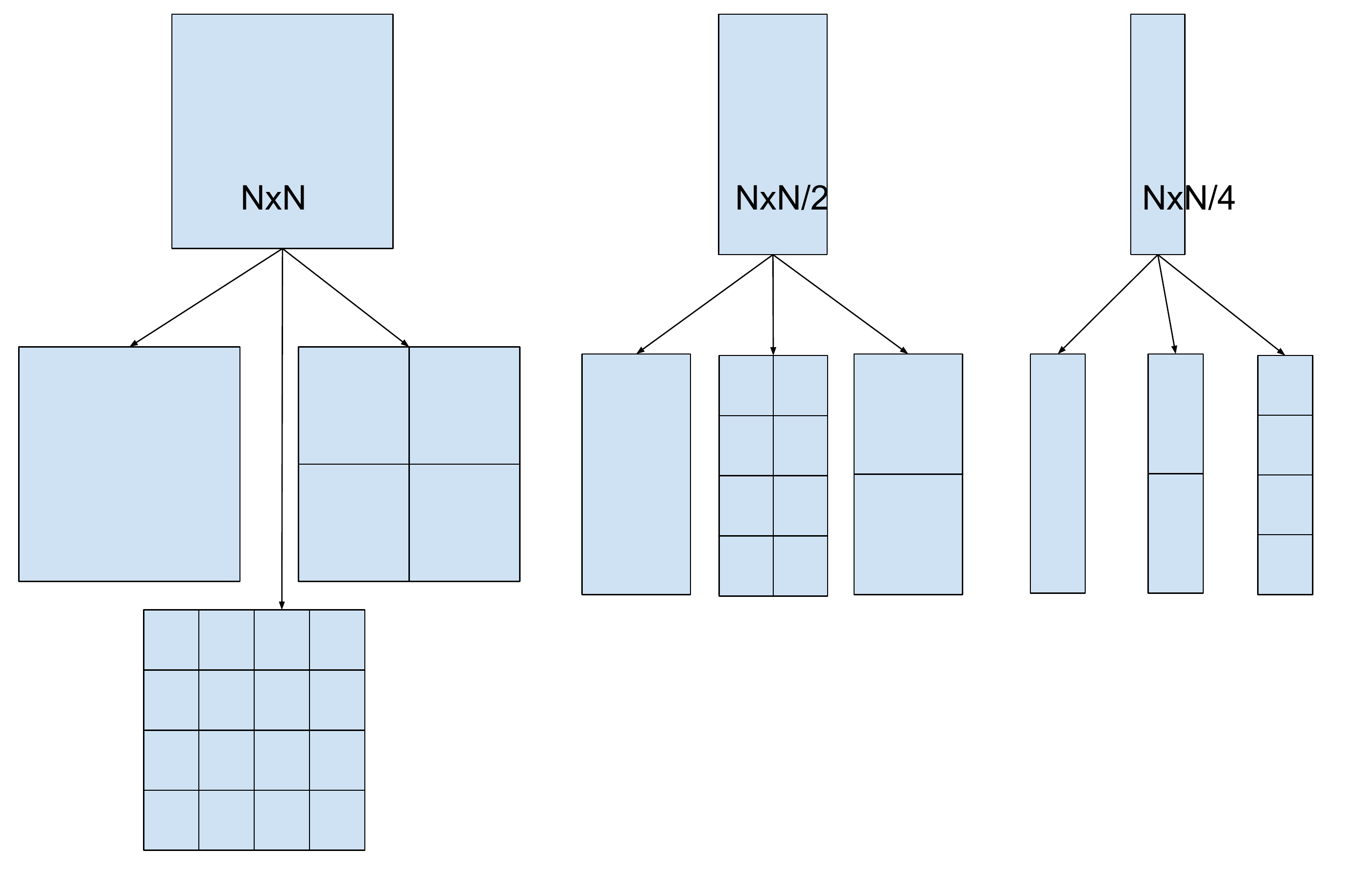}
\caption{The transform block size options for square and rectangular intra blocks.}
\label{fig:txfm_intra}
\end{figure}

The intra coded block inherits the uniform transform block size approach, i.e. all transform blocks have the same size. Similar to the inter block case, the maximum transform block size matches the coding block size, and can go up to 2 levels down for the luma component. The available options for square and rectangular coding block sizes are shown in Figure \ref{fig:txfm_intra}.

The chroma components tend to have much less variations in their statistics. Therefore the transform block is set to use the largest available size.

\subsubsection{Transform Kernels}
\label{sec:txfm_kernels}
\begin{figure*}[!htb]
\centering
\includegraphics[width=0.98\textwidth]{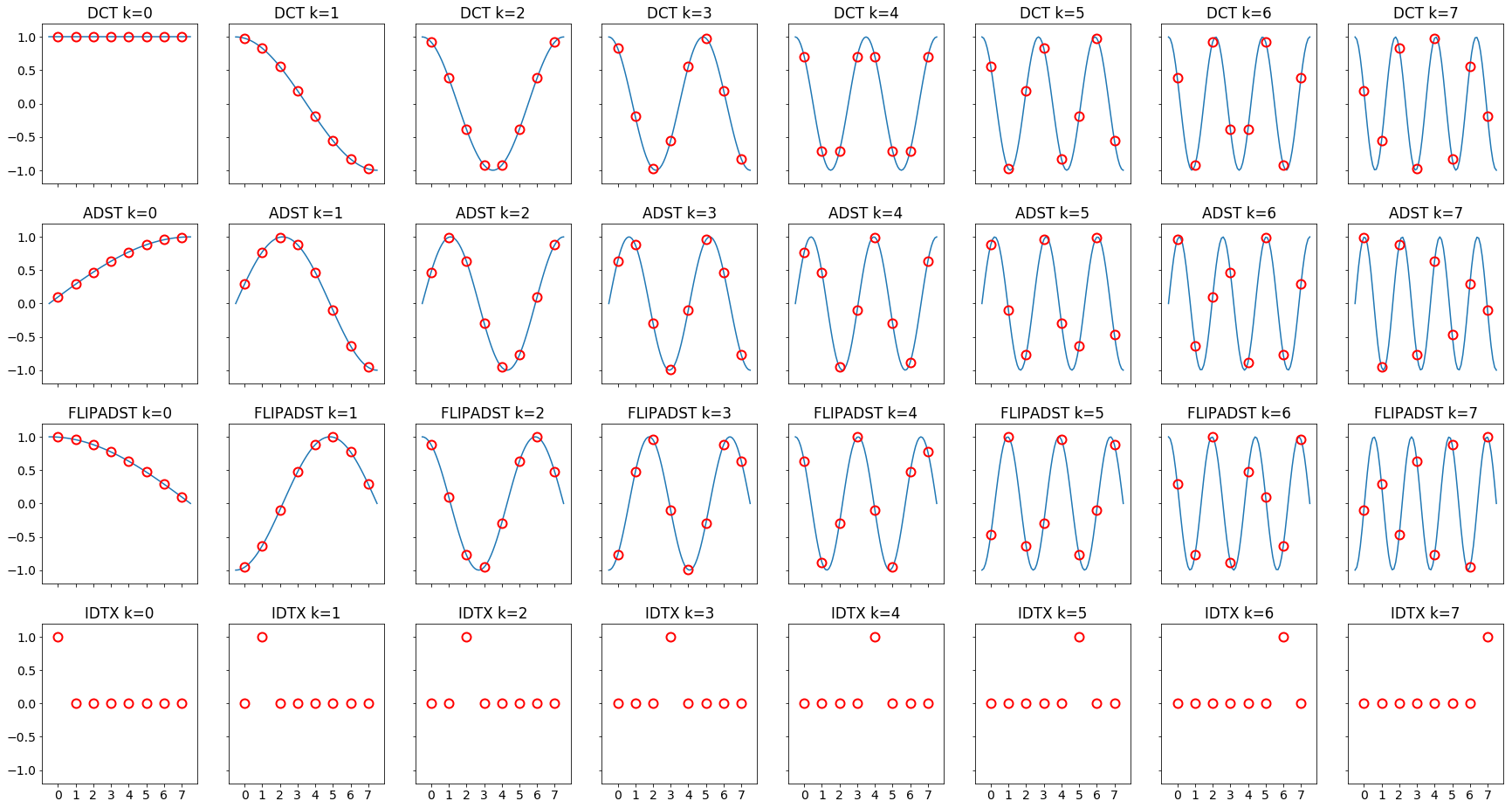}
\caption{Transform kernels of DCT, ADST, FLIPADST and IDTX for dimension $N = 8$. The discrete basis values are displayed as red circles, with blue lines indicating the associated sinusoidal function.  The bases of DCT and ADST (a variant with a fast butterfly structured implementation) take the form of $cos(\frac{(2n+1)k\pi}{2N})$ and $sin(\frac{(2n+1)(2k+1)\pi}{4n})$ respectively, where $n$ and $k$ denote time index and the frequency index, taking values from $\{0, 1, ...,  N-1\}$. FLIPADST utilizes the reversed ADST bases, and IDTX denotes the identity transformation.}
\label{fig:tx_basis}
\end{figure*}

Unlike VP9 where each coding block has only one transform type, AV1 allows each transform block to choose its own transform kernel independently. The 2-D separable transform kernels are extended to combinations of four 1-D kernels: DCT, ADST, flipped ADST (FLIPADST), and identity transform (IDTX), resulting in a total of 16 2-D transform kernels. The FLIPADST is a reverse of the ADST kernel. The kernels are selected based on statistics and to accommodate various boundary conditions. The DCT kernel is widely used in signal compression and is known to approximate the optimal linear transform, Karhunen-Loeve transform (KLT), for consistently correlated data. The ADST, on the other hand, approximates the KLT where one-sided smoothness is assumed, and therefore is naturally suitable for coding some intra prediction residuals. Similarly the FLIPADST captures one-sided smoothness from the opposite end. The IDTX is further included to accommodate situations where sharp transitions are contained in the block and neither DCT nor ADST are effective. Also, the IDTX, combined with other 1-D transforms, provides the 1-D transforms themselves, therefore allowing for better compression of  horizontal and vertical patterns in the residual \cite{transform1D}. The waveforms corresponding to the four 1-D transform kernels are presented in Figure \ref{fig:tx_basis} for dimension $N = 8$.

Even with modern single instruction multiple data (SIMD) architectures, the inverse transform accounts for a significant portion of the decoder computational cost. The butterfly structure \cite{fastDCT} allows substantial reduction in multiplication operations over plain matrix multiplication, i.e., a reduction from $O(N^2)$ to $O(NlogN)$, where $N$ is the transform dimension. Hence it is highly desirable for large transform block sizes. Note that since the original ADST derived in \cite{ADST}  cannot be decomposed for the butterfly structure, a variant of it, as introduced in \cite{btfADST} and also as shown in Figure \ref{fig:tx_basis}, is adopted by AV1 for transform block sizes of $8\times8$ and above.

When the transform block size is large, the boundary effects are less pronounced, in which setting the transform coding gains of all sinusoidal transforms largely converge \cite{ADST}. Therefore only the DCT and IDTX are employed for transform blocks at dimension $32\times32$ and above.

\subsection{Quantization}
The transform coefficients are quantized and the quantization indexes are entropy coded. The quantization parameter (QP) in AV1 ranges between 0 and 255.

\subsubsection{Quantization Step Size}
\begin{figure}[!t]
\includegraphics[width=0.98\columnwidth]{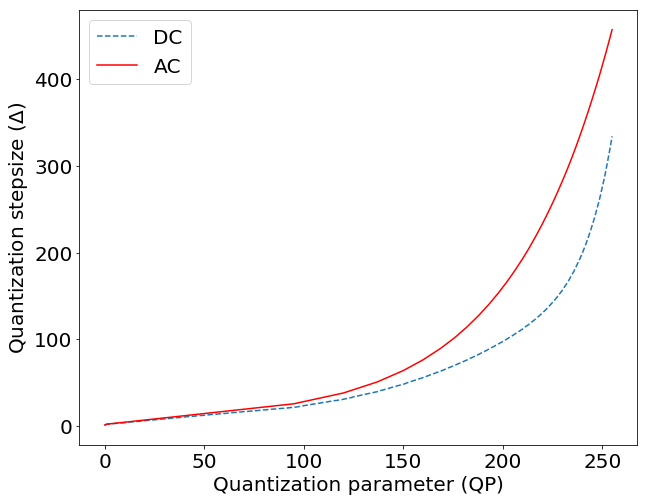}
\centering
\caption{The quantization parameter and quantization step size maps for DC and AC coefficients.}
\label{fig:qp_delta}
\end{figure}

 At a given QP, the quantization step size for DC coefficient is smaller than that for AC coefficient. The mapping from QP to quantization step size for both DC and AC coefficients is drawn in Figure \ref{fig:qp_delta}. The lossless coding mode is achieved when QP is 0. By default, all the AC coefficients will use the same quantization step size. Since human visual system tends to have different tolerance to distortions at various frequencies, AV1 also supports 15 sets of pre-defined quantization weighting matrices, where the quantization step size for each individual frequency component is further scaled differently. Each frame can optionally select a quantization weighting matrix set for luma and chroma planes, respectively. 

\subsubsection{Quantization Parameter Modulation}
AV1 assigns a base QP for a coded frame, denoted by $QP_{base}$. The QP values for the DC and AC coefficients in both luma and chroma components are shown in Table \ref{table:frame_qps}. $\Delta QP_{p, b} $ are additional offset values transmitted in the frame header, where $p \in \{Y, U, V\}$ denotes the plane and $b \in \{DC, AC\}$ denotes the DC or the AC transform coefficients. 

Recognizing the coding blocks within a frame may have different rate-distortion trade-offs, AV1 further allows QP offset at both superblock and coding block levels. The resolution of QP offset at superblock level is signaled by the frame header. The available options are 1, 2, 4, and 8. The coding block level QP offset can be achieved through segmentations. AV1 allows a frame to classify its coding blocks into up to 8 segments, each has its own QP offset decided by the frame header. The segment index associated with each coding block is sent through the bit-stream to the decoder.

Therefore, the effective QP for AC coefficients in a coding block, $QP_{cb}$, is given by
\begin{equation}
QP_{cb} = clip(QP_{frame}+ \Delta QP_{sb} + \Delta QP_{seg}, 1, 255),
\end{equation}
where $\Delta QP_{sb}$ and $\Delta QP_{seg}$ are the QP offsets from the superblock and the segment, respectively. The clip function ensures it stays within a valid range. The QP is not allowed to change from a non-zero value to zero, since zero is reserved for lossless coding. 

\begin{table}[]
\begin{center}
\caption{Frame level QP values ($QP_{frame}$) for Y/U/V planes.}
\label{table:frame_qps}
\begin{tabular}{| c | c | c |}
\hline
 & AC & DC     \\
 \hline
 Y & $QP_{base}$ & $QP_{base}$ + $\Delta QP_{Y, DC} $   \\
 U & $QP_{base}$ + $\Delta QP_{U, AC} $ &  $QP_{base}$ + $\Delta QP_{U, DC} $   \\
 V & $QP_{base}$ + $\Delta QP_{V, AC} $ &  $QP_{base}$ + $\Delta QP_{V, DC} $   \\
 \hline
\end{tabular}
\end{center}
\end{table}

%

The decoder rebuilds the quantized samples using a uniform quantizer. Given the quantization step size $\Delta$ and the quantization index $k$, the reconstructed sample is $k\Delta$.

\section{Entropy Coding System}
AV1 employs an M-ary symbol arithmetic coding method that was originally developed for the Daala video codec \cite{daala} to compress the syntax elements, where integer $M\in [2,~14]$. The probability model is updated per symbol coding.

\subsection{Probability Model}
Consider an M-ary random variable whose probability mass function (PMF) at time $n$ is defined as
\begin{equation}
\bar{P}_n =
\begin{bmatrix}
p_1(n), p_2(n), \cdots, p_{M}(n)
\end{bmatrix},
\end{equation}
and the cumulative distribution function (CDF) given by
\begin{equation}
\bar{C}_n =
\begin{bmatrix}
c_1(n), c_2(n), \cdots, c_{M-1}(n), 1
\end{bmatrix},
\label{eq:cdf}
\end{equation}
where $c_k(n) = \sum_{i=1}^{k}p_i(n)$. When the symbol is coded, a new outcome $k\in\{1,2,\cdots, M\}$ is observed. The probability model is then updated as
\begin{equation}
\bar{P}_n = \bar{P}_{n-1} (1 - \alpha) + \alpha \bar{e}_k,
\label{eq:pmf_update}
\end{equation}
where $\bar{e}_k$ is an indicator vector whose k-th element is 1 and the rest are 0, and $\alpha$ is the update rate. 

To update the CDF, we first consider  $c_m(n)$ where $m<k$:
\begin{align*}
c_m(n) = \sum_{i=1}^{m}p_i(n) &= \sum_{i=1}^{m}p_i(n - 1) \cdot (1 - \alpha) \\
& = c_m(n - 1) \cdot (1 - \alpha).
\end{align*}
For $m \ge k$ cases, we have
\begin{align*}
1 - c_m(n) = \sum_{i = m+1}^{M}p_i(n) &= \sum_{i = m+1}^{M}p_i(n - 1)  \cdot (1 - \alpha) \\
&= (1 - c_m(n-1)) \cdot (1 - \alpha),
\end{align*}
where the second equation follows \eqref{eq:pmf_update} and $m+1 > k$. Rearranging the terms, we have
\begin{equation}
c_m(n) = c_m(n - 1) + \alpha \cdot (1 - c_m(n-1)).
\end{equation}
In summary, the CDF is updated as
\begin{equation}
c_m(n) =
\begin{cases}
 c_m(n - 1) \cdot (1 - \alpha), ~m < k \\
 c_m(n - 1) + \alpha \cdot (1 - c_m(n-1)), ~m\ge k
\end{cases}
\end{equation}
AV1 stores M-ary symbol probabilities in the form of CDFs. The elements in \eqref{eq:cdf} are scaled by $2^{15}$ for integer precision. The arithmetic coding directly uses the CDFs to compress symbols \cite{arithmetic-coding}.

The probability update rate associated with a symbol adapts based on the count of this symbol's appearance within a frame:
\begin{equation}
\alpha = \frac{1}{2^{3 + I(\text{count}>15) + I(\text{count}>32) + min(log_2(M), 2)}},
\end{equation}
where $I(event)$ is $1$ if the event is true, and $0$ otherwise. It allows higher adaptation rate at the beginning of each frame. The probability models are inherited from one of the reference frames whose index is signaled in the bit-stream.

\subsection{Arithmetic Coding}
\begin{figure}[!t]
\centering
\includegraphics[width=0.98\columnwidth]{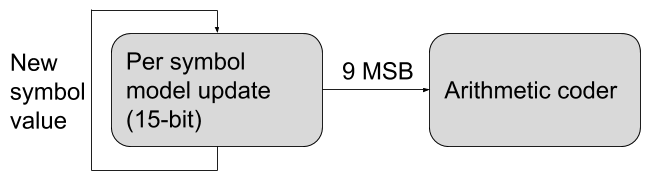}
\caption{The probability model is updated and maintained in 15-bit precision, whilst only the most significant 9 bits are used by the arithmetic coder.}
\label{fig:prob_model_prec}
\end{figure}

The M-ary symbol arithmetic coding largely follows \cite{arithmetic-coding} with all the floating-point data scaled by $2^{15}$ and represented by 15-bit unsigned integers. 
To improve hardware throughput, AV1 adopts a dual model approach to make the involved multiplications fit in 16 bits. The probability model CDF is updated and maintained an 15-bit precision, but when it is used for entropy coding, only the most significant 9 bits are fed into the arithmetic coder, as shown in Figure \ref{fig:prob_model_prec}. 

Let $R$ denote the arithmetic coder's current interval length, and $Value$ denote the code string value. The decoding processing is depicted as Algorithm \ref{alg:arth_decode2}. Note that the interval length $R$ is scaled down by $1/256$ prior to the multiplication, which makes the the product $(R >> 8) \times f)$ fits into 16 bits.

\begin{algorithm}
\caption{The modified arithmetic decoder operations.}
\label{alg:arth_decode2}
\begin{algorithmic}
\STATE $low \leftarrow R$
\FOR{$k=1;~Value<low;~k=k+1$}

\STATE $up \leftarrow low$
\STATE $f \leftarrow 2^{9} - (c_k  >>  6)$

\STATE $low \leftarrow ((R >> 8) \times f) >> 1$
\ENDFOR
\STATE $R \leftarrow up - low$
\STATE $Value \leftarrow Value - low$
\end{algorithmic}
\end{algorithm}

\subsection{Level Map Transform Coefficient Coding System}
\label{sec:entropy}
The transform coefficient entropy coding system is an intricate and performance critical component in video codecs. We discuss its design in AV1 that decomposes it into a series of symbol codings.

\subsubsection{Scan Order}
\begin{figure}[!t]
\centering
\includegraphics[width=0.98\columnwidth]{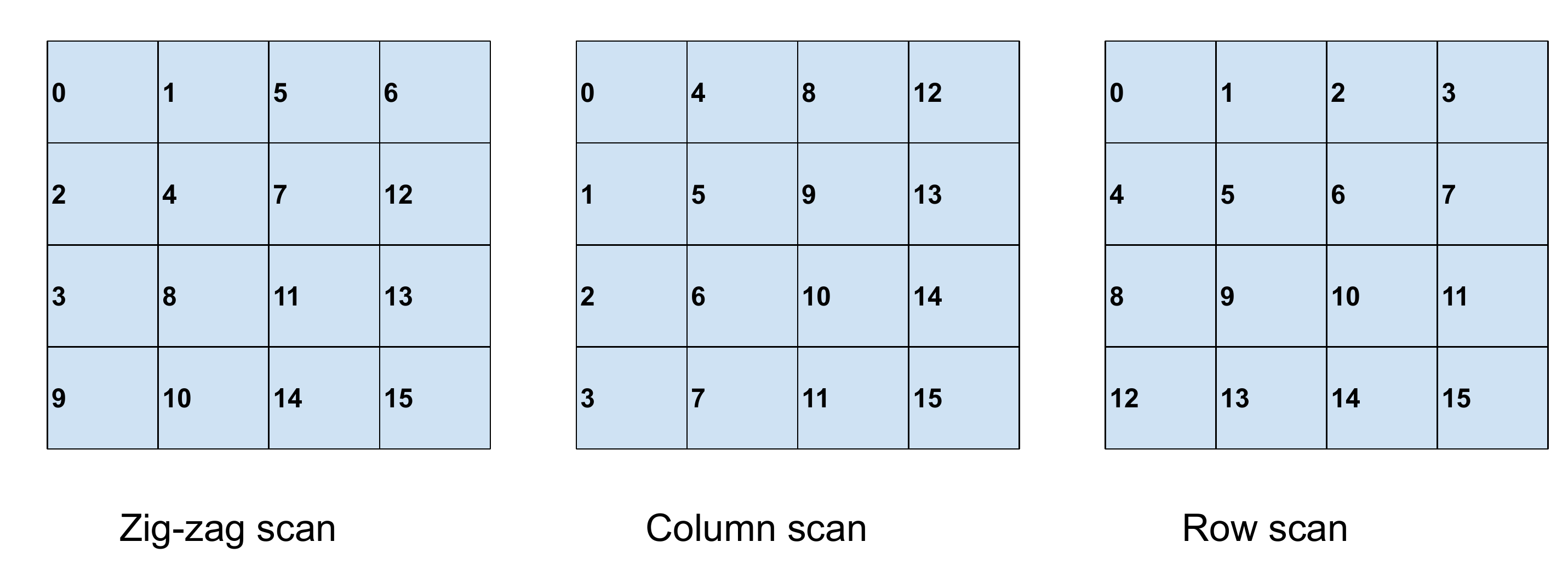}
\caption{The scan order is decided by the transform kernel. An example is drawn for $4\times4$ transform blocks. The index represents the scan order. Left: zig-zag scan for 2-D transform block. Middle: column scan for 1-D vertical transform. Right: row scan for 1-D horizontal transform.}
\label{fig:scan_order}
\end{figure}

A 2-D quantized transform coefficient matrix is first mapped into an 1-D array for sequential processing. The scan order depends on the transform kernel (see Section \ref{sec:txfm_kernels}). A column scan is used for 1-D vertical transform and a row scan is used for 1-D horizontal transform. In both settings, we consider that the use of 1-D transform indicates strong correlation along the selected direction and weak correlation along the perpendicular direction. A zig-zag scan is used for both 2-D transform and identity matrix (IDTX), as shown in Figure \ref{fig:scan_order}.

\subsubsection{Symbols and Contexts}
The index of the last non-zero coefficient in the scan order is first coded. The coefficients are then processed in reverse scan order. The range of a quantized transform coefficient is $[-2^{15}, 2^{15})$. In practice, the majority of quantized transform coefficients are concentrated close to the origin. Hence AV1 decomposes a quantized transform coefficients into 4 symbols:
\begin{itemize}
  \item Sign bit: When it is 1, the transform coefficient is negative; otherwise it is positive.
  \item Base range (BR): The symbol contains 4 possible outcomes $\{0, 1, 2, >2\}$, which are the absolute values of the quantized transform coefficient. An exception is for the last non-zero coefficient, where BR$\in \{1, 2, >2\}$, since $0$ has been ruled out.
  \item Low range (LR): It contains 4 possible outcomes $\{0, 1, 2, >2\}$ that correspond to the residual value over the previous symbols' upper limit.
  \item High range (HR): The symbol has a range of $[0, 2^{15})$ and corresponds to the residual value over the previous symbols' upper limit.
\end{itemize}

\begin{figure}[!t]
\centering
\includegraphics[width=0.98\columnwidth]{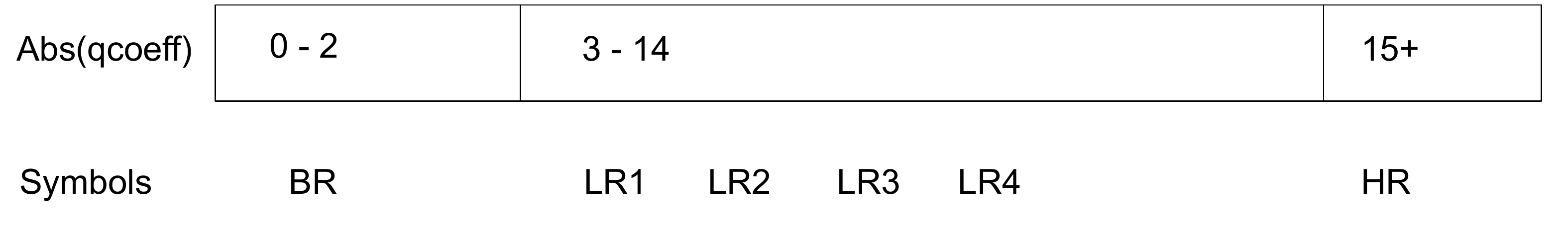}
\caption{The absolute value of a quantized transform coefficient $V$ is decomposed into BR, LR, and HR symbols. }
\label{fig:qcoeff_symbols}
\end{figure}

To code a quantized transform coefficient $V$, one first processes its absolute value. As shown in Figure \ref{fig:qcoeff_symbols}, if $|V| \in [0, 2]$, the BR symbol is sufficient to signal it and the coding of $|V|$ is terminated. Otherwise the outcome of the BR symbol will be ``$>2$'', in which case an LR symbol is used to signal $|V|$. If $V \in [3, 5]$, this LR symbol will be able to cover its value and complete the coding. If not, a second LR is used to  further code $|V|$. This is repeated up to 4 times, which effectively covers the range $[3, 14]$. If $|V|>14$,  an additional HR symbol is coded to signal $(|V|-14)$.

\begin{figure}[!t]
\centering
\includegraphics[width=0.98\columnwidth]{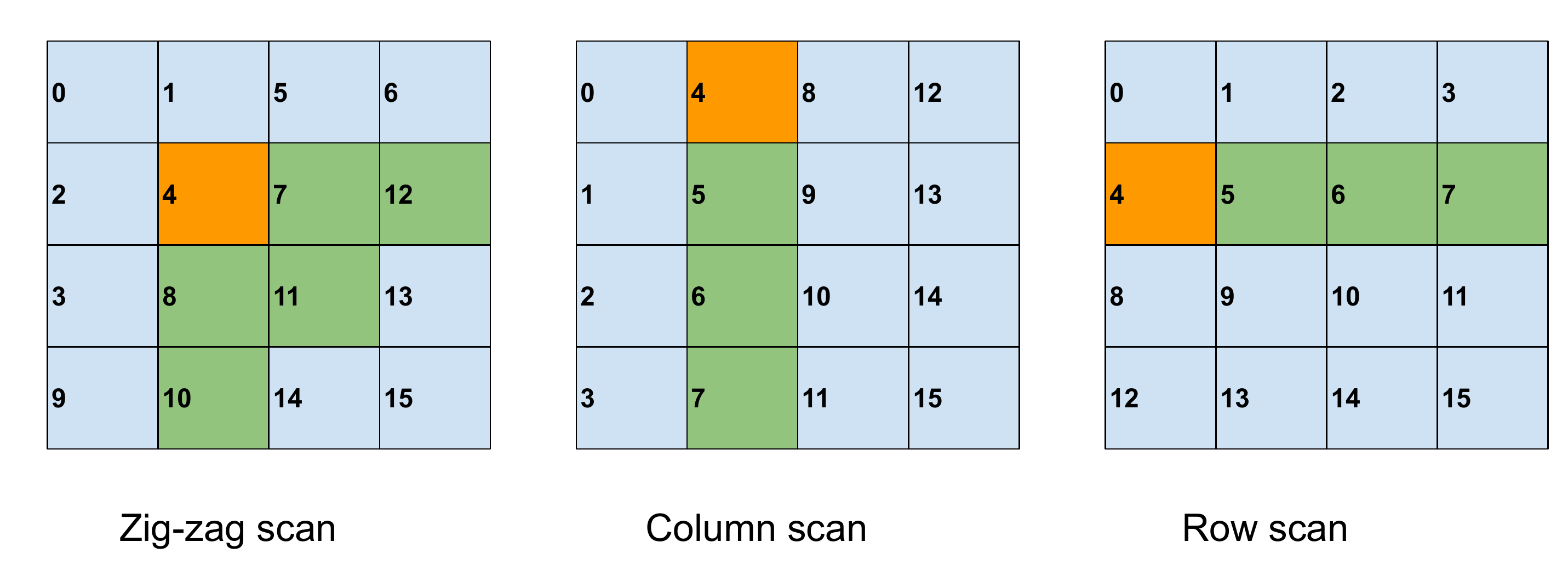}
\caption{Reference region for symbol BR. Left: A coefficient (in orange) in a 2-D transform block uses 5 previously processed coefficients (in green) to build the context for its conditional probability model. Middle and Right: A coefficient (in orange) in a 1-D transform block uses 3 previously processed coefficients (in green) along the transform direction to build the context for its conditional probability model.}
\label{fig:txcoeff_ref}
\end{figure}

The probability model of symbol BR is conditioned on the previously coded coefficients in the same transform block. Since a transform coefficient can have correlations with multiple neighboring samples \cite{level-map}, we extend the reference samples from two spatially nearest neighbors in VP9 to a region that depends on the transform kernel as shown in Figure \ref{fig:txcoeff_ref}. For 1-D transform kernels, it uses 3 coefficients after the current sample along the transform direction. For 2-D transform kernels, up to 5 neighboring coefficients in the immediate right-bottom region are used. In both cases, the absolute values of the reference coefficients are added and the sum is considered as the context for the probability model of BR. 

\begin{figure}[!t]
\centering
\includegraphics[width=0.98\columnwidth]{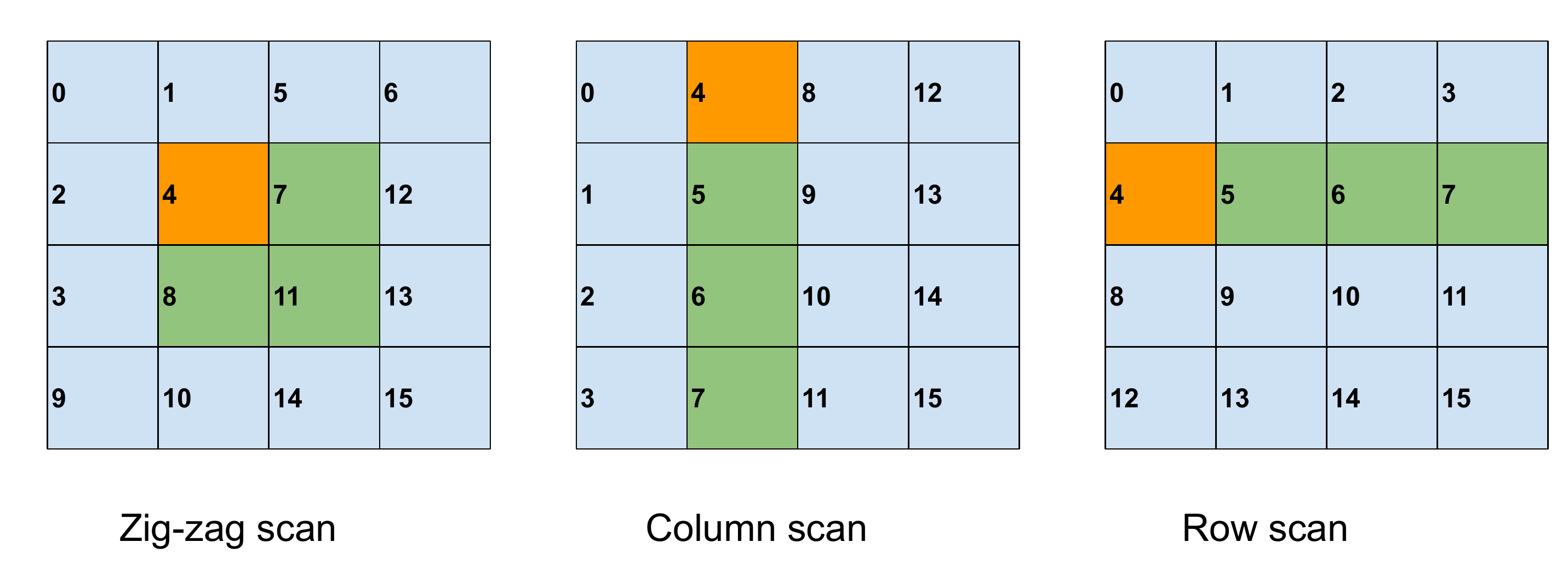}
\caption{Reference region for symbol LR. Left: A coefficient (in orange) in a 2-D transform block uses 3 previously processed coefficients (in green) to build the context for its conditional probability model. Middle and Right: A coefficient (in orange) in 1-D transform block uses 3 previously processed coefficients (in green) along the transform direction to build the context for its conditional probability model.}
\label{fig:txcoeff_ref_lw}
\end{figure}

Similarly, the probability model of symbol LR is designed as shown in Figure \ref{fig:txcoeff_ref_lw}, where the reference region for 2-D transform kernels is reduced to the nearest 3 coefficients. The symbol HR is coded using Exp-Golomb code \cite{Golomb}.

The sign bit is only needed for non-zero quantized transform coefficients. Since the sign bits of AC coefficients are largely uncorrelated, they are coded in raw bits. To improve hardware throughput, all the sign bits of AC coefficients within a transform block are packed together for transmission in the bit-stream, which allows a chunk of data to bypass the entropy coding route in hardware decoders. The sign bit of the DC coefficient, on the other hand, is entropy coded using a probability model conditioned on the sign bits of the DC coefficients in the above and left transform blocks.

\section{Post-Processing Filters}
\begin{figure}[!t]
\centering
\includegraphics[width=0.98\columnwidth]{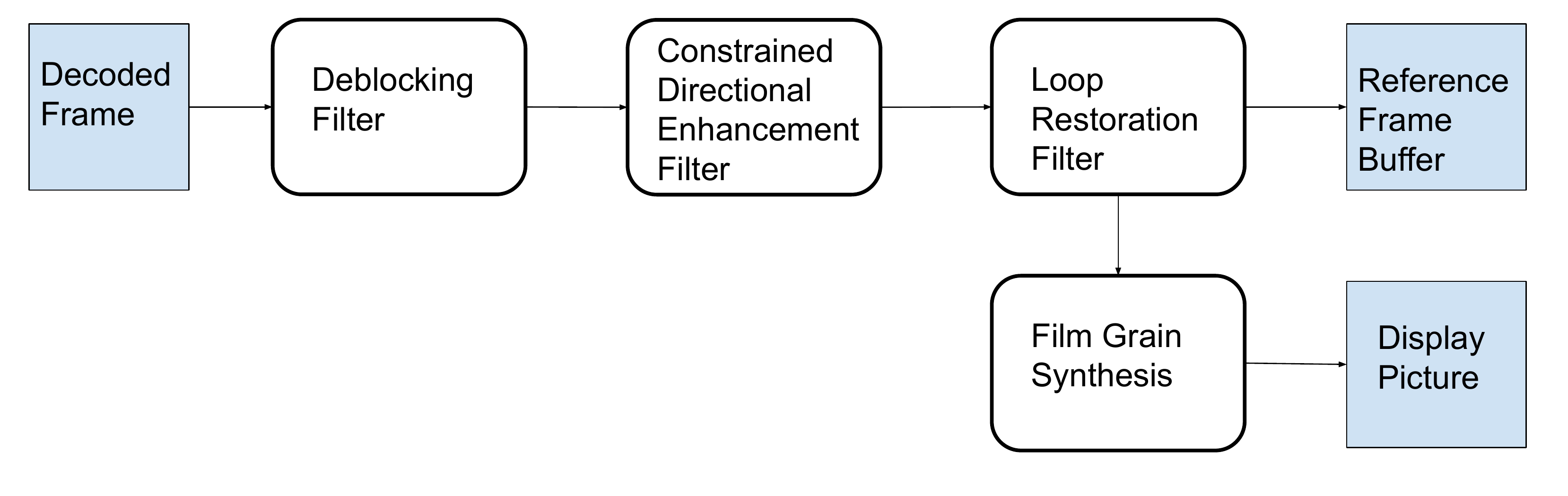}
\caption{AV1 allows 3 optional in-loop filter stages including a deblocking filter, a constrained directional enhancement filter, and a loop restoration filter. A normative film grain synthesis stage is supported for the displayed picture.}
\label{fig:post_filters}
\end{figure}

AV1 allows 3 optional in-loop filter stages: a deblocking filter, a constrained directional enhancement filter (CDEF), and a loop restoration filter, as illustrated in Figure \ref{fig:post_filters}. The filtered output frame is used as a reference frame for later frames. A normative film grain synthesis stage can be optionally applied prior to display. Unlike the in-loop filter stages, the results of the film grain synthesis stage do not influence the prediction for subsequent frames. It is hence referred to as out-of-loop filter.

\subsection{Deblocking Filter}
The deblocking filter is applied across the transform block boundaries to remove block artifacts caused by the quantization error. The logic for the vertical and horizontal edges is fairly similar. We use the vertical edge case to present the design principles.

\subsubsection{Filter Length}
\begin{figure}[!t]
\includegraphics[width=0.75\columnwidth]{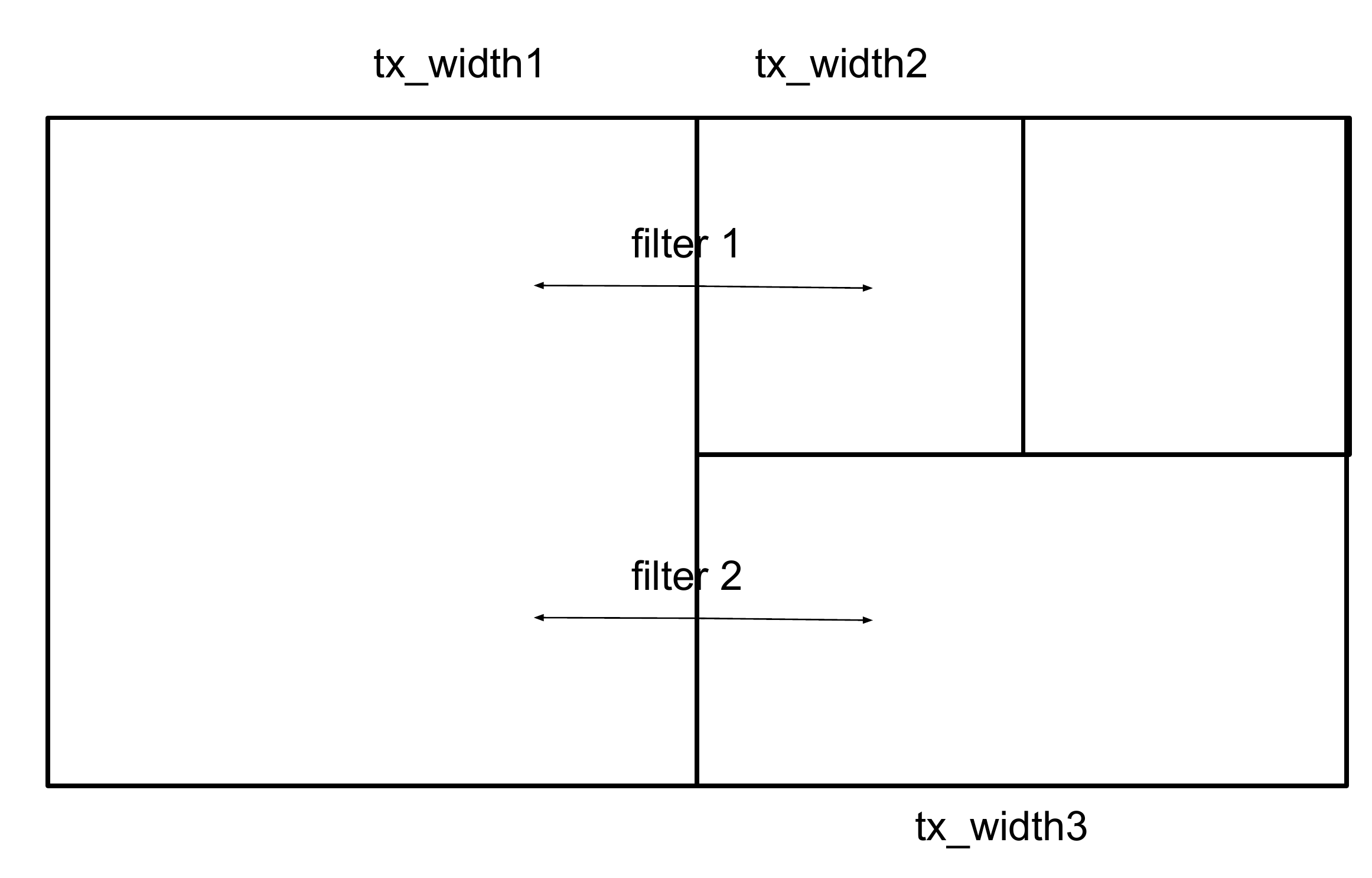}
\centering
\caption{The filter length is decided by the minimum transform block sizes on both sides.}
\label{fig:filter_length}
\end{figure}

AV1 supports 4-tap, 8-tap, and 14-tap FIR filters for the luma components, and 4-tap and 6-tap FIR filters for chroma components. All the filter coefficients are preset in the codec. The filter length is decided by the minimum transform block sizes on both sides. For example, in Figure \ref{fig:filter_length} the length of filter1 is given by $min(tx\_width1, tx\_width2)$, whereas the length of filter2 is given by $min(tx\_width1, tx\_width3)$. If the transform block dimension is 16 or above on both sides, the filter length is set to be 14.

Note that this selected filter length is the maximum filter length allowed for a given transform block boundary. The final filter further depends on a flatness metric discussed next.

\subsubsection{Boundary Conditions}
\begin{figure}[!t]
\includegraphics[width=0.98\columnwidth]{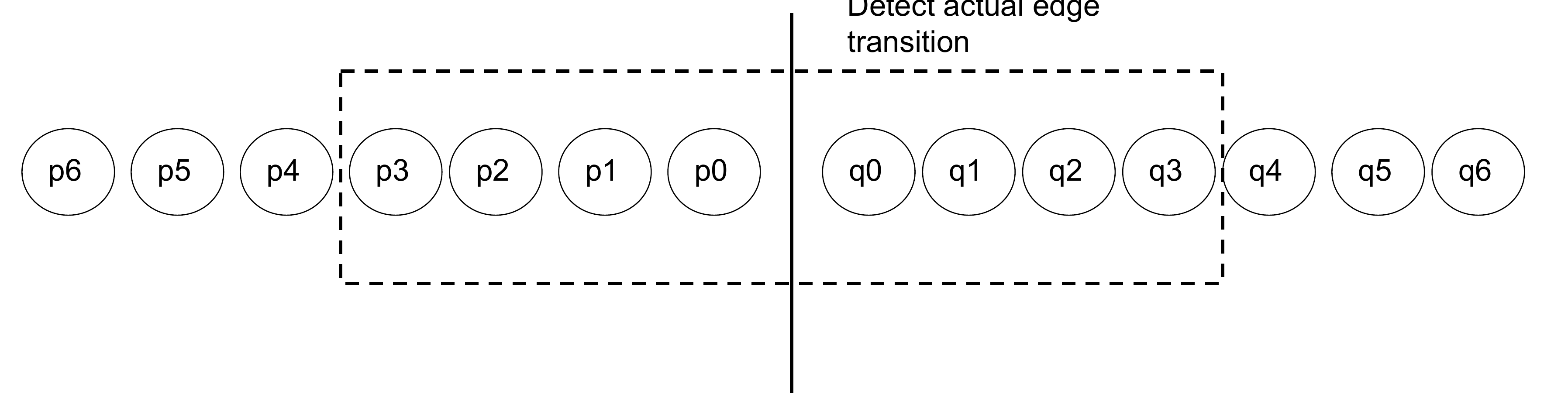}
\centering
\caption{Pixels at a transform block boundary. The dashed line shows the pixels near the transform block boundary. $p_0$-$p_6$ and $q_0$-$q_6$ are the pixels on the two sides.}
\label{fig:txb_boundary}
\end{figure}

The FIR filters used by the deblocking stage are low-pass filters. To avoid blurring an actual edge in the original image, an edge detection is conducted to disable the deblocking filter at transitions that contain a high variance signal. We use notations shown in Figure \ref{fig:txb_boundary}, where the dashed line shows the pixels near the transform block boundary. Denote the pixels on the two sides $p_0$-$p_6$ and $q_0$-$q_6$. We consider the transition along the lines $p_6$ to $q_6$ high variance and hence disable the deblocking filter, if any of the following conditions is true:
\begin{itemize}
  \item $|p_1 - p_0| > T_0$
  \item $|q_1 - q_0| > T_0$
  \item $2|p_0 - q_0| + \frac{|p_1 - q_1|}{2} > T_1$
\end{itemize}
If the filter length is 8 or 14, two additional samples are checked to determine if the transition contains a high variance signal:
\begin{itemize}
  \item $|p_3 - p_2| > T_0$
  \item $|q_3 - q_2| > T_0$
\end{itemize}
The thresholds $T_0$ and $T_1$ can be decided on a superblock by superblock basis. A higher threshold allows more transform block boundaries to be filtered. In AV1 these thresholds can be independently set in the bit-stream for the vertical and horizontal edges in the luma component and for each chroma plane.

To avoid the ringing artifacts, AV1 further requires that a long filter is only used when both sides are ``flat''. For the 8-tap filter, this requires $|q_k-q_0| \leq 1$ and $|p_k - p_0|\leq 1$ where $k\in\{1, 2, 3\}$. For the 14-tap filter, the condition extends to $k\in\{1, 2, \cdots, 6\}$. If any flatness condition is false, the codec reverts to a shorter filter for that boundary.

\subsection{Constrained Directional Enhancement Filter}
\begin{figure}[!t]
\includegraphics[width=0.98\columnwidth]{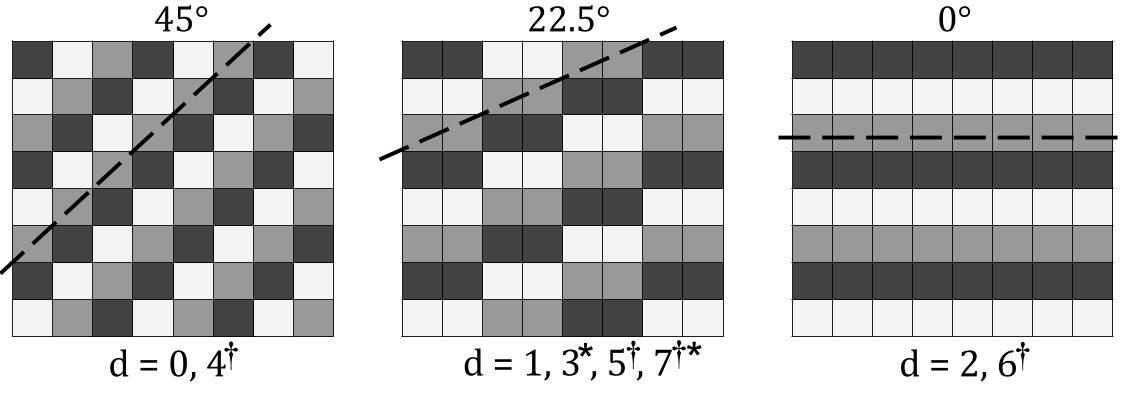}
\centering
\caption{ The templates of preset directions and their associated directions. The templates correspond to directions of 45\textdegree, 22.5\textdegree~and 0\textdegree, as shown by the dash lines. Each preset direction $d\in\{0,\cdots,7\}$ can be obtained by using the template directly, rotating the template by 90\textdegree~clockwise (marked by $\dagger$) or reflecting the template along the horizontal axis (marked by $\star$).}
\label{fig:cdef_directions}
\end{figure}

The constrained directional enhancement filter (CDEF) allows the codec to apply a non-linear deringing filter along certain (potentially oblique) directions \cite{cdef}. It operates in $8\times8$ units. 
As presented in Figure \ref{fig:cdef_directions}, 8 preset directions are defined by rotating and reflecting the three shown templates.  
The decoder uses the reconstructed pixels to select the prevalent direction index by minimizing
\begin{equation}
E_d^2 = \sum_k \sum_{p\in P_{d, k}} (x_p - \mu_{d, k})^2,
\end{equation}
where $x_p$ is the value of pixel $p$, $P_{d,k}$ are the pixels in line $k$ following direction $d$, and $\mu_{d,k}$ is the mean value of $P_{d,k}$:
\begin{equation}
\mu_{d,k} = \frac{1}{|P_{d, k}|}\sum_{p\in P_{d, k}}x_p.
\end{equation}

\begin{figure}[!t]
\includegraphics[width=0.98\columnwidth]{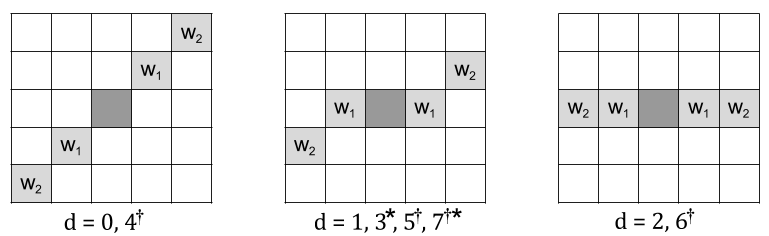}
\centering
\caption{The primary filter templates associated with direction $d\in\{0, \cdots, 7\}$ (subject to rotation and reflection), where $w_1=4/16$ and $w_2=2/16$ for even strength indexes, and $w_1=w_2=3/16$ for odd strength indexes.}
\label{fig:cdef_primary}
\end{figure}

\begin{figure}[!t]
\includegraphics[width=0.98\columnwidth]{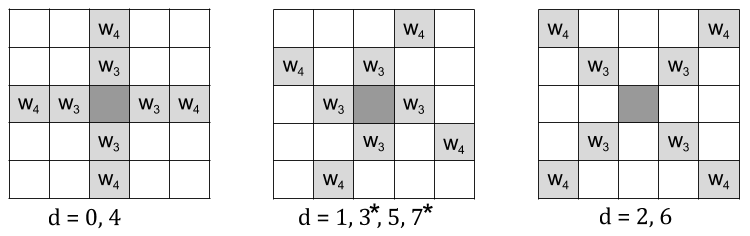}
\centering
\caption{The secondary filter templates associated with each direction (subject to reflection), where  $w_3 = 2/16$ and $w_4=1/16$. The secondary filter is applied along the direction 45\textdegree off the corresponding primary direction $d$.}
\label{fig:cdef_secondary}
\end{figure}

A primary filter is applied along the selected direction, whilst a secondary filter is applied along the direction oriented 45\textdegree~off the primary direction. The filter operation for pixel $p(x, y)$ is formulated by
\begin{align*}
\hat{p}(x, y) = p(x, y) &+ \sum_{m,n}w^p_{d, m, n} f(p(m, n) - p(x, y), S^p, D) \\
&+ \sum_{m,n}w^s_{d, m, n} f(p(m, n) - p(x, y), S^s, D),
\end{align*}
where $w^p_{d, m, n}$ and $w^s_{d, m, n}$ are the filter coefficients associated with the primary and secondary filters, respectively, as shown in Figure \ref{fig:cdef_primary} and \ref{fig:cdef_secondary}. $S_p$ and $S_s$ are the strength indexes for the primary and secondary filters, and $D$ is the damping factor. The $f()$ is a piece-wise linear function:
\begin{align*}
&f(diff, S, D) = \\
&\begin{cases}
min(diff, max(0, S - \lfloor \frac{diff}{2^{D - \lfloor log_2 S \rfloor}} \rfloor)), ~if~ diff > 0 \\
max(diff, min(0, \lceil \frac{diff}{2^{D - \lceil log_2 S \rceil}} \rceil)) - S, ~otherwise
\end{cases}
\end{align*}
that rules out reference pixels whose values are far away from $p(x, y)$. Note that the reference pixels $p(m, n)$ are the reconstructed pixels after the deblocking filter is applied, but before application of the CDEF filter.

Up to 8 groups of filter parameters, which include the primary and secondary filter strength indexes of luma and chroma components, are signaled in the frame header. Each $64\times64$ block selects one group from the presets to control its filter operations.

\subsection{Loop Restoration Filter}
\label{sec:lrf}
The loop restoration filter is applied to units of either $64\times 64$, $128\times128$, or $256\times256$ pixel blocks, named loop restoration units (LRU). Each unit can independently select either to bypass filtering, to use a Wiener filter, or to use a self-guided filter \cite{lrf}. It is applied to the reconstructed pixels after any prior in-loop filtering stages.

\subsubsection{Wiener Filter}
\begin{figure}[!t]
\includegraphics[width=0.7\columnwidth]{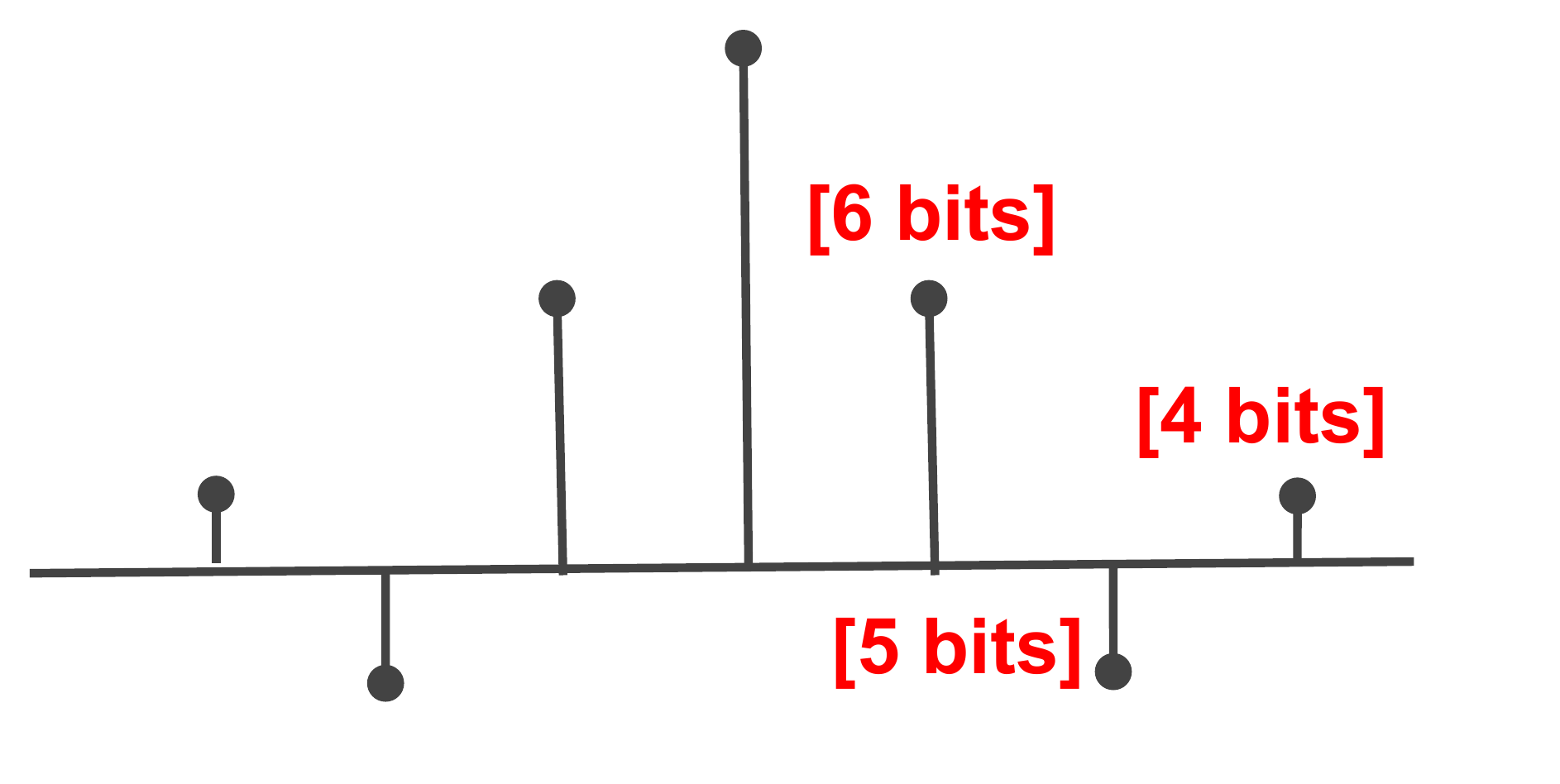}
\centering
\caption{The bit precision for Wiener Filter parameters.}
\label{fig:wiener_filter_params}
\end{figure}

A $7\times7$ separable Wiener filter is applied through the LRU. The filter parameters for the vertical and horizontal filters are decided by the encoder and signaled in the bit-stream. Due to symmetric and normalization constraints, only 3 coefficients need to be sent for each filter. Also note that the Wiener filters are expected to have a higher weight magnitude towards the origin, so the codec reduces the number of bits spent on higher tap coefficients, as shown in Figure \ref{fig:wiener_filter_params}.

\subsubsection{Self-Guided Filter}
\begin{figure}[!t]
\includegraphics[width=0.98\columnwidth]{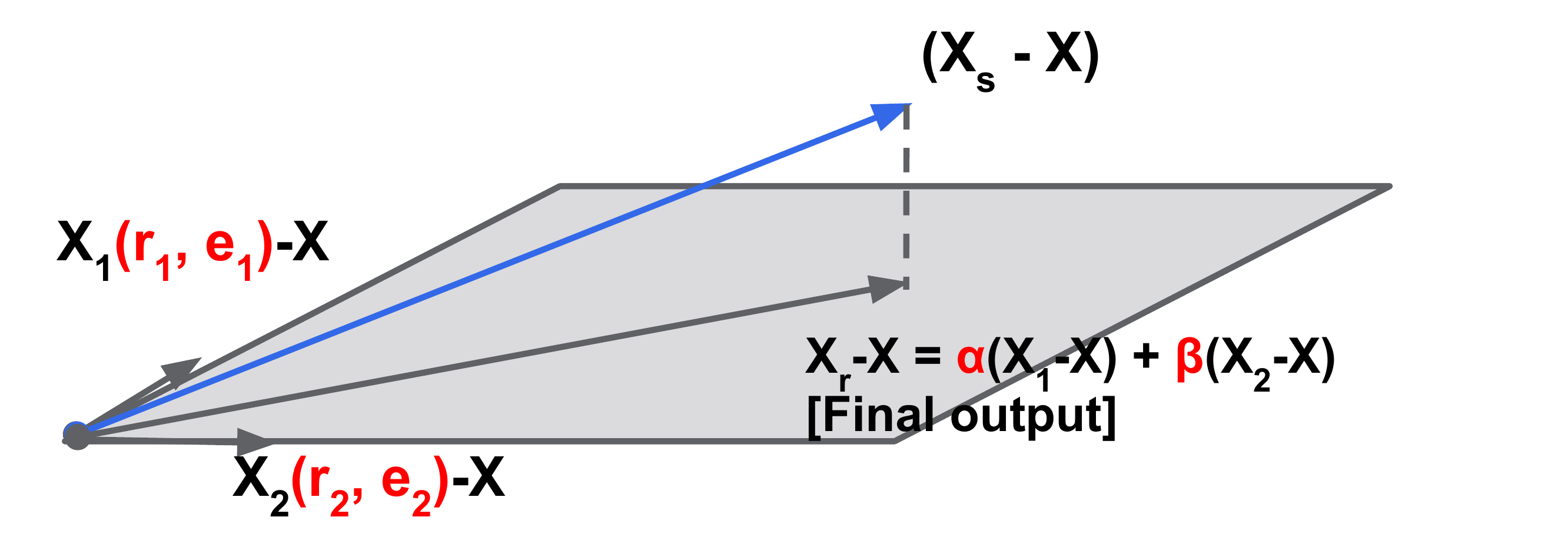}
\centering
\caption{Project the gap between the source pixels $X_s$ and reconstructed pixels $X$ on to a sub-space spanned by simple denoising results, $X_1 - X$ and $X_2 - X$. The parameters in red are the ones configurable through bit-stream syntax.}
\label{fig:self_guided_filter}
\end{figure}

The scheme applies simple filters to the reconstructed pixels, $X$, to generate two denoised versions, $X_1$ and $X_2$, which largely preserve the edge transition. Their differences from the reconstructed pixels, $(X_1 - X)$ and $(X_2 - X)$, are used to span a sub-space, upon which we project the differences between the reconstructed pixels and the original pixels, $(X_s - X)$, as shown in Figure \ref{fig:self_guided_filter}. The least-square regression parameters obtained by the encoder are signaled to the decoder, which are used to build a linear approximation of $(X_s - X)$ based on the known bases $(X_1 - X)$ and $(X_2 - X)$.

In particular, a radius $r$ and a noise variance $e$ are used to generated the denoised versions of the LRU as follows:
\begin{enumerate}
  \item Obtain the mean $\mu$ and variance $\sigma^2$ of pixels in a \mbox{$(2r+1)\times(2r+1)$} window around every pixel x.
  \item Compute the denoised pixel as
  \begin{equation}
    \hat{x} = \frac{\sigma^2}{\sigma^2 + e} x + \frac{e}{\sigma^2 + e} \mu.
  \end{equation}
\end{enumerate}
The pair $(r, e)$ effectively controls the denoising filter strength. Two sets of denoised pixels, denoted in the vector form $X_1$ and $X_2$, are generated using $(r_1, e_1)$ and $(r_2, e_2)$, which are selected by the encoder and are signaled in the bit-stream. Let $X$ denote the vector formed by the reconstructed pixels and $X_s$ the vector of source pixels. The self-guided filter is formulated by
\begin{equation}
X_r = X + \alpha(X_1 - X) + \beta(X_2 - X).
\label{eq:lrf}
\end{equation}
The parameters ($\alpha$, $\beta$) are obtained by the encoder using least square regression:
\begin{equation}
\begin{bmatrix}
\alpha \\
\beta
\end{bmatrix}
= (A^TA)^{-1}A^Tb,
\end{equation}
where
\begin{equation*}
A = 
\begin{bmatrix}
X_1 - X \\
X_2 - X
\end{bmatrix} 
\text{and~}
b = X_s - X.
\end{equation*}
The parameters $(\alpha, \beta)$ are sent to the decoder to formulate \eqref{eq:lrf}.

\subsection{Frame Super-Resolution}
\label{sec:superres}
\begin{figure}[!t]
\includegraphics[width=0.98\columnwidth]{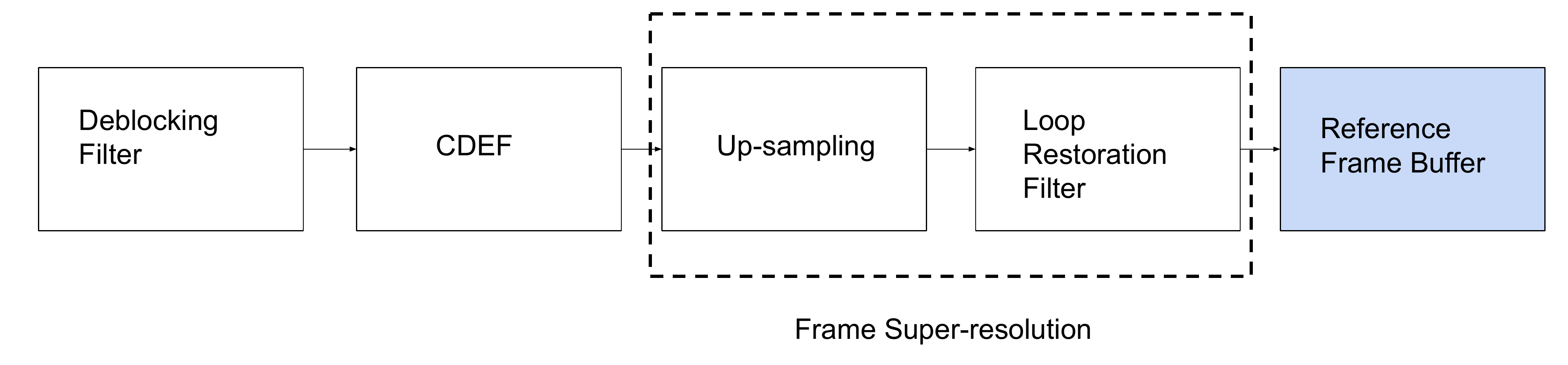}
\centering
\caption{The frame super-resolution up-samples the reconstructed frame to the original dimension. It comprises a linear up-sampling and a loop restoration filter.}
\label{fig:frame_superres_flow}
\end{figure}

When the source input is down scaled from the original video signal, a frame super-resolution is natively supported as part of the post-processing filtering that converts the reconstructed frame to the original dimension. As shown in Figure \ref{fig:frame_superres_flow}, the frame super-resolution consists of an up-sampling stage and a loop restoration filter \cite{superres}.

\begin{figure}[!t]
\includegraphics[width=0.98\columnwidth]{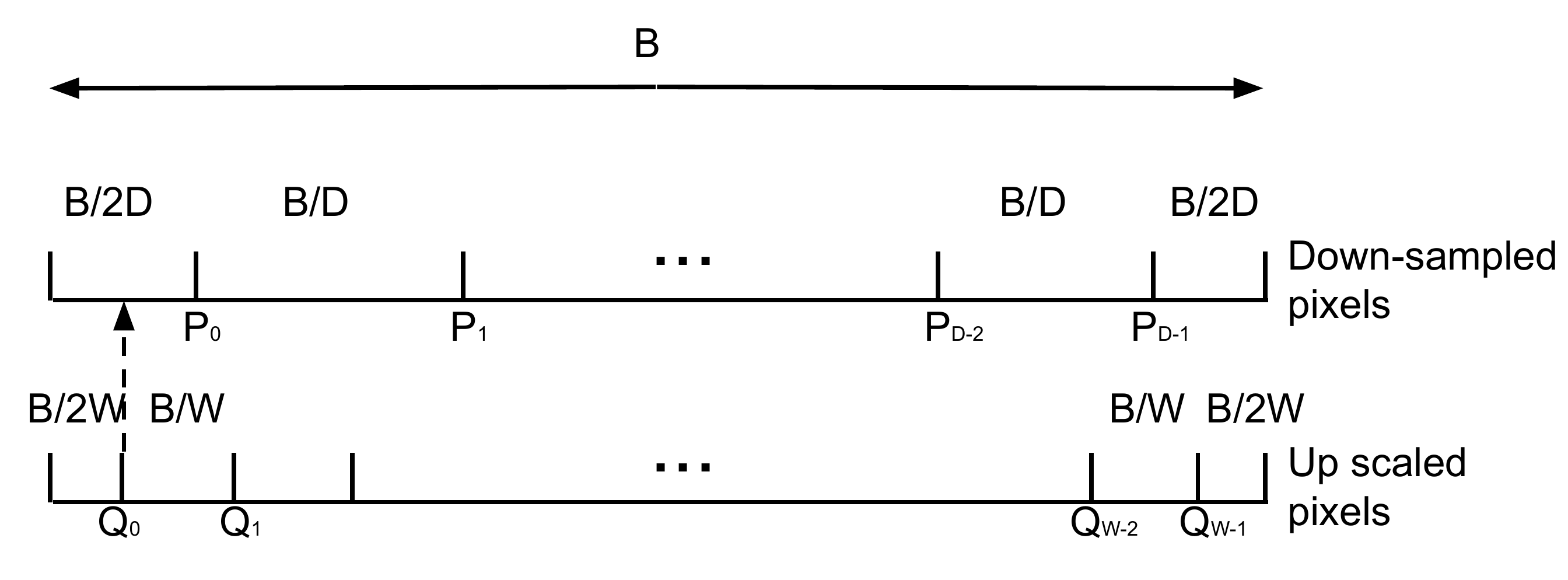}
\centering
\caption{Frame super-resolution sampling positions. The analog frame width is denoted by B. The down-sampled frame contains $D$ pixels in a row, which are used to interpolate $W$ pixels for a row in the up-scaled frame.}
\label{fig:frame_superres}
\end{figure}

The up-sampling stage is applied to the reconstructed pixels after the CDEF filter. As mentioned in Section \ref{sec:frame_scaling}, the down-sampling and up-sampling operations only apply to the horizontal direction. The up-sampling process for a row of pixels in a frame is shown in Figure \ref{fig:frame_superres}. Let B denote the analog frame width. The down-sampled frame contains D pixels in a row, and the up-scaled frame contains W pixels in a row. Their sampling positions are denoted by $P_k$ and $Q_m$ respectively, where $k\in\{0, 1, \cdots, D-1\}$ and $m\in\{0, 1, \cdots, W-1\}$. Note that $P_0$ and $Q_0$ are located at $\frac{B}{2D}$ and $\frac{B}{2W}$, respectively. After normalizing the relative distance by $\frac{B}{D}$, which corresponds to one full-pixel offset in the down-sampled frame, it is straight-forward to show that the offset of $Q_m$ from $P_0$ is:
\begin{equation}
Q_m - P_0 = \frac{D-W}{2W} + m \Delta_Q,
\end{equation}
where $\Delta_Q = \frac{D}{W}$. 

In practice, these offsets are calculated at $\frac{1}{16384}$ pixel precision. They are rounded to the nearest $\frac{1}{16}$-pixel position for interpolation filter. An 8-tap FIR filter is used to generate the sub-pixel interpolation. Note that the rounding error
\begin{equation}
e = \text{round}(\Delta_Q) - \Delta_Q
\end{equation}
is built up in the offset for $Q_m$, i.e., $\frac{D - W}{2W} + m(\Delta_Q + e)$, as $m$ increases from 0 to $W-1$. Here the function round() maps a variable to the nearest sample in $\frac{1}{16384}$ resolution. This would make the left-most pixel in a row have minimum rounding error in the offset calculation, whereas the right-most pixel has the maximum rounding error. To resolve such spatial bias, the initial offset for $Q_0$ is further adjusted by $-\frac{eW}{2}$, which makes the left- and right-most pixels have equal magnitude of rounding error, and the middle pixel $Q_{W/2}$ close to zero rounding error. In summary the adjusted offset of $Q_m$ from $P_0$ is
\begin{equation}
Q_m~\text{offset} = \frac{D - W}{2W} - \frac{eW}{2} + m~\text{round}(\Delta_Q).
\end{equation}

The loop restoration filter in Section \ref{sec:lrf} is then applied to the upsampled frame to further recover the high frequency components. It is experimentally shown in \cite{superres} that the loop restoration filter whose parameters are optimized by the encoder can substantially improve the objective quality of the up-sampling frame.

\subsection{Film Grain Synthesis}
Film grain is widely present in creative content, such as movie and TV materials. Due to its random nature, the film grain is very difficult to compress using conventional coding tools that exploit signal correlations. AV1 provides a film grain synthesis option that builds a synthetic grain and adds it to the decoded picture prior to its display. This allows one to remove the film grain from the source video signal prior to compression. A set of model parameters are sent to the decoder to create a synthetic grain that mimics the original film grain.

\begin{figure}[!t]
\includegraphics[width=0.8\columnwidth]{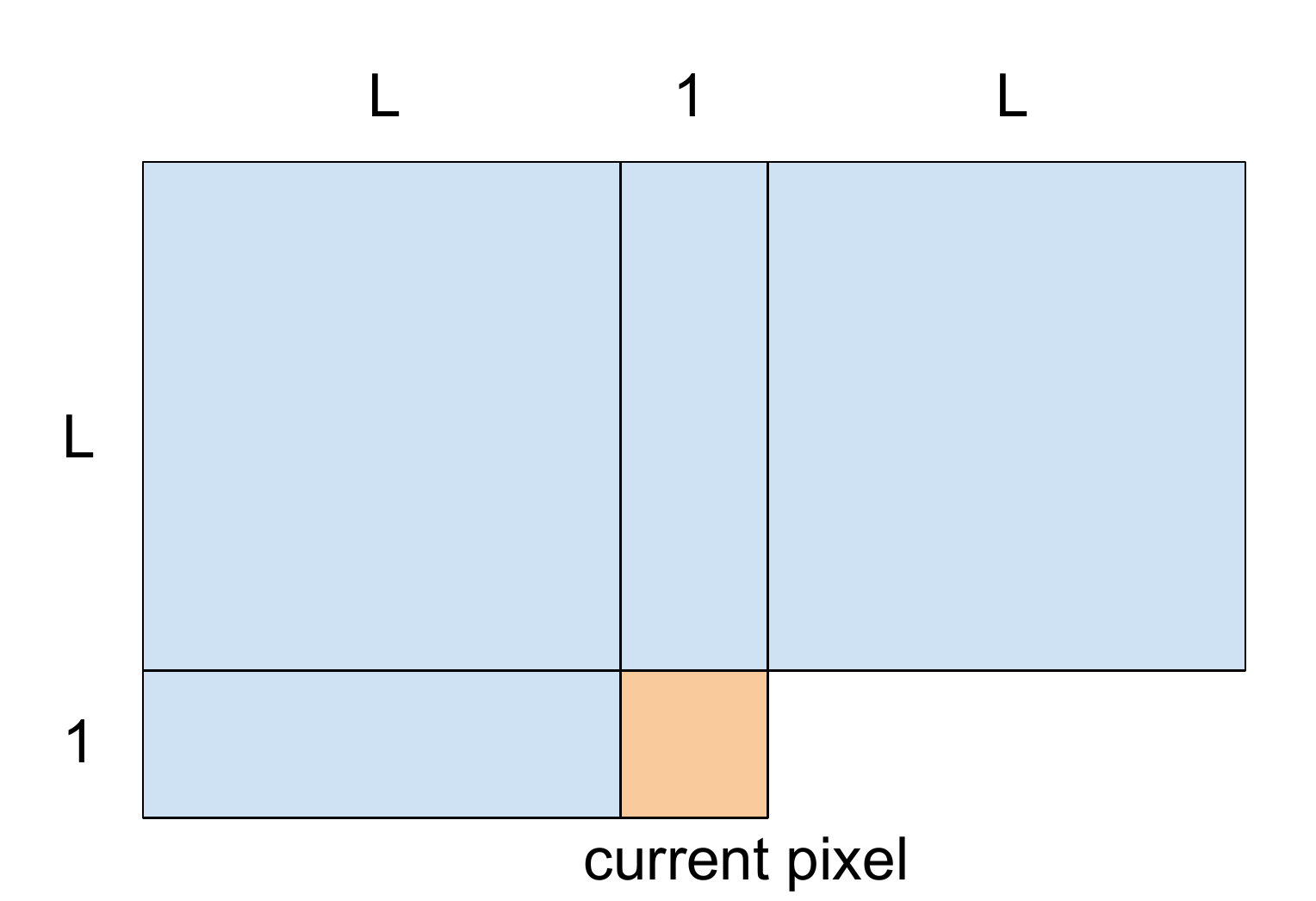}
\centering
\caption{The reference region (in blue) is used by the AR model to generate the grain at a current sample (in orange). The reference region includes a $(2L + 1)\times L$ block above and an $L\times 1$ block to the left. The total number of reference samples is $2L(L + 1)$.}
\label{fig:film_grain_ref}
\end{figure}

AV1 adopts an auto-regressive (AR) model to build the grain signal \cite{film_grain}. The grain samples are generated in raster scan order. A grain sample in luma plane is generated using a $(2L + 1)\times L$ block above and an $L\times 1$ block to the left, as shown in Figure \ref{fig:film_grain_ref}, which involves $2L(L+1)$ reference samples, where $L\in\{0, 1, 2, 3\}$. The AR model is given by
\begin{equation}
G(x, y) = \sum_{m, n \in S_{ref}}a_{m, n}G(x - m, y - n) + z,
\label{eq:film_grain_ar}
\end{equation}
where $S_{ref}$ is the reference region and $z$ is a pseudo random variable that is drawn from a zero-mean unit-variance Gaussian distribution. The grain samples for chroma components are generated similar to \eqref{eq:film_grain_ar} with one additional input from the collocated grain sample in the luma plane. The model parameters associated with each plane are transmitted through the bit-stream to formulate the desired grain patterns. 

The AR process is used to generate a template of grain samples corresponding to a $64\times64$ pixel block. Patches whose dimensions correspond to a $32\times32$ pixel block are drawn at pseudo random positions within this template and are applied to the reconstructed video signal. 

The final luma pixel at position $(x, y)$ is given by
\begin{equation}
\hat{P}(x, y) = P(x, y) + f(P(x, y)) G(x, y),
\end{equation}
where $P(x, y)$ is the decoded pixel value and $f(P(x, y))$ scales the grain sample according to the collocated pixel intensity. The $f()$ is a piece-wise linear function and is configured by the parameters sent through the bit-stream. The grain samples applied to the chroma components are scaled based on the chroma pixel value as well as the collocated luma pixel values. A chroma pixel is given by
\begin{align}
\hat{P_u}(x, y) &= P_u(x, y) + f(t) G_u(x, y), \\
t &= b_uP_u(x, y) + d_u\bar{P}(x,y) + h_u,
\end{align}
where $\bar{P}(x,y)$ denotes the average of the collocated luma pixels. The parameters $b_u$, $d_u$, and $h_u$ are signaled in the bit-stream for each chroma plane.

The film grain synthesis model parameters are decided on a frame by frame basis and are signaled in the frame header. AV1 also allows a frame to re-use the previous frame's model parameter set and bypass sending a new set in the frame header.

\section{Profile and Level Definition}
\label{sec:leveldef}
AV1 defines profiles and levels to specify the decoder capability. Three profiles define support for various bit-depth and chroma sampling formats, namely \textit{Main}, \textit{High} and \textit{Professional}. The capability required for each profile is listed in Table \ref{table:profiles}.

\begin{table}[!ht]
\renewcommand{\arraystretch}{1.3}
\caption{Capability Comparisons of AV1 Profiles}
\label{table:profiles}
\centering
\begin{tabular}{|c|c|c|c|c|c|c|c|}
\hline
Proflile& \multicolumn{3}{c|}{Bit-depth} & \multicolumn{4}{c|}{Chroma sampling}\\
\cline{2-8}
 & 8 & 10 & 12 &  4:0:0 & 4:2:0 & 4:2:2 & 4:4:4\\
\hline
Main & \checkmark & \checkmark & & \checkmark & \checkmark & & \\
\hline
High & \checkmark & \checkmark & & \checkmark & \checkmark & & \checkmark \\
\hline
Professional & \checkmark & \checkmark & \checkmark & \checkmark & \checkmark & \checkmark & \checkmark \\
\hline
\end{tabular}
\end{table}

Levels are defined to quantify the decoder performance in terms of maximum bit rate, maximum samples per picture, and other characteristics, as shown in \cite{bitstream-spec}. A decoder that supports a given level should be capable of processing all bit-streams that conform to the specifications provided by the level definition. To account for various coding structure and rate allocation strategies that might be used to create a bit-stream, a decoder model that describes the smoothing buffer, decoding process, frame buffering, and display process is provided to verify that a bit-stream meets the level definitions \cite{bitstream-spec} \cite{decoder_model}.

\section{Performance Evaluation}
We compared the compression performance of libvpx VP9 \cite{libvpx} and libaom AV1 \cite{libaom}. The source code of libvpx VP9 can be accessed at \cite{libvpx}. The experiment used the version with git hash ebac57ce. The source code of libaom AV1 can be found at \cite{libaom}. The experiment used the version with git hash ac2c30ba. 

Both codecs used the default 2-pass encoding mode and variable bit-rate control, and ran at the highest compression performance mode, i.e., --cpu-used=0. To achieve the compression performance, both VP9 and AV1 encoder allowed adaptive GOP size, where the decisions were made based on the first pass encoding statistics. The quantization parameter offsets between different frames within a GOP were also adaptively optimized based on the first pass coding statistics. The test sets included video resolutions ranging from  480p to 1080p. All the clips were coded using their first 150 frames. The BD-rate reductions in overall PSNR and SSIM are shown in Tables \ref{table:midres}-\ref{table:hdres}. 

To evaluate the relative encoding and decoding complexity, we gathered the instruction counts for both the encoding and decoding processes on single thread at each operating point. The average ratios between AV1 and VP9 are shown in column ``Enc Ins. Count'' and ``Dec Ins. Count'' in Tables \ref{table:midres}-\ref{table:hdres} to reflect the relative encoding and decoding complexity, respectively. The average AV1 encoding complexity is roughly 34.6 -- 39.6 times of the VP9 encoding complexity, both at their high compression performance. The average AV1 decoding complexity on average is about 3 times of VP9 decoding complexity. Note that we use the instruction count to evaluate the codec complexity, since it closely tracks the actual runtime on the same computer and is largely invariant in a cloud computing environment.

We next evaluated the intra coding performance, where all the 150 frames were coded as intra frames with the same quantization parameter. The same quantization parameter set was used for both all intra and video coding modes. The compression efficiency in overall PSNR and SSIM BD-rate reductions are shown in Tables \ref{table:kf_midres}-\ref{table:kf_hdres}. Similarly the relative encoding and decoding complexity between AV1 and VP9 in intra coding mode are provided in column ``Enc Ins. Count'' and ``Dec Ins. Count'' in Tables \ref{table:kf_midres}-\ref{table:kf_hdres}, respectively.

Note that the results are intended for reference on the scale of the relative coding gains and complexity. Different encoder implementations might have different compression and complexity trade off considerations, and different performance results. An extensive codec performance evaluation under various encoder constraints is beyond the scope of this paper. Readers are referred to \cite{AV1-ATSIP} for more comparison results under encoder constraints. Also note that a dedicated decoder implementation might be further optimized for decoding complexity reduction.

\section{Conclusion}
This paper provides a technical overview of the AV1 codec. It outlines the design theories of the compression techniques and the considerations for hardware feasibility, which together define the AV1 codec.

\section{Acknowledgement}
The AV1 codec includes contributions from the entire AOMedia teams \cite{AOMedia} and the greater eco-system around the globe. An incomplete contributor list can be found at \cite{libaom-contributor}. The authors sincerely appreciate the constructive feedback from the anonymous reviewers and the associated editors, which substantially improves the quality of this manuscript.

\begin{table}[!t]
\renewcommand{\arraystretch}{1.3}
\caption{Compression performance comparison - mid resolution. The average ratios between AV1 and VP9 instruction counts for encoding and decoding are shown in column “Enc Ins. Count” and “Dec Ins. Count”, respectively.}
\label{table:midres}
\centering
\begin{varwidth}{0.75\textwidth}
\begin{tabular}{|c|c|c|c|c|}
\hline
Clip & Overall  & SSIM & Enc Ins. & Dec Ins. \\
       &  PSNR     &          & Count & Count \\ \hline
BQMall\_832x480 &	-34. &	-36.7 &	32.2 &	2.8 \\
BalloonFestival\_854x480 &	-35.3 &	-39.4 &	68.9 &	2.5 \\
BasketballText\_832x480 &	-35.8 &	-38.0 &	30.2 &	3.0 \\
Basketballl\_832x480 &	-36.1 &	-40.1 &	31.4 &	2.9 \\
Campfire\_854x480 &	-30.0 &	-38.3 &	81.9 &	3.0 \\
CatRobot\_854x480 &	-37.7 &	-38.0 &	31.9 &	2.8 \\
DaylightRoad2\_854x480 &	-29.4 &	-28.9 &	36.2 &	3.0 \\
Drums\_854x480 &	-36.8 &	-38.6 &	28.5 &	2.5 \\
Flowervase\_832x480 &	-35.2 &	-37.0 &	38.2 &	3.2 \\
Keiba\_832x480 &	-30.2 &	-32.7 &	27.2 &	2.9 \\
Market3Clipr2\_854x480 &	-36.8 &	-38.1 &	32.4 &	2.8 \\
Mobisode2\_832x480 &	-39.8 &	-39.4 &	73.3 &	3.0 \\
NetflixNarrator\_850x480 &	-35.3 &	-38.1 &	31.4 &	2.4 \\
PartyScene\_832x480 &	-33.4 &	-37.3 &	30.6 &	2.6 \\
RaceHorses\_832x480 &	-28.7 &	-32.2 &	29.9 &	2.9 \\
ShowGirl2\_854x480 &	-27.7 &	-30.1 &	33.1 &	2.8 \\
Tango\_854x480 &	-32.0 &	-32.1 &	30.9 &	3.1 \\
ToddlerFountain854x480 &	-17.8 & 	-21.0 &	41.0 &	3.0 \\
TrafficFlow\_854x480 & -38.6 &	-39.8 &	29.2 &	2.6 \\
aspen\_480p &	-31.0 &	-31.3 &	29.7 &	2.9 \\
blue\_sky\_480 &	-39.8 &	-43.7 &	27.1 &	3.5 \\
city\_4cif &	-33.4 &	-33.8 &	29.2 &	2.8 \\
controlled\_burn\_480p &	-36.7 &	-35.4 &	28.5 &	2.3 \\
crew\_4cif &	-24.7 &	-25.2 &	35.3 &	3.5 \\
crowd\_run\_480p &	-24.9 &	-30.9 &	30.5 &	2.6 \\
ducks\_take\_off\_480p &	-36.3 &	-38.7 &	36.4 &	2.6 \\
harbour\_4cif &	-29.0 &	-32.1 &	33.5 &	2.8 \\
ice\_4cif &	-27.0 &	-31.8 &	30.5 &	3.2 \\
into\_tree\_480p &	-35.5 &	-33.6 &	47.9 &	2.7 \\
netflix\_aerial &	-47.3 &	-47.5 &	29.8 &	2.7 \\
netflix\_barscene &	-37.6 &	-37.2 &	30.7 &	2.4 \\
netflix\_driving &	-32.9 &	-33.0 &	34.6 &	2.6 \\
netflix\_foodmarket &	-29.2 &	-29.6 &	33.4 &	2.8 \\
netflix\_ritualdance &	-27.0 &	-32.5 &	49.3 &	3.1 \\
netflix\_rollercoaster &	-33.7 &	-34.1 &	31.8 &	3.4 \\
netflix\_square & 	-32.7 &	-32.3 &	30.5 &	2.5 \\
netflix\_tunnelflag &	-37.2 &	-45.6 &	22.7 &	2.7 \\
old\_town\_480p &	-34.7 &	-34.9 &	41.8 &	2.4 \\
park\_joy\_480p &	-27.1 &	-30.0 &	31.2 &	2.5 \\
red\_kayak\_480p &	-14.0 &	-14.4 &	36.6 &	3.4 \\
rush\_field\_480p &	-30.1 &	-28.9 &	34.2 &	3.0 \\
shields\_640x360 & -37.2 &	-37.2 &	30.3 &	2.7 \\
sintel\_shot\_854x480 &	-30.6 &	-33.9 &	37.8 &	3.6 \\
snow\_mnt\_480p &	-33.9 &	-30.2 &	60.1 &	2.3 \\
soccer\_4cif & 	-34.7 &	-40.7 &	27.6 &	2.8 \\
speed\_bag\_480p & 	-44.7 &	-49.0 &	28.6 &	3.7 \\
station2\_480p &	-57.6 &	-56.7 &	32.4 &	2.9 \\
tears\_of\_steel\_480p &	-27.5 &	-30.5 &	34.0 &	3.2 \\
touchdown\_pass\_480p & 	-25.3 &	-30.2 &	34.9 &	3.0 \\
west\_wind\_480p & 	-23.6 & 	-20.8 & 	34.2 &	3.3 \\
\hline
OVERALL & 	-33.0 &	-34.8 & 	34.6 &	2.9  \\
\hline
\end{tabular}
\end{varwidth}
\end{table}

\begin{table}[!t]
\renewcommand{\arraystretch}{1.3}
\caption{Compression performance comparison - high resolution. The average ratios between AV1 and VP9 instruction counts for encoding and decoding are shown in column “Enc Ins. Count” and “Dec Ins. Count”, respectively.}
\label{table:hdres}
\centering
\begin{varwidth}{0.75\textwidth}
\begin{tabular}{|c|c|c|c|c|}
\hline
Clip & Overall  & SSIM & Enc Ins. & Dec Ins. \\
       &  PSNR     &          & Count &  Count \\
\hline
BalloonFestival\_1280x720 &	-36.2	 & -43.0	& 88.2 &	2.5 \\
CSGO\_1080p &	-29.2 &	-30.7 &	36.9 &	3.9 \\
Campfire\_1280x720 &	-31.8	 & -40.7 &	102.7 &	3.2 \\
CatRobot\_1280x720 &	-39.5	 & -39.4 &	37.3 & 	3.0 \\
DaylightRoad2\_1280x720 &	-30.1 & 	-29.8	 & 42.5 & 	3.1 \\
Drums\_1280x720 & 	-37.0 & 	-38.9	 & 34.3 & 	2.6 \\
Market3Clipr2\_1280x720 &	-38.4	 & -40.6 & 	37.6 & 	2.6 \\
Netflix\_Aerial\_2048x1080 & 	-31.4 & 	-33.0 &	36.8 & 	2.8 \\
Netflix\_Boat\_2048x1080 &	-37.6	 & -38.6 & 	33.2 &	2.4 \\
Netflix\_Crosswalk\_2048x1080 &	-33.2 & 	-35.3	 & 37.7 &	2.5 \\
Netflix\_Dancers\_1280x720 & 	-38.2	 & -41.8 &	44.6	 & 3.2 \\
Netflix\_DrivingPOV\_2048x1080 &	-29.8 & -30.1 & 	43.2	 & 2.9 \\
Netflix\_FoodMarket2\_1280x720 &	-33.4 &	-35.0 &	35.8 &	2.5 \\
Netflix\_FoodMarket\_2048x1080 &	-30.3	 & -33.0 & 	43.2	 & 3.1 \\
Netflix\_PierSeaside\_2048x1080 &	-46.4 &	-42.1 &	38.3	 & 2.9 \\
Netflix\_Square\_2048x1080 &	-34.0	 & -34.8 & 	29.6 & 	2.8 \\
Netflix\_TunnelFlag\_2048x1080 &	-39.3	 & -42.4 & 	25.5 & 	3.0 \\
RollerCoaster\_1280x720 & 	-32.4	 & -34.9 & 	35.8 & 	3.2 \\
ShowGirl2\_1280x720 &	-28.6	 & -32.1 &	41.9 & 	3.0 \\
Tango\_1280x720 &	-32.7 & 	-32.7	 & 36.8 & 	3.3 \\
ToddlerFountain1280x720 &	-17.9	 & -20.1 &	47.5	 & 3.2 \\
TrafficFlow\_1280x720 &	-39.7 &	-40.5	 & 36.7 &	2.7 \\
aspen\_1080p &	-41.5	 & -45.0 & 	32.9 & 	3.3 \\
basketballdrive\_1080p &	-32.5	 & -36.2	 & 41.4 & 	3.4 \\
cactus\_1080p &	-38.2	 & -38.3 & 	43.8 & 	3.0 \\
city\_720p &	-38.0 &	-37.6	 & 38.4 & 	2.6 \\
controlled\_burn\_1080p &	-39.3	 & -42.6 &	41.2 & 	2.7 \\
crew\_720p &	-25.4	 & -28.2 & 	44.3 & 	3.7 \\
crowd\_run\_1080p &	-27.0	 & -30.8 & 	31.4 & 	2.8 \\
dinner\_1080p30 &	-37.5 & 	-40.2	 & 34.0 &	3.4 \\
ducks\_take\_off\_1080p &	-33.1 &	-35.2 &	40.8	 & 2.8 \\
factory\_1080p &	-33.6 &	-40.6 &	29.6 & 	3.2 \\
in\_to\_tree\_1080p &	-26.7	 & -22.8 & 	56.2	 & 2.9 \\
johnny\_720p & 	-42.9 & 	-43.8 & 	43.9 & 	2.4 \\
kristenandsara\_720p &	-40.6	 & -38.4	& 44.0 & 	2.8 \\
night\_720p &	-34.8	 & -36.6 & 	41.1 & 	3.0 \\
old\_town\_cross\_720p & 	-38.4 & 	-37.6	 & 55.5 & 	2.5 \\
parkjoy\_1080p &	-27.2 	& -34.0 &  	31.9 & 	2.7 \\
ped\_1080p &	-32.9	 & -38.2 & 	35.8 & 	3.3 \\
red\_kayak\_1080p &	-16.1	 & -14.7 & 	37.5 & 	3.5 \\
riverbed\_1080p &	-17.4	 & -16.9	& 35.5 & 3.6 \\
rush\_field\_1080p &	-27.2	 & -30.0 & 	34.5 & 	3.2 \\
rush\_hour\_1080p & 	-26.2	 & -34.5 & 	47.9 & 	3.7 \\
shields\_720p &	-45.5	 & -44.5 & 	44.8 & 	2.8 \\
station2\_1080p &	-56.4 & 	-56.4	 & 45.6 &	3.9 \\
sunflower\_720p &	-45.5	 &  -49.6 & 	26.6 & 	3.2 \\
tennis\_1080p & 	-33.4	 & -37.9 & 	32.8 & 	3.3 \\
touchdown\_1080p &	-29.6	 & -35.8 & 	37.8 & 	3.4 \\
tractor\_1080p &	-34.9	 & -39.6 & 	31.6 &	2.9 \\
vidyo4\_720p & 	-38.8 & 	-42.1 &	41.7 &	2.9 \\
\hline
OVERALL &	-34.2	 & -36.4 & 	39.6 &	3.1 \\
\hline
\end{tabular}
\end{varwidth}
\end{table}

\begin{table}[!t]
\renewcommand{\arraystretch}{1.3}
\caption{Intra frame compression performance comparison - mid resolution. The average ratios between AV1 and VP9 instruction counts for encoding and decoding are shown in column “Enc Ins. Count” and “Dec Ins. Count”, respectively.}
\label{table:kf_midres}
\centering
\begin{varwidth}{0.75\textwidth}
\begin{tabular}{|c|c|c|c|c|}
\hline
Clip & Overall  & SSIM & Enc Ins. & Dec Ins. \\
       &  PSNR     &          & Count &  Count \\
\hline
BQMall\_832x480 &	-21.5	 & -21.6 &	9.6 & 	1.9 \\
BalloonFestival\_854x480 &	-21.5	 & -23.4 & 	14.1 &	2.0 \\
BasketballText\_832x480 &	-30.5	 & -26.9 & 	9.6 & 	2.0 \\
Basketball\_832x480 &	-31.6	 & -28.1 & 	9.7 & 	1.8 \\
Campfire\_854x480 & 	-31.6 & -28.1 & 	12.6 & 	1.9 \\
CatRobot\_854x480 & 	-23.5 &	-23.6	 & 10.3 &	1.9 \\
DaylightRoad2\_854x480 &	-23.2	 & -23.2 &	10.1	 & 1.9 \\
Drums\_854x480 &	-22.7	 & -21.6 & 	10.5 & 	1.9 \\
Flowervase\_832x480 &	-22.1	 & -22.8	 & 9.0 & 	2.3 \\
Keiba\_832x480 &	-24.2 & 	-21.5	 & 8.9 & 	2.0 \\
Market3r2\_854x480 &	-21.9	 & -20.8 & 	10.2	 & 1.8 \\
Mobisode2\_832x480 &	-32.8	 & -32.4 & 	9.3 & 	2.5 \\
NetflixNarrator\_850x480 &	-21.9 & 	-23.0 & 	9.9 & 	2.1 \\
PartyScene\_832x480 &	-15.5	 & -14.7 & 	11.9 & 	1.7 \\
RaceHorses\_832x480 &	-16.5	 & -13.5 & 	13.4	& 1.7 \\
ShowGirl2\_854x480 &	-20.2	 & -20.7 & 	10.0 & 	2.0 \\
Tango\_854x480 &	-24.5 & 	-24.2	 & 10.4 & 	2.0 \\
ToddlerFountain854x480 &	-16.6 &	-12.8	 & 11.5 & 	1.7 \\
TrafficFlow\_854x480 & 	-31.4	 & -32.0 & 	11.7 & 	1.9 \\
aspen\_480p &	-19.5	 & -18.9 & 	12.6 & 	2.0 \\
blue\_sky\_480p &	-19.5	 & -21.3 & 	14.0 & 	2.0 \\
city\_4cif &	-19.8 & 	-18.1 & 	11.1	& 1.9 \\
controlled\_burn\_480p &	-15.1 & 	-12.5	 & 12.5 & 	1.8 \\
crew\_4cif &	-22.9	 & -21.3 & 	9.9 & 	1.9 \\
crowd\_run\_480p & -13.3	 & -9.9 & 	13.6	 & 1.7 \\
ducks\_480p &	-23.9 & 	-27.4	 & 12.9 & 	1.6 \\
harbour\_4cif &	-19.6	 & -17.8 & 	11.4 & 	2.1 \\
ice\_4cif & 	-27.2 &	-26.4	 &  8.5	 & 2.4 \\
into\_tree\_480p &	-24.1	 & -19.7 & 	9.3 & 	1.7 \\
netflix\_aerial & 	-15.5 & 	-11.5 & 	12.3	 & 1.6 \\
netflix\_barscene &	-22.3	 & -21.0 & 	9.0 & 	2.1 \\
netflix\_driving &	-20.0	 & -20.3 & 	10.5 & 	1.8 \\
netflix\_foodmarket &	-20.2	 & -17.1 & 	10.5	& 1.9 \\
netflix\_ritualdance & 	-22.5 & 	-20.2	 & 11.6 & 	1.9 \\
netflix\_rollercoaster & 	-24.2	 & -24.8 & 	9.7 & 	2.0 \\
netflix\_square &	-16.0	 & -15.0 & 	12.8	& 1.8 \\
netflix\_tunnelflag &	-31.1 &	-39.1	 & 10.9	 & 2.0 \\
old\_town\_480p & 	-19.1 & 	-18.5	 & 10.7 & 	1.7 \\
park\_joy\_480p & -15.4  & 	-13.6	 & 13.9	 & 1.6 \\
red\_kayak\_480p & 	-16.6 & 	-15.5	 & 11.4 & 	2.0 \\
rush\_field\_480p & 	-15.4	 & -14.1	& 12.8	 & 1.8 \\
shields\_640x360 & 	-19.9 & 	-21.5	 & 14.1 &	1.7 \\
sintel\_shot\_854x480 &	-29.9	 & -31.2 & 	10.8 & 	3.1 \\
snow\_mnt\_480p &	-12.4 & 	-8.9 & 	18.9 & 	2.0 \\
soccer\_4cif &	-21.5	 & -17.9 & 	10.5 & 	1.9 \\
speed\_bag\_480p & 	-28.2 &	-33.8	 & 9.2 & 	2.9 \\
station2\_480p & 	-24.3	 & -21.7	& 10.2 & 	1.8 \\
tears\_of\_steel\_480p &	-25.6 & 	-24.4	 & 10.4 & 	1.9 \\
touchdown\_pass\_480p & 	-18.8 & 	-13.7	 & 10.2 & 	2.2 \\
west\_wind\_480p &	-20.2	 & -16.8 & 	11.8 & 	2.1 \\
\hline
OVERALL &	-22.0	 & -21.0 & 	11.1 & 	2.0 \\
\hline
\end{tabular}
\end{varwidth}
\end{table}

\begin{table}[!t]
\renewcommand{\arraystretch}{1.3}
\caption{Intra frame compression performance comparison - hd resolution. The average ratios between AV1 and VP9 instruction counts for encoding and decoding are shown in column “Enc Ins. Count” and “Dec Ins. Count”, respectively.}
\label{table:kf_hdres}
\centering
\begin{varwidth}{0.75\textwidth}
\begin{tabular}{|c|c|c|c|c|}
\hline
Clip & Overall  & SSIM & Enc Ins. & Dec Ins. \\
       &  PSNR     &          & Count &  Count \\
\hline
BalloonFestival\_1280x720 &	-23.4 & 	-25.3	 & 12.2 & 	2.1 \\
CSGO\_1080p &	-23.4 & 	-19.7	 & 9.0	& 2.3 \\
Campfire\_1280x720 &	-32.3 & 	-28.6 & 	11.6 & 	1.9 \\
CatRobot\_1280x720 &	-25.6	 & -26.0 & 	8.9	 & 2.0 \\
DaylightRoad2\_1280x720 &	-24.2	 & -24.1 &	8.5 &	1.9 \\
Drums\_1280x720 &	-23.4	  & -21.7 &	8.8 & 	1.9 \\
Market3r2\_1280x720 &	-21.8 & 	-19.5 & 	8.8	 & 1.8 \\
Netflix\_Aerial\_2048x1080 &	-20.2 & 	-17.5	 & 11.1 & 	2.0 \\
Netflix\_Boat\_2048x1080 & 	-19.6	 & -19.9 & 	12.8 & 	2.0 \\
NetflixCrosswalk\_2048x1080 &	-24.8	 & -25.7 & 	9.0 & 	2.6 \\
Netflix\_Dancers\_1280x720 &	-31.4 & 	-30.5	 & 7.7	 & 3.1 \\
NetflixDrivingPOV2048x1080 &	-22.4	 & -20.4 & 	9.1 & 	2.0 \\
NetflixMarket2\_1280x720 &	-20.1 &	-18.9 & 	10.7 & 	1.8 \\
NetflixMarket\_2048x1080 &	-21.1	 & -17.9 & 	9.5 & 	2.0 \\
NetflixPierSeaside2048x1080 &	-24.7 & 	-21.8	 & 9.1 & 	2.0 \\
Netflix\_Square\_2048x1080 &	-19.7 & 	-18.5 & 	10.5	 & 2.0 \\
NetflixTunnelFlag\_2048x1080 &	-34.4 & 	-35.8	 & 8.7	 & 2.2 \\
RollerCoaster\_1280x720 & 	-24.9	 & -23.9	 & 9.2 & 	2.0 \\
ShowGirl2\_1280x720 &	-21.7	 & -22.1 & 	8.6 & 	2.2 \\
Tango\_1280x720 &	-26.0	 & -25.4 & 	8.9 & 	2.2 \\
ToddlerFountain1280x720 &	-17.1	 & -13.2 & 	9.9 & 	1.8 \\
TrafficFlow\_1280x720 &	-32.6	 & -32.7	& 9.8 & 	2.0 \\
aspen\_1080p &	-22.4 &	-21.5 &	10.1	 & 2.8 \\
basketballdrive\_1080p & 	-30.9	 & -27.8	 & 7.6 & 	1.9 \\
cactus\_1080p &	-24.6 & 	-22.0	 & 8.9	 & 1.7 \\
city\_720p &	-19.7 	& -17.8 &	9.8	 & 2.2 \\
controlled\_burn\_1080p &	-15.9	 & -14.0 &	12.2 &	2.2 \\
crew\_720p &	-24.7 & 	-22.9 & 	8.3 & 	2.2 \\
crowd\_run\_1080p &	-17.0	 & -12.9 & 	11.1	& 1.7 \\
dinner\_1080p & 	-29.6 & 	-28.9 & 	8.7 & 	2.4 \\
ducks\_1080p &	-25.9 &	-27.1 &	9.6 &	 1.6 \\
factory\_1080p &	-23.8 &	-20.3	 & 11.8 &	2.2 \\
in\_to\_tree\_1080p &	-25.0	 & -18.3 & 	8.8	& 1.6 \\
johnny\_720p & 	-28.5	 & -26.7	 & 8.1	 & 2.3 \\
kristenandsara\_720p &	-26.6	 & -25.6 & 	8.3 &	 2.4 \\
night\_720p &	-20.5	 & -19.3	 & 9.6 & 	2.0 \\
old\_town\_720p &	-20.2	 & -18.2 & 	9.6 & 	1.6 \\
parkjoy\_1080p &	-17.7	 & -14.9	 & 12.4 & 	1.7 \\
ped\_1080p &	-29.7	 & -29.2 & 	7.5 & 	2.1 \\
red\_kayak\_1080p &	-18.4	 & -16.4	& 11.4	 & 2.4 \\
riverbed\_1080p &	-18.1 & 	-15.9 & 	10.6 & 	2.2 \\
rush\_field\_1080p & 	-15.4 & 	-14.3 & 	10.7 & 	2.2 \\
rush\_hour\_1080p & 	-26.8 &	-29.7 &	8.3 & 	2.5 \\
shields\_720p &	-20.2 & 	-19.5	 & 10.3 & 	1.7 \\
station2\_1080p &	-24.8	 & -22.3	& 9.2	 & 2.1 \\
sunflower\_720p & 	-29.2 & 	-32.2 & 	10.3 & 	2.6 \\
tennis\_1080p &	-27.5 &	-27.0	 & 7.7 & 	2.1 \\
touchdown\_1080p &	-19.0	 & -15.7 & 	9.2 & 	2.5 \\
tractor\_1080p &	-28.5 & 	-28.9	 & 9.8 & 	2.0 \\
vidyo4\_720p & 	-25.1	 & -24.4	 & 8.3 & 	2.3 \\
\hline
OVERALL &	-23.8	 & -22.5 & 	9.5	 & 2.1  \\
\hline
\end{tabular}
\end{varwidth}
\end{table}

\ifCLASSOPTIONcaptionsoff
  \newpage
\fi

\bibliographystyle{IEEEtran}
\bibliography{av1overview}


\end{document}